%% file: Achernar-binary-v3r2.tex
\documentclass[]{aa}
\usepackage{graphics}
\usepackage{graphicx}
\usepackage{multicol}
\usepackage{longtable}
\usepackage{supertabular}
\usepackage{amssymb}
\usepackage{txfonts}
\usepackage{placeins}
\usepackage{multirow}
\usepackage{babel}

\usepackage{natbib,twoopt}
\usepackage{amsmath} 
\usepackage[breaklinks=true]{hyperref} 
\bibpunct{(}{)}{;}{a}{}{,} 
\makeatletter
\newcommandtwoopt{\citeads}[3][][]{\href{http://adsabs.harvard.edu/abs/#3}%
{\def\hyper@linkstart##1##2{}%
\let\hyper@linkend\@empty\citealp[#1][#2]{#3}}}
\newcommandtwoopt{\citepads}[3][][]{\href{http://adsabs.harvard.edu/abs/#3}%
{\def\hyper@linkstart##1##2{}%
\let\hyper@linkend\@empty\citep[#1][#2]{#3}}}
\newcommandtwoopt{\citetads}[3][][]{\href{http://adsabs.harvard.edu/abs/#3}%
{\def\hyper@linkstart##1##2{}%
\let\hyper@linkend\@empty\citet[#1][#2]{#3}}}
\newcommandtwoopt{\citeyearads}[3][][]%
{\href{http://adsabs.harvard.edu/abs/#3}
{\def\hyper@linkstart##1##2{}%
\let\hyper@linkend\@empty\citeyear[#1][#2]{#3}}}
\makeatother

\newcommand{\half}{H$_{\rm \alpha}$}

\newcommand{\eg}{{e.g.}}
\newcommand{\ie}{{i.e.}}

\newcommand{\m}{$\mu$m}

\newcommand{\kms}{km\,s$^{\rm -1}$}

\newcommand{\Vdiff}{$V_\textrm{diff}$}

\usepackage{color}
\definecolor{mygreen}{RGB}{0,128,0}

\hypersetup{colorlinks=true,linkcolor=blue,citecolor=blue,urlcolor=blue}

\begin{document}

\title{The binary system of the spinning-top Be star Achernar
\thanks{The series of high resolution spectra of Achernar (with the continuum normalized to unity) are available in FITS format at \url{http://dx.doi.org/10.5281/zenodo.6977303} and at the CDS via anonymous ftp to cdsarc.u-strasbg.fr (130.79.128.5) or via \url{http://cdsweb.u-strasbg.fr/cgi-bin/qcat?J/A+A/}.}}
\titlerunning{The binary system of the Be star Achernar}
\authorrunning{P. Kervella et al.}
\author{
Pierre~Kervella\inst{1}
\and
Simon~Borgniet\inst{1}
\and
Armando~Domiciano de Souza\inst{2}
\and
Antoine~M\'erand\inst{3}
\and
Alexandre~Gallenne\inst{4,5}
\and
Thomas~Rivinius\inst{6}
\and
Sylvestre~Lacour\inst{1,3}
\and
Alex~Carciofi\inst{7}
\and
Daniel~Moser~Faes\inst{8}
\and
Jean-Baptiste~Le~Bouquin\inst{9}
\and
Monica~Taormina\inst{10}
\and
Bogumił Pilecki\inst{10}
\and
Jean-Philippe~Berger\inst{9}
\and
Philippe~Bendjoya\inst{2}
\and
Robert~Klement\inst{11}
\and
Florentin~Millour\inst{2}
\and
Eduardo~Janot-Pacheco\inst{7}
\and
Alain~Spang\inst{2}
\and
Farrokh~Vakili\inst{2}
}
\institute{
LESIA, Observatoire de Paris, Universit\'e PSL, CNRS, Sorbonne Universit\'e, Universit\'e Paris Cit\'e, 5 place Jules Janssen, 92195 Meudon, France,
\email{pierre.kervella@observatoiredeparis.psl.eu}
\and
Universit\'e C\^ote d'Azur, Observatoire de la C\^ote d'Azur, CNRS, Lagrange UMR 7293, CS 34229, 06304, Nice Cedex 4, France
\and
European Southern Observatory, Karl-Schwarzschild-Str. 2, 85748 Garching, Germany
\and
Universidad de Concepci\'on, Departamento de Astronom\'ia, Casilla 160-C, Concepci\'on, Chile
\and
Unidad Mixta Internacional Franco-Chilena de Astronom\'ia (CNRS UMI 3386), Departamento de Astronom\'ia,
Universidad de Chile, Camino El Observatorio 1515, Las Condes, Santiago, Chile
\and 
European Southern Observatory, Alonso de Córdova 3107, Casilla 19001, Santiago 19, Chile
\and
Instituto de Astronomia, Geof\'isica e Ci\'encias Atmosf\'ericas, Universidade de S\~ao Paulo, Brazil
\and
National Radio Astronomy Observatory, 1003 Lopezville Road, Socorro, NM 87801, USA
\and
Universit\'e Grenoble Alpes, CNRS, IPAG, 38000 Grenoble, France
\and
Centrum Astronomiczne im. Miko{\l}aja Kopernika, PAN, Bartycka 18, 00-716 Warsaw, Poland
\and
The CHARA Array of Georgia State University, Mount Wilson Observatory, Mount Wilson, CA 91023, USA
}
\date{Received ; Accepted}
\abstract
{Achernar, the closest and brightest classical Be star, presents rotational flattening, gravity darkening, occasional emission lines due to a gaseous disk, and an extended polar wind.
It is also a member of a close binary system with an early A-type dwarf companion. 
}
{We aim to determine the orbital parameters of the Achernar system and to estimate the physical properties of the components.
}
{We monitored the relative position of Achernar B using a broad range of high angular resolution instruments of the VLT/VLTI (VISIR, NACO, SPHERE, AMBER, PIONIER, GRAVITY, and MATISSE) over a period of 13 years (2006-2019).
These astrometric observations are complemented with a series of $\approx 750$ optical spectra for the period from 2003 to 2016.
}
{We determine that Achernar~B orbits the primary Be star on a seven-year period, eccentric orbit ($e=0.7258 \pm 0.0015$) which brings the two stars within 2\,au at periastron.
The mass of the Be star is found to be $m_A = 6.0 \pm 0.6\,M_\sun$ for a secondary mass of $m_B = 2.0 \pm 0.1\,M_\sun$ (the latter was estimated from modeling).
We find a good agreement of the parameters of Achernar A with the evolutionary model of a critically rotating star of $6.4\,M_\sun$ at an age of 63\,Ma.
The equatorial plane of the Be star and the orbital plane of the companion exhibit a relative inclination of $30^\circ$.
We also identify a resolved comoving low-mass star, which leads us to propose that Achernar is a member of the Tucana-Horologium moving group.
}
{The proximity of Achernar makes this star a precious benchmark for stellar evolution models of fast rotators and intermediate mass binaries.
Achernar~A is presently in a short-lived phase of its evolution following the turn-off, during which its geometrical flattening ratio is the most extreme.
Considering the orbital parameters, no significant interaction occurred between the two components, demonstrating that Be stars may form through a direct, single-star evolution path without mass transfer.
Since component A will enter the instability strip in a few hundred thousand years, Achernar appears to be a promising progenitor of the Cepheid binary systems.
}

\maketitle


\section{Introduction}

Classical Be stars are fast-rotating intermediate-mass stars that are occasionally surrounded by an equatorial gaseous disk \citepads{2003PASP..115.1153P, 2013A&ARv..21...69R, 2011AJ....142..149N}.
The formation mechanism of these very fast-rotating stars is however largely uncertain.
As most massive stars, Be stars are commonly found in binary or multiple systems, and two main formation scenarios are considered.
Scenario (1) is based on single-star evolution from an initially fast rotating main sequence hot dwarf. This case also includes the binary stars with companions orbiting with periods larger than a few years, as the wide physical separation from the Be star prevents mass transfer during their evolution.
Scenario (2) explains the spin-up of the Be star progenitor through mass transfer from a close-in companion. In this configuration, the orbital period of the companion is typically a few months. The stripping of the outer layers of the companion leads to its transformation into a hot O subdwarf (sdO). Due to the accretion process, the Be star gains a considerable rotational angular momentum.

A first question to examine is how often Be stars are members of binary systems, and when they are, which fraction of them have wide (noninteracting) and close-in (mass transfer) companions.
A survey of 31 classical Be stars conducted by \citetads{2021ApJS..257...69H} using the NPOI interferometer found a multiplicity frequency of 45\%, including both very wide and close-in systems.
\citetads{2010MNRAS.405.2439O} determined from an adaptive optics survey with the VLT/NACO instrument that approximately one-third of the Be stars in the Milky Way have wide, noninteracting companions.
\citetads{2019ApJ...885..147K} observed a sample of 57 classical Be stars in the radio domain to search for the turndown in spectral energy distribution caused by the truncation of their circumstellar disk by a close-in companion on a short-period orbit.
These authors detect the signature of companions in all the 26 stars with sufficient data coverage, and conclude that many if not all Be stars host close-in companions that influence the outer regions of their circumstellar disk.
Such tight companions may spin up the Be stars through scenario (2), which could therefore be the dominant mechanism for the formation of Be stars.
In support of this hypothesis,  \citetads{2015A&A...577A..51M} determined the orbital elements and masses of the sdO star orbiting the classical Be star $\phi$\,Per on a 4-month period orbit. In this system, the gas disk surrounding the Be star is coplanar with the companion's orbit and the Be star is rotating very close to its critical velocity. The authors conclude that the transfer of material from the companion to the Be star in the past is at the origin of its very high present rotation velocity.
For three similar Be+sdB/sdO subdwarf systems, \citetads{2022ApJ...926..213K} determined the orbital parameters using optical interferometry. With orbital periods between 60 and 250\,days, mass transfer occurred from the subdwarf progenitors to the Be star component, resulting in the acceleration of its rotation velocity as predicted by scenario (2).

At a distance of $42.8 \pm 1.0$\,pc ($\varpi = 23.39 \pm 0.57$\,mas; \citeads{2007ASSL..350.....V}), \object{Achernar} ($\alpha$\,Eri, \object{HD 10144}, \object{HIP 7588}) is the nearest Be star.
\citetads{1966Obs....86..108A} discovered the Be nature of Achernar through the detection of Hydrogen lines in emission.
The shape of the photosphere of Achernar A was first measured using optical interferometry by \citetads{2003A&A...407L..47D}, revealing its extreme rotational distorsion (see also \citeads{2012A&ARv..20...51V} for a review of early interferometric observations of rapidly rotating stars).
The initially reported flattening ratio larger than 3/2 triggered theoretical modeling efforts (e.g., \citeads{2010A&A...517A...7Z, 2008ApJ...676L..41C, 2004ApJ...606.1196J}).
\citetads{2012A&A...545A.130D} used spectro-interferometry with the VLTI/AMBER instrument to spatially resolve the Doppler effect on the photosphere of Achernar (see also \citeads{2012A&A...538A.110M}) and to determine its de-projected equatorial rotation velocity ($v_\mathrm{eq} = 298 \pm 9$\,km\,s$^{-1}$).
The photosphere was subsequently studied in detail by \citetads{desouza2014} using VLTI/PIONIER interferometric data, refining its physical parameters, including its equatorial radius ($R_\mathrm{eq} = 9.16 \pm 0.23\ R_\odot$), inclination ($i = 60.6^{+7.1}_{-3.9}\ \deg$), orientation on the sky of the visible pole ($PA_\mathrm{rot} = 216.9 \pm 0.4\ \deg$), equatorial velocity ($v_\mathrm{eq} = 298.8^{+6.9}_{-5.5}$\,km\,s$^{-1}$) and Von Zeipel gravity-darkening coefficient $\beta = 0.166^{+0.012}_{-0.010}$.
In its close circumstellar environment, Achernar exhibits a polar wind which was resolved by interferometry both at near-infrared \citepads{2006A&A...453.1059K} and thermal infrared \citepads{2009A&A...493L..53K} wavelengths.
%
Using thermal infrared imaging with VLT/VISIR, \citetads{2007A&A...474L..49K} discovered a companion star of Achernar A, that \citetads{kervella2008} characterized to be an early A-type main sequence star with an orbital period of several years.
This configuration is particularly interesting as such a long period orbit points at the possibility of scenario (1) for the spin up of Achernar A, without invoking mass-transfer from a donor star.
The main objectives of our long-term program reported in the present article are to characterize the orbital and physical parameters of the Achernar AB system, including the masses of the two stars and their evolutionary state.
Based on these results, we aim at testing the viability of scenario (1) for the formation of Achernar A as a Be star.
In addition, during this long-term monitoring, Achernar A went through several Be emission-line phases, which we aim at correlating with the periastron passage of its companion star.

The collected observations are described in Sect.~\ref{observations}, and we present their analysis in Sect.~\ref{analysis}.
We examine the implications of the determined properties of the Achernar system in Sect.~\ref{discussion}, in particular on its evolutionary status and its membership in the Tuc-Hor association, and we summarize our conclusions in Sect.~\ref{conclusion}.

\section{Observations}\label{observations}

We report in this paragraph the observations of Achernar collected using the VLT/VISIR (Sect.~\ref{visir}), NACO and SPHERE imaging instruments (Sect.~\ref{sect:nacosphere}), as well as the interferometric data collected with VLT/NACO-SAM (Sect.~\ref{sect:NACO-SAM}) and the VLTI instruments AMBER (Sect.~\ref{sect:AMBER}), PIONIER (Sect.~\ref{sect:pionier}), GRAVITY (Sect.~\ref{sect:gravity}) and MATISSE (Sect.~\ref{sect:matisse}).
The radial velocity measurements obtained by spectroscopy are described in Sect.~\ref{sect:radvelocityA}.
The complete list of relative position and photometric flux measurements of Achernar B with respect to Achernar A are listed in Table~\ref{tab:allastrometry1}, and the radial velocity measurements of Achernar A in Table~\ref{tab:allRV1}.

\subsection{VISIR imaging\label{visir}}

In the thermal infrared domain, we adopted the VISIR astrometric and photometric measurements of Achernar obtained by \citetads{kervella2008} (see also \citeads{2007A&A...474L..49K}) at wavelengths of 8.59 and $11.25\,\mu$m.
We added quadratically a 15\,mas uncertainty to the statistical measurement error bars on the differential position of B with respect to A in each right ascension and declination axes.
This corresponds to one-fifth of the pixel scale of the VISIR images (75\,mas\,pix$^{-1}$).
This additional uncertainty is intended to take into account the perturbations of the PSF fitting that are introduced by the residual speckles.
We considered for the photometric flux ratio $f_B/f_A$ the ratio of the peak intensities of the two stars in the VISIR images as listed by \citetads{kervella2008}. We included an additional uncertainty of 0.4\% to conservatively take the difference between the peak intensity and aperture photometry photometric ratios into account.

\subsection{NACO and SPHERE adaptive optics imaging\label{sect:nacosphere}}

\begin{figure*}
     \centering
         \includegraphics[width=16cm]{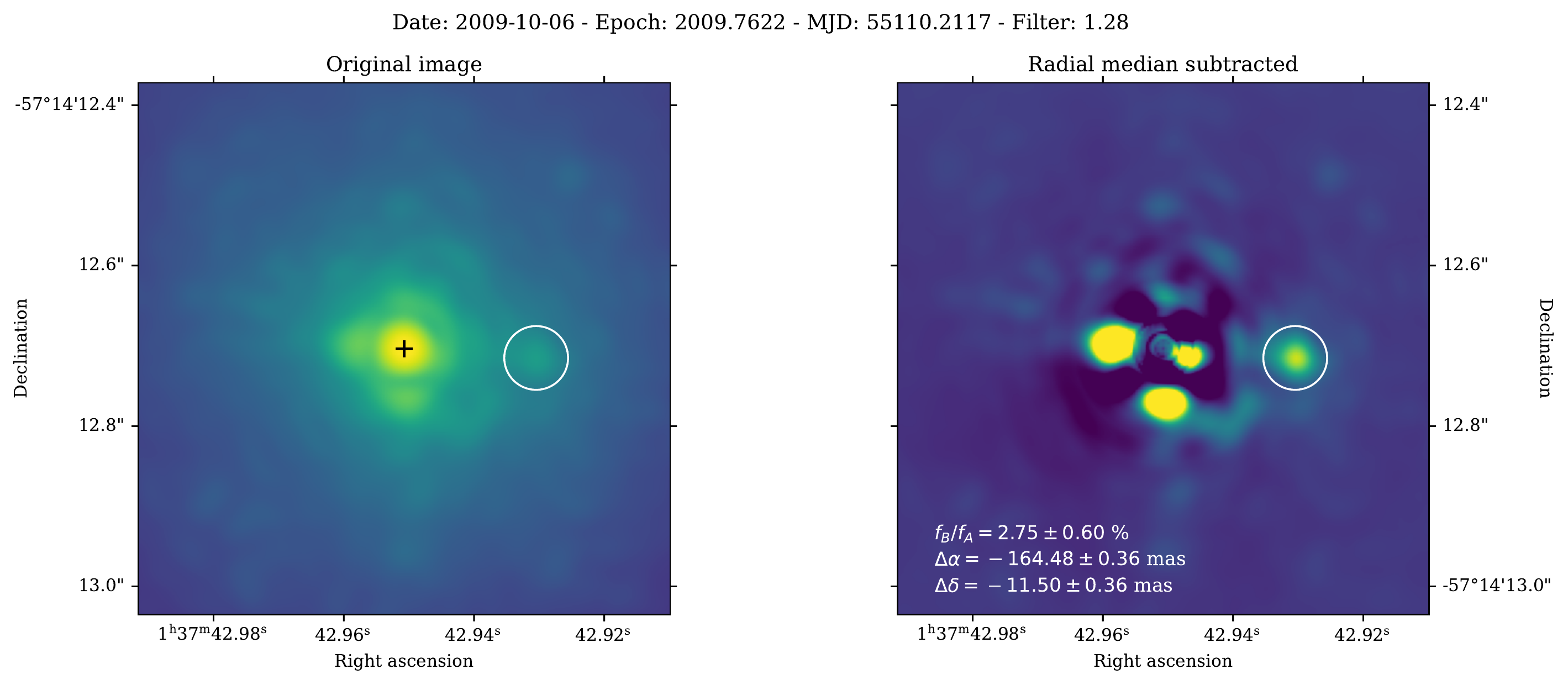}
         \includegraphics[width=16cm]{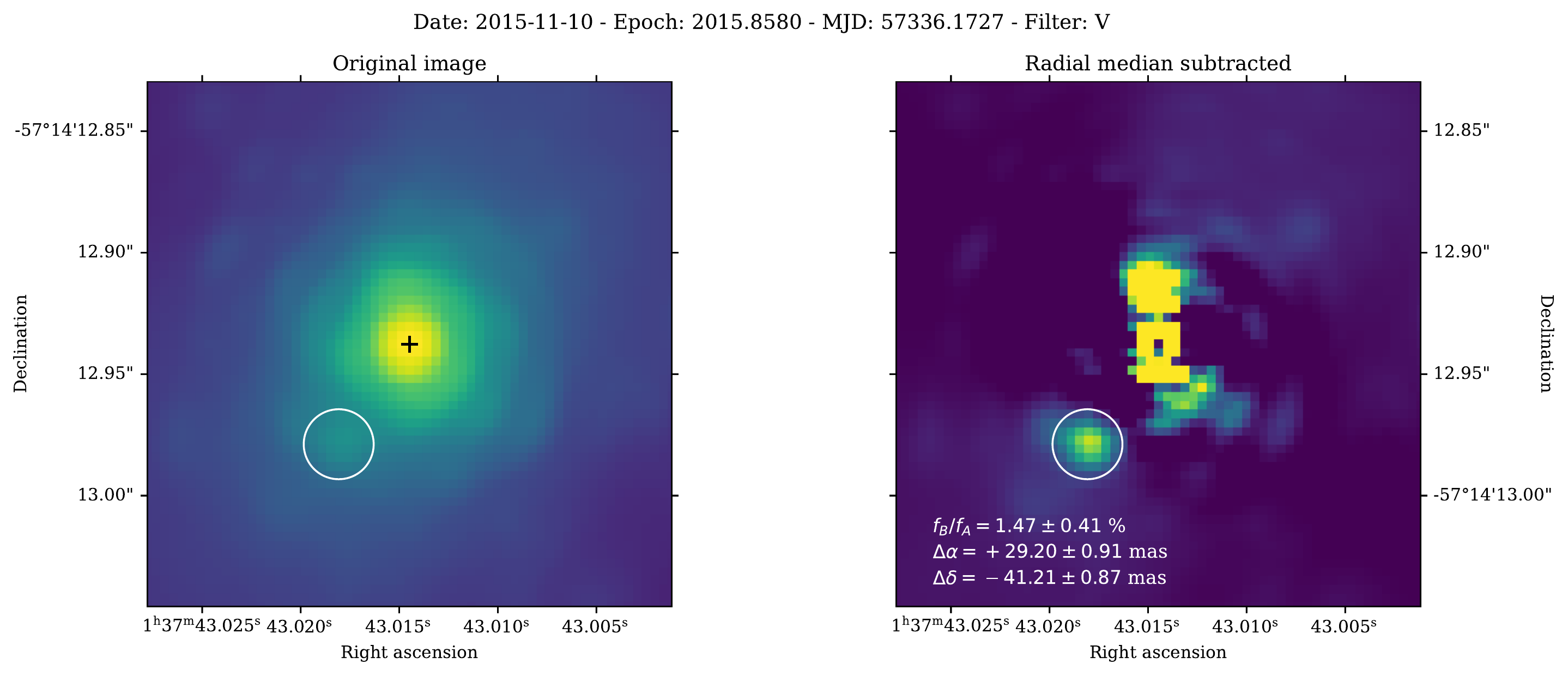}
     \caption{Sample NACO $1.28\,\mu$m (top row) and SPHERE/ZIMPOL V band (bottom row) images of Achernar A and B. The  panels show both stars with A centered in the field of view and a logarithmic intensity scale. The right images show the result of the subtraction of the radial median profile of the left image. The NACO image has been resampled by a factor four, resulting in the same pixel scale as the ZIMPOL image (3.315\,mas\,pix$^{-1}$). The position of A is marked with a black ``+''  symbol in the left images and the position of B is shown with a white circle. \label{fig:nacosample}}
\end{figure*}

We observed Achernar using the VLT/NACO near-infrared adaptive optics system \citepads{1998SPIE.3354..606L, 2003SPIE.4841..944L, 2003SPIE.4839..140R} in its regular imaging mode in 2007 and using the short-exposure cube mode between 2009 and 2012. The cube mode presents the advantage of producing a higher Strehl ratio in the images as well as a better sampling of the point spread function (PSF) through a shift-and-add approach after resampling.
The images were recorded using the S13 camera, whose plate scale is $13.26 \pm 0.06$\,mas\,pix$^{-1}$ \citepads{2008A&A...484..281N}. We obtained images using 10 narrow-band filters\footnote{\url{https://www.eso.org/sci/facilities/paranal/decommissioned/naco/inst/filters.html}} (\texttt{NB\_1.04, NB\_1.08, NB\_1.09, NB\_1.24, NB\_1.26, NB\_1.28, NB\_1.64, NB\_1.75, NB\_2.12, NB\_2.17}) with central wavelengths ranging from 1.040 to $2.166\,\mu$m.

We also collected SPHERE/ZIMPOL \citepads{2019A&A...631A.155B} polarimetric images of Achernar at visible wavelengths in October-November 2015 (Prog. ID: 096.D-0353(A), PI: Kervella), close to the periastron passage which occurred on 25 October 2015 (epoch 2015.81). We used four different medium- and narrow-band filters\footnote{\url{https://www.eso.org/sci/facilities/paranal/instruments/sphere/inst/filters.html}} (\texttt{V, Cnt\_Ha, B\_Ha, N\_I}) with central wavelengths between 0.554 and $0.8168\,\mu$m.We employed the field-stabilized derotator P2 mode and we measured the Q and U Stokes parameters. To avoid the saturation of the detector, we inserted a ND2.0 neutral density filter with an exposure time per frame of 1.2\,second. The pixel scale of the pipeline processed images from ZIMPOL is 3.315\,mas\,pix$^{-1}$. The separation of the two stars at the time of our SPHERE observations was smaller than 50\,mas. We therefore did not use the classical Lyot coronagraph of ZIMPOL whose minimum mask diameter is too large for this small separation. We limited our present analysis of the SPHERE data to the total intensity images, and we postpone the interpretation of the polarimetric quantities to a future work.

To analyse both the NACO and SPHERE images, we first precisely centered  Achernar A on a fixed reference pixel, and measured its flux using aperture photometry (over a radius corresponding to the core of the Airy pattern). To mitigate the strong Airy ring pattern and scattered light wings of the image of Achernar A at the location of B, we subtracted a radial median profile from the image. Thanks to its faintness, Achernar B does not affect this radial profile. We then measured the position of Achernar B using a Gaussian fit over a small window (the size of the core of the PSF) and aperture photometry using the same aperture size as A (Fig.~\ref{fig:nacosample}). 
Component B is generally well detectable in the images, but its position and photometry is affected by residual speckles which create a variable background. The uncertainty on the photometric measurement was determined from the dispersion of aperture photometry at 36 positions spread over a ring at the same angular separation from the primary as Achernar B.
For the differential astrometry we added quadratically the same uncertainty of 1/5th of the pixel scale as for the VISIR data set (Sect.~\ref{visir}) to the statistical measurement error bars.
This reflects the inhomogeneity of the images, while simultaneously accounting for the range of angular resolutions of the different observing wavelengths as the pixel scale is different for NACO and ZIMPOL.

We rejected the NACO and SPHERE astrometric measurements of Achernar B for which the detection is too uncertain, that is, with a photometric $S/N<2$ on the flux of B.
In five additional cases, a visual inspection of the adaptive optics images showed that the position of Achernar B was strongly biased by the presence of large speckles, and we also excluded these measurements.
The data points which we excluded from the orbital and spectral energy distribution fits are marked with a ``$\bullet$'' symbol in the ``Flag'' column of Table~\ref{tab:allastrometry1}.

\subsection{NACO aperture masking\label{sect:NACO-SAM}}

Aperture masking is an interferometric technique that enables to reach a higher angular resolution than the diffraction limit of a filled-pupil telescope through the insertion of a nonredundant mask in the optical beam. The sparse aperture masking (SAM) mode of NACO \citepads{2010SPIE.7735E..1OT} implements this technique in combination with adaptive optics, effectively creating a Fizeau interferometer \citepads{2000PASP..112..555T}.
We collected observations of Achernar on 7 January 2009 using the 18-hole mask together with three intermediate- and narrow-band filters in the near-infrared domain (\texttt{NB\_1.64, IB\_2.12, IB\_2.18}). The star $\delta$\,Phe (\object{HIP 7083}) was observed to calibrate the visibilities obtained on Achernar.
The data reduction to convert the images to visibilities was done as described by \citetads{2011A&A...532A..72L} and \citetads{2016A&A...595A..31G}.
The differential astrometry and flux ratio of Achernar B were computed using the LITPro\footnote{\url{http://www.jmmc.fr/litpro/}} software \citepads{2008SPIE.7013E..1JT} from the JMMC, adopting a simple model of two point sources (the angular diameter of Achernar A is unresolved by a single 8-meter telescope). The resulting positions are listed in Table \ref{tab:allastrometry1}.

\subsection{AMBER interferometry\label{sect:AMBER}}

We observed Achernar with the AMBER interferometer \citep{Petrov2007_v464p1} at the Very Large Telescope Interferometer  \citep[VLTI;][]{Haguenauer2010_v7734p1} on four nights in 2008 (3, 7, 9 and 13 November 2008), close to the periastron passage which occurred on 10 October 2008 (epoch 2008.78).
The log of the AMBER observations is given in Table~\ref{tab:interferometrylog}.
Each night, data were recorded with a different baseline configuration of the VLTI array, using three 1.8-meter Auxiliary telescopes (AT).
Interference fringes were obtained in the LR-HK mode (low spectral resolution LR: $\lambda/\delta\lambda\approx30$).
The observations of Achernar were interspersed with those of a calibration star ($\delta$ Phe). 
The data reduction was performed with \textit{amdlib} version~3.0.9 \citepads{Chelli2009_v502p705, Tatulli2007_v464p29}. To improve the quality of the final reduced data the best observed frames were selected according to the following criteria: fringe and flux $S/N >1$ and absolute optical path difference $\mathrm{OPD} <20$\,\m.
After the data reduction one obtains several calibrated observables for each individual data recording (data from a few files with fringes forming an observing block, recorded over a few minutes) and and each baseline and instrumental configuration: three visibilities $V^2$, three (spectrally) differential visibilities \Vdiff\, and phases $\phi_\textrm{diff}$, one closure phases $CP$, and one flux, all simultaneously in the H and K bands.
Since the S/N of AMBER LR data is very degraded in the borders of the spectral bands, we only analyzed data within the spectral ranges $1.51 \leq \lambda \leq 1.84$\,\m\ (H band) and $2.00 \leq \lambda \leq 2.43$\,\m\ (K band).

For the present study of the binary system, we use the interferometric data to measure the separation and flux ratio between Achernar A and B. The physical parameters of the primary (the Be star), which is partially resolved by the interferometer on the longest baselines, can be found in \citetads{desouza2014}.
From the low IR flux ratio (secondary to primary flux $\sim2\%$) previously measured for this system \citepads{kervella2008}, the interferometric observables are expected to show essentially the primary signal, modulated by the presence of the weak companion, with the amplitude of the modulation depending basically on its flux and the spatial frequency of the modulation depending basically on the apparent separation. Thus, in order to maximize the S/N when measuring the binary separation and flux ratio from AMBER data, we have chosen to analyze the differential visibilities \Vdiff\footnote{Visibility as a function of the wavelength, normalized by the average visibility in the spectral band considered.}, which are less subject to atmospheric and instrumental errors than the absolute $V^2$. Moreover, we also use the closure phases $CP$, which gather all the useful information from the differential phases $\phi_\textrm{diff}$ in LR, with an additional cancellation effect of atmospheric and instrumental perturbations to the interference phases.

The model employed to measure the separation and flux ratio of Achernar AB from the AMBER LR-HK \Vdiff\, and $CP$ is composed of two parts. 
Firstly, the bright primary is represented by the same fast-rotator numerical model and corresponding physical parameters presented by \citetads{desouza2014}.
Next, we add to this model a faint companion represented by a small unresolved uniform disk (angular diameter of 0.5 mas) with a flux $f_B/f_A$ in H and K bands (relative to the primary), and located at an angular separation $(\Delta \alpha,\Delta \delta)$ with respect to A.

This model was then fit to AMBER data with the \textit{emcee} Python package \citepads{Foreman-Mackey2013_v125p306}.
The parameters of the fast rotating Be star being fixed to the previously determined values, the remaining free parameters to be fit are directly related to binary: the flux ratio $f_B/f_A$ in the H and K bands, the angular separation $(\Delta \alpha, \Delta \delta)$ at the start of the observing series ($\mathrm{MJD}=54774.1213$), and the temporal rate of evolution of this separation $(\Delta \mu_\alpha, \Delta \mu_\delta)$.
We adopted a progressive fitting approach, starting with the higher S/N K-band data and ending with the full AMBER LR-HK data set.
The best-fit model has a satisfactory reduced $\chi^2=0.51$, indicative of a moderate overestimation of the original data error bars.
From the fit parameters we computed four astrometric positions of the companion corresponding to the four AMBER observing nights. The individual relative positions of B with respect to A at these four epochs, are listed in Table~\ref{tab:allastrometry1}. Single values of the H and K relative fluxes of Achernar B were fit to the AMBER data, as we assumed the flux ratio to be constant over the 10 day period covering the four pointings. These two flux ratios are also listed in Table~\ref{tab:allastrometry1}.

\begin{table}
\caption{List of the AMBER, PIONIER, GRAVITY and MATISSE interferometric observations of Achernar.
N$_{\lambda}$ is the number of spectral channels. \label{tab:interferometrylog}}
\centering
  \begin{tabular}{l c c c c}
    \hline
    \hline
    Date & N$_{\rm obs}$ & Band & N$_{\lambda}$ & VLTI  \\
         &               &  &               & Configuration  \\
    \hline  \noalign{\smallskip}
    AMBER \\
    2008-11-03 & 16 & H+K & 38 & D0-G1-H0 \\
    2008-11-07 & 32 & H+K & 38 & A0-D0-H0 \\
    2008-11-09 & 22 & H+K & 38 & E0-G0-H0 \\
    2008-11-13 & 30 & H+K & 38 & A0-G1-K0 \\    
    \hline  \noalign{\smallskip}
    PIONIER \\
    2016-10-12 & 4 & H & 6 & A0-B2-C1-D0 \\
    2016-10-13 & 4 & H & 6 & A0-B2-C1-D0 \\
    2016-11-11 & 2 & H & 6 & A0-B2-C1-D0 \\
    2016-11-12 & 2 & H & 6 & A0-B2-C1-D0 \\
    \hline  \noalign{\smallskip}
    GRAVITY \\
    2016-11-15 & 6 & K & 1742 & A0-B2-C1-D0 \\
    2016-11-21 & 6 & K & 1742 & A0-B2-C1-D0 \\
    2016-11-22 & 3 & K & 1742 & A0-B2-C1-D0 \\
    2018-12-25 & 2 & K & 1742 & A0-B2-C1-D0 \\
    \hline  \noalign{\smallskip}
    MATISSE \\
    2019-09-20 &  & L & 118 & A0-B2-C1-D0 \\
    2019-09-21 &  & L & 118 & A0-B2-C1-D0 \\
    2019-09-25 &  & L & 118 & A0-B2-C1-D0 \\
    \hline
    \end{tabular}
\end{table}

\subsection{PIONIER interferometry\label{sect:pionier}}

We observed Achernar with the PIONIER instrument \citep{2010SPIE.7734E..99B,lebouquin2011} at the Very Large Telescope Interferometer (VLTI, \citeads{2014SPIE.9146E..0JM, 2018SPIE10701E..03W, 2020SPIE11446E..06H}) on four different nights in 2016, so as to resolve its companion. PIONIER recombines the light beams coming from the four Auxiliary Telescopes (ATs) of the VLTI and provides the corresponding interferometric observables, that is, six simultaneous squared visibility measurements and four closure phases in the $H$-band ($\sim$1.65~nm). We provide the log of our observations in Table~\ref{tab:interferometrylog}.
The VLTI was set in the compact telescope configuration, that is, with baselines from 11 to 34\,m.
This allowed us to obtain a large interferometric field of view (up to $\sim 180-220$\,mas) at the expense of angular resolution, as our focus was on detecting the binary companion and not resolving the disk of Achernar A.
Six spectral channels were used to increase the coverage of the ($u$, $v$) plan. The interferometric data were automatically reduced using the standard \texttt{pndrs} procedure for PIONIER \citepads{lebouquin2011}.

\begin{figure}
     \centering
         \includegraphics[width=\hsize]{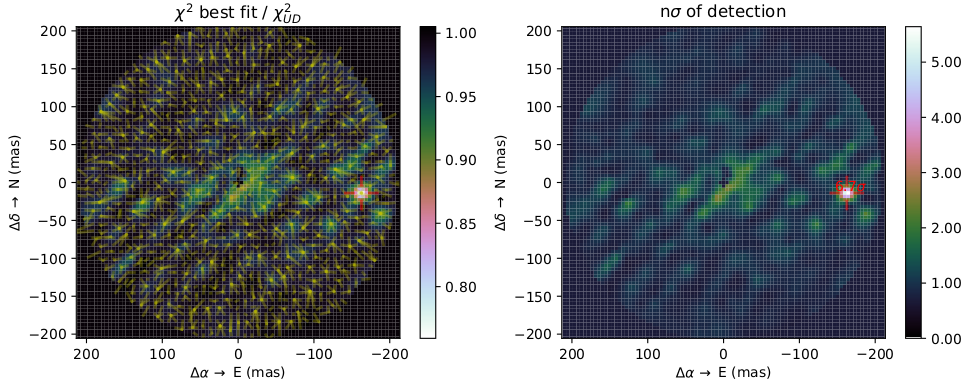}
         \includegraphics[width=\hsize]{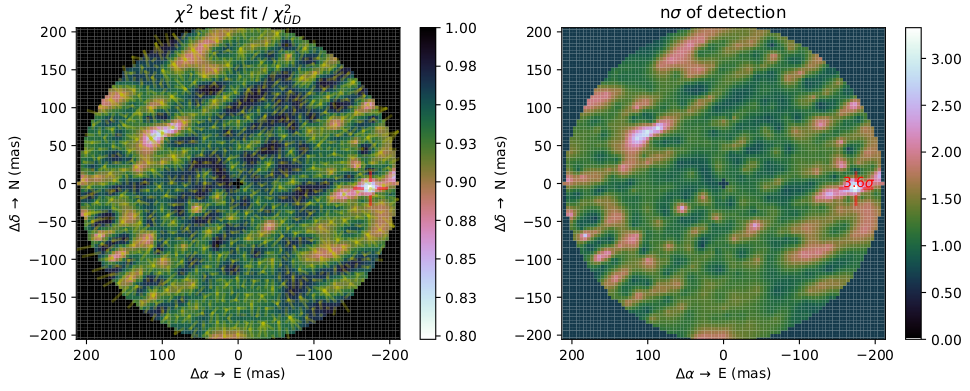}
     \caption{Detection of Achernar B from PIONIER observations with \texttt{CANDID}. The left and right panels show the $\chi^{2}_{r}$~maps of local minima and detection level maps, respectively, derived with \texttt{CANDID} from our PIONIER 2016 observations. The top panels correspond to the October 2016 epoch and the bottom panels to the November 2016 epoch.}\label{fig:candid_map}
\end{figure}

We used the Python-based companion analysis tool \citep[\texttt{CANDID},][]{gallenne2015} to systematically search for the binary companion to Achernar A in PIONIER interferometric data. Briefly, \texttt{CANDID} uses a 2D grid of fit (corresponding to the position of the putative companion) as a starting point for a multiparameter fit based on a least-squares algorithm.
At each point of the grid, the following parameters are adjusted: the companion position ($\Delta \alpha$, $\Delta \delta$), the angular diameter of the primary ($\theta_{\rm UD}$), and the flux ratio $f$ of the companion (which we assume to be point-like). Both interferometric observables (square visibilities and closure phases) are explored simultaneously. The deduced map of local minima is converted into a companion detection level map (expressed in a number of sigmas, see Fig.~\ref{fig:candid_map}).
As a remark, on the short baselines that were used for these observations, the apparent disk of Achernar A \citepads{desouza2014, vedova2017} is unresolved by the interferometer.
The elliptical shape and inhomogeneous gravity darkening of Achernar A thus have no influence of the relative astrometry of the secondary. 

As the 1-day position shift of the companion to Achernar A is small at the epoch of observation (about 1\,year after periastron), we combined our two pairs of consecutive-night PIONIER observations into two separate epochs (October and November 2016) to increase the companion signal-to-noise ratio within \texttt{CANDID}.
We clearly detected Achernar B at our two PIONIER observation epochs with a detection level S/N>3 (Fig.~\ref{fig:candid_map}).
A secondary minimum of the $\chi^2$ is visible at the November 2016 epoch, in the northeast quadrant (top-left). Its significance is however lower than the peak corresponding to the position of Achernar B, and it is absent from the October 2016 epoch. Its elongated shape and lack of a peaked maximum points at a speckle of noise.
The astrometric positions of the companion are detailed in Table~\ref{tab:allastrometry1}. The motion of Achernar B between October and November 2016 is noticeable ($\sim$14~mas).
We note that we also detect Achernar B at both observing epochs when using only the interferometric closure phases within \texttt{CANDID} (though with a reduced S/N), which attests to the robustness of the detections.

\subsection{GRAVITY interferometry\label{sect:gravity}}

A series of observations of Achernar were collected with the GRAVITY instrument \citepads{2017A&A...602A..94G} in November 2016 (one year after periastron) and December 2018, using the VLTI 1.8\,m Auxiliary Telescopes on the A0-B2-C1-D0 configuration.
As for the PIONIER observations, this compact quadruplet provides baselines up to a ground length of 34\,meters.
The list of the observations selected for the present study is presented in Table~\ref{tab:interferometrylog}.
GRAVITY was configured in high spectral resolution mode, with 1742 spectral channels over the K band for a spectral resolution of $R \approx 4000$ ($\Delta \lambda = 0.264$\,nm).
The raw data were processed using the standard GRAVITY pipeline \citepads{2014SPIE.9146E..2DL}.
The observations were collected in the split polarization mode, and both polarizations were analyzed jointly to determine the position and relative flux of Achernar B using the \texttt{PMOIRED}\footnote{\url{https://github.com/amerand/PMOIRED}} library \citepads{2022arXiv220711047M, 2022ascl.soft05001M} and a grid fit.

\begin{figure}
     \centering
         \includegraphics[width=\hsize]{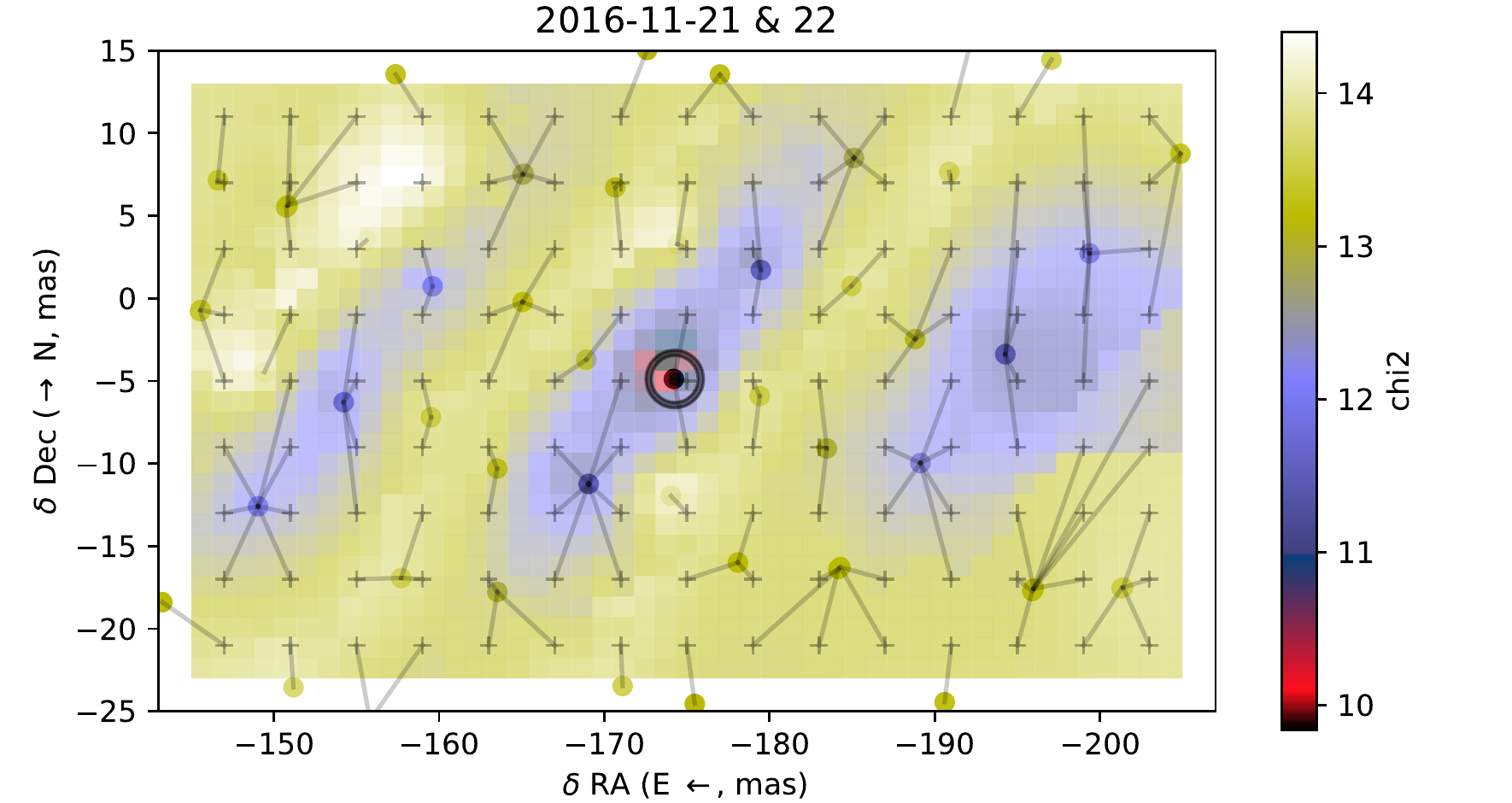}
     \caption{Map of the $\chi^2$ minimization fit to the GRAVITY observations of 21-22 November 2016 using the \texttt{PMOIRED} algorithm. The explored angular separation $(\delta \mathrm{RA}, \delta \mathrm{Dec})$ represents the position of Achernar~B relative to A.
The starting positions of the $\chi^2$ minimum searches are marked with gray ``+'' symbols. The best-fit position is marked with a circle close to the center of the diagram.}\label{fig:pmoired_gridsearch}
\end{figure}

\begin{figure*}
     \centering
         \includegraphics[width=17cm]{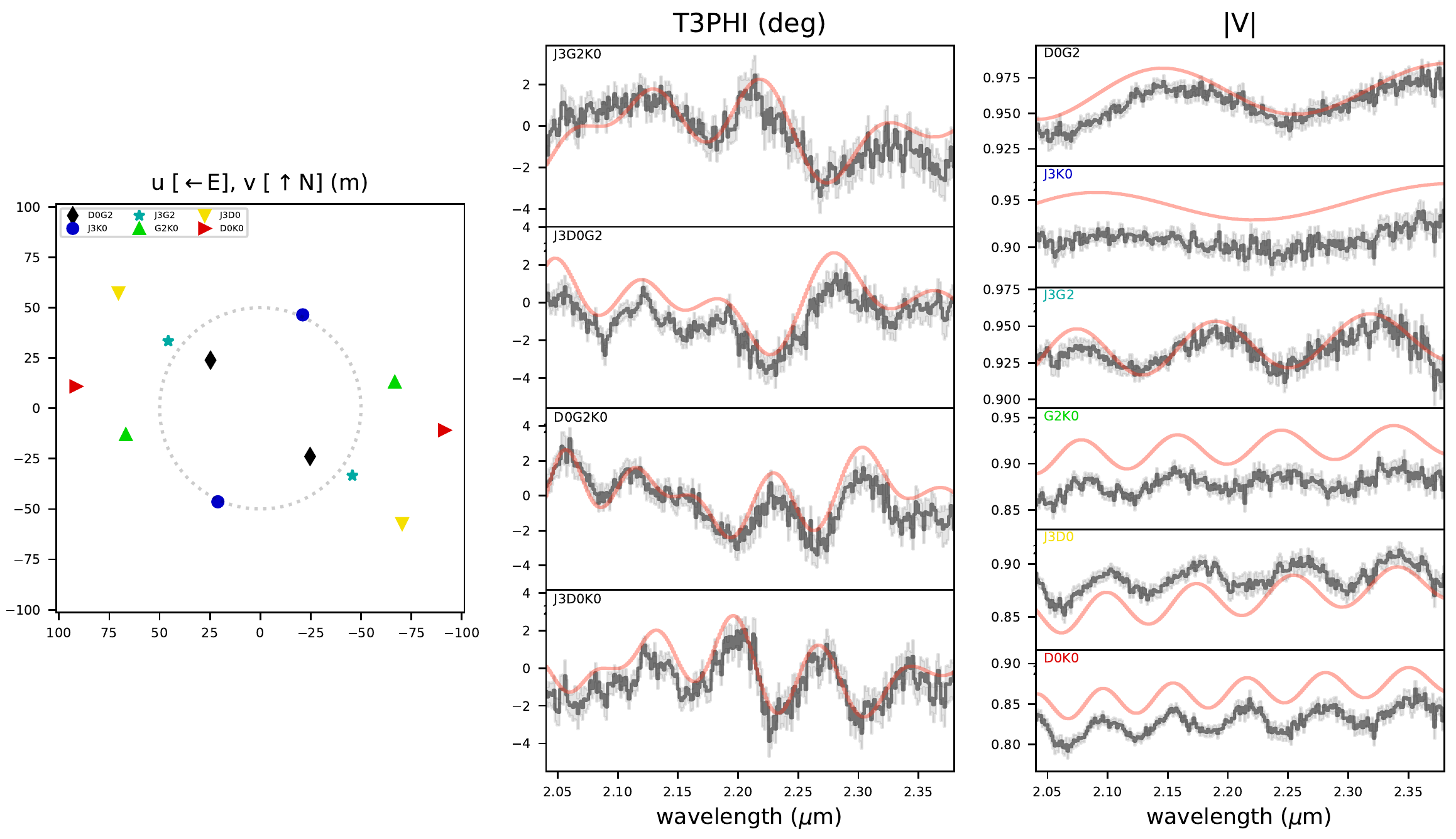}
     \caption{Detection of Achernar B from the GRAVITY observation of 22 November 2016 at UT02:58:15 with \texttt{PMOIRED}. $T3\phi$ is the closure phase for each of the four telescope triplets, and $|V|$ the amplitude of the fringe visibility for each of the six baselines. For comparison, the data are shown as black curves, and the best-fit binary star model for all observing epochs as red curves. This diagram represents one of the GRAVITY data set used for the global fit, and the individual diagrams corresponding to the other data sets are presented in Fig.~\ref{fig:GRAVITYfits1}}.\label{fig:pmoired_bestfit}
\end{figure*}

The $\chi^2$ minimization grid search map explored by \texttt{PMOIRED} is shown in Fig.~\ref{fig:pmoired_gridsearch} for the combination of the nights of 21-22 November 2016. A comparison of the global best-fit model compared to one of the GRAVITY observations is presented in Fig.~\ref{fig:pmoired_bestfit}. Similar figures for the other GRAVITY observing epochs are presented in Appendix~\ref{GRAVITYplots}.
The left panel of Fig.~\ref{fig:pmoired_bestfit} shows the coverage of the $(u,v)$ plane, and the central column shows the four measured closure phases (black curves) as a function of wavelength together with the best-fit model (red curves).
The right column shows the visibility modulus $|V|$ measured on the six baselines (black curves), also with the best fit model (in red).
The sinusoidal oscillations which are clearly visible in the $T3\phi$ and $|V|$ signals are the signature of the presence of Achernar B.
The best-fit binary star model has a reduced $\chi^2 \approx 11$. The resulting uncertainties on the parameters (position, flux ratio) have been renormalized to take into account the relatively important residual dispersion of the data with respect to the binary model (visible in Fig.~\ref{fig:pmoired_bestfit}).
The resulting relative position and flux of Achernar B with respect to Achernar A are presented in Table~\ref{tab:allastrometry1}.
Achernar B was not detected in the GRAVITY data collected on 15 November 2016 and 25 December 2018.

\subsection{MATISSE interferometry\label{sect:matisse}}

MATISSE is a spectro-interferometric imaging instrument combining the light from four telescope of the VLT Interferometer in the infrared L, M and N bands \citepads{matisse22}.
Achernar was observed with MATISSE on 20, 21 and 25 September 2019, during the commissioning of the GRA4MAT fringe tracker, in the L band ($2.8-4.2\,\mu$m) using the medium spectral resolution setting ($R \approx 500$). A description of the observations is presented in Sect. 5.6.1 of \citetads{matisse22}.
The separation between Achernar A and B was found to be $\rho = 293$\,mas at a position angle $PA = 309^\circ$.
The flux ratio between B and A is poorly constrained, and we therefore do not include it in the fit of the spectral energy distribution of Achernar B (Sect.~\ref{sect:massB}).

\subsection{Radial velocity of Achernar A\label{sect:radvelocityA}}

Detecting Achernar B and monitoring its orbit in spectroscopic data and radial velocities (RV) is a daunting task.
First, Achernar A being fast-rotating Be star, its spectrum exhibits only very few, extremely rotationally broadened absorption lines that are very shallow ($v_\mathrm{rot} \sin i \simeq 260$\,km\,s$^{-1}$; \citeads{desouza2014}).
Second, the contrast of Achernar B relative to its primary ($\sim 50$) complicates its disentanglement from the primary features in spectra.
Additionally, due to its probable early A spectral type \citepads{kervella2008}, Achernar B is also likely to be a fast rotating star itself  \citepads{2007A&A...463..671R}, with correspondingly shallow spectral features.
Third, the primary RV curve that is expected from the approximate mass ratio of B with respect to A is of relatively low amplitude ($\lesssim$10~\kms), and is flat over most of its 7-year period due to the eccentricity of the orbit.
Nonetheless, long-term RV data of Achernar A have the potential to detect or at least constrain the relatively short radial velocity peak expected at the time of passage at periastron of the companion (which happened in late 2008 and late 2015). It is also important to determine the barycentric radial velocity of the system, that is required to assess its space motion (see Sect.~\ref{comoving_star}).
To this end, we gathered all the Achernar mid- to high-resolution (from $R \simeq 10\,000$ to $R \simeq 130\,000$) optical (390 to 880\,nm) spectra available to our knowledge from various echelle spectrographs, and complemented them with new observations collected with the CHIRON instrument during periastron crossing in late 2015.
We thus assembled a data set of $\sim 750$~spectra covering a time span of 4460\,days ($\sim 12$\,years), that is, more than one and a half time the orbital period of Achernar B.
Our data set is detailed in Table~\ref{tab:spectro}.

\subsubsection{Archival data and new observations}

We first used the public archive of the European Southern Observatory (ESO)\footnote{\url{http://archive.eso.org/wdb/wdb/adp/phase3_main/form}} to retrieve $\sim$240 spectra mainly acquired with FEROS in 2006 and in 2013-2014, as well as with HARPS and UVES between 2002 and 2004. Then, we looked into the Be Stars Observation Survey \citep[BeSOS,][]{arcos2017} public catalogue\footnote{\url{http://besos.ifa.uv.cl}} to gather more than 450 Achernar spectra acquired with the PUCHEROS echelle spectrograph ($R \simeq 18\,000$; Table~\ref{tab:spectro}) between 2012 and 2014.
Finally, we also retrieved one CORALIE archive spectrum of Achernar from 2015, and one echelle spectrum acquired in 2011 from the BeSS database\footnote{\url{http://basebe.obspm.fr}}.
All the archive spectra were retrieved in the ESO \texttt{s1d} standard or equivalent, \ie~processed by the spectrographs' data reduction systems (DRS) and in a 1d format (flux versus calibrated wavelength with all echelle orders already reconnected. 

We carried out a dedicated spectroscopic observing campaign from August 2015 to January 2016 with the high-resolution CHIRON spectrograph \citepads{tokovinin2013} to monitor intensely the expected passage at periastron of Achernar B (Table~\ref{tab:spectro}). We obtained more than seventy CHIRON spectra over 22 epochs, in the narrow-slit mode allowing for the highest spectral resolution ($R \simeq 130\,000$; \citeads{jones2017}).
We used the CHIRON pipeline to obtain the spectra in an order-by-order, wavelength-calibrated, 1d format (not de-blazed).
Finally, we acquired a few additional spectra between 2009 and 2013 with the BESO spectrograph \citepads{steiner2006} at Cerro Armazones (Table~\ref{tab:spectro}) 

We additionally retrieved thirty narrow-band, medium-resolution ($6\,000 \leq R \leq 15\,000$) \half~spectra of Achernar from the BeSS database, to better track the absorption and emission phases in the \half~Balmer line and their potential link to the passage at periastron of Achernar B \citepads{vedova2017}. These spectra were acquired between 2010 and 2020 (observers: Romeo, Heathcote, Locke, Powles, Luckas, Bohlsen, and Jeffredo).

As a remark, an extensive series of high-resolution and high S/N spectra of Achernar has been collected from 1992 to the early 2000s at the Brazilian National Astrophysical Observatory \citepads{2000ASPC..214..272L}. We did not include them in the present analysis as they cover a Be phase of Achernar, and the presence of strong and variable hydrogen and helium emission lines makes the determination of radial velocities too uncertain.

\subsubsection{Spectrum processing}
We corrected the spectra for cosmic ray impacts and for the barycentric Earth radial velocity (BERV), if necessary.
Apart from the PUCHEROS spectra which were already given in a normalized flux format, we normalized the 1d spectra by interpolating a continuum function or envelope over the spectral wavelength range. Our spectrum normalization process is described in details in \citetads{borgniet2019}.
In the case of CHIRON data, we first normalized each spectral order separately by approximating the blaze function either with a fourth-degree polynomial model or by using the iodine calibration per order. We then reconnected all the normalized spectral orders together. Finally, we selected the Achernar spectra to use in our further analysis based on a minimum spectral S/N threshold of 60 at 550\,nm. We thus rejected almost one hundred low S/N spectra.

\subsubsection{Radial velocity from cross correlation\label{sect:RV}}
The methodology that we employed to determine the radial velocity of Achernar A from our sample of spectra is presented in details in Appendix~\ref{sect:radvelA}.
We adopted a classical cross-correlation approach, focusing on the two reddest He I lines common to our complete spectral data set ($\lambda = 5875.66$ and $6678.15\,\AA$).
Due to the extreme rotational broadening of the spectral lines of Achernar A and the high contrast with the secondary, obtaining radial velocities (RV) from spectroscopy proved to be a difficult task.
While we secured a long series of observations of Achernar, the scatter of the resulting RV measurements is too large to constrain its orbital velocity.
In addition, we did not detect the spectrum of Achernar B, due to the high contrast with A and also because it is probably itself a rapid rotator.
As a consequence, RVs are insufficiently accurate to directly constrain the mass ratio of the two stars from its astrometric orbit.

\begin{figure}
     \centering
         \includegraphics[width=\hsize]{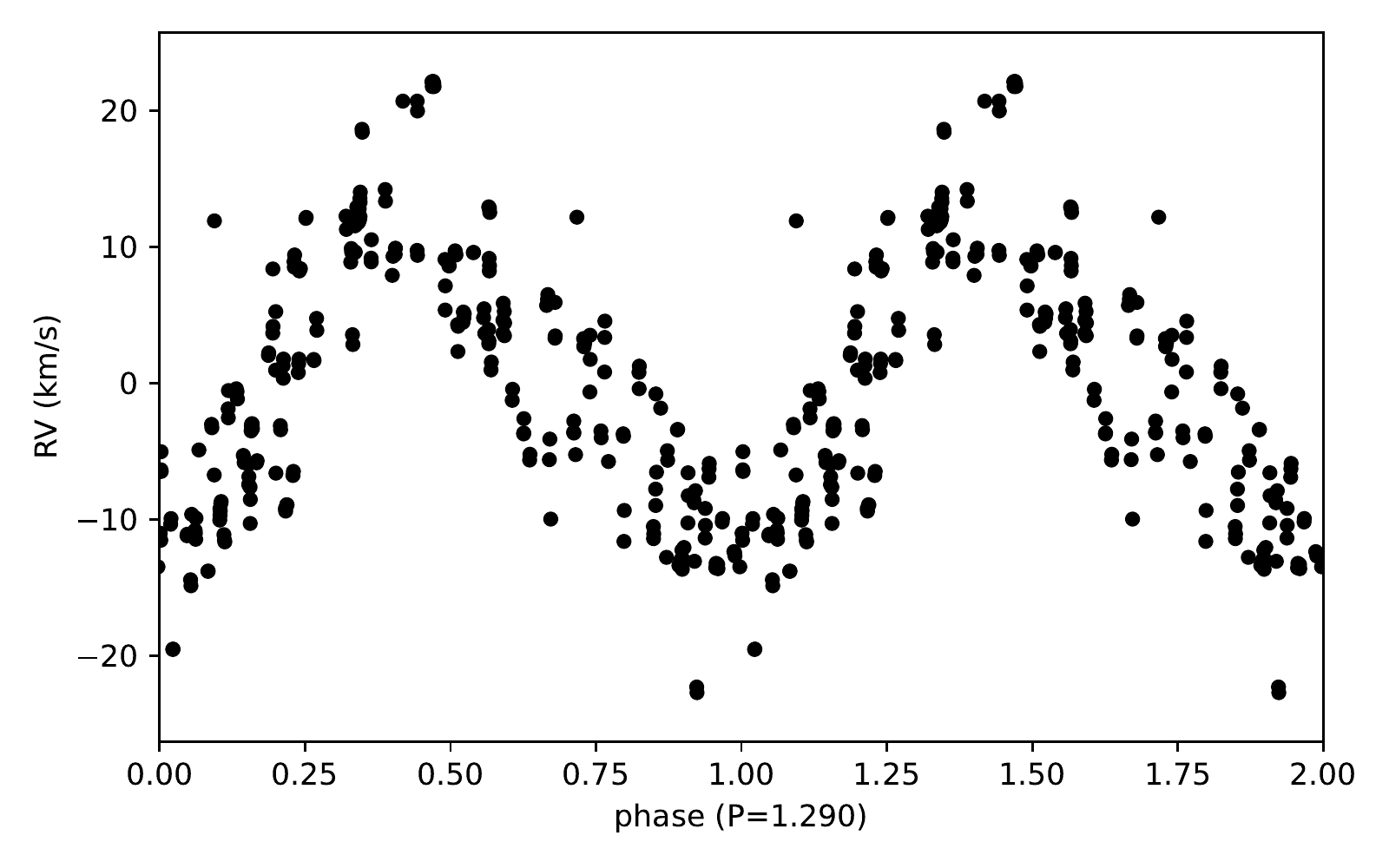}
     \caption{Short-period radial velocity variations observed on Achernar, phased with a period of 1.29\,day (frequency of 0.775 day$^{-1}$).}\label{fig:rv_oscillation}
\end{figure}

\begin{figure}
     \centering
         \includegraphics[width=\hsize]{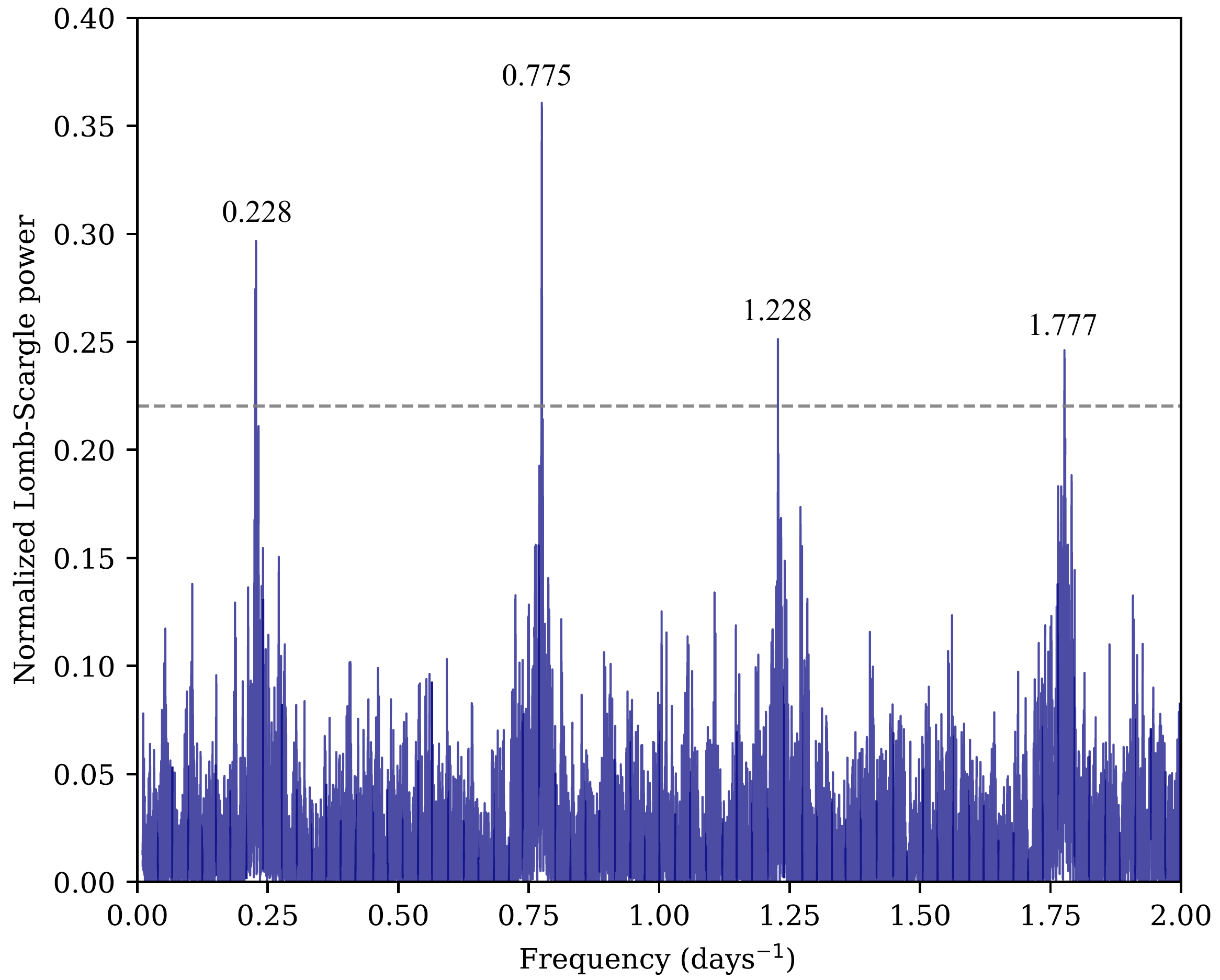}
     \caption{Lomb-Scargle periodogram of the radial velocity time series of Achernar. The 0.01 normalized power false alarm limit (equivalent to $3\sigma$ significance) is shown as a gray dashed line. The frequencies of the main peaks above this threshold are indicated in day$^{-1}$.}\label{fig:RVperiodogram}
\end{figure}

Short-period variations of the radial velocity of Achernar A and other Be stars have been reported, for example, by \citetads{2003A&A...411..229R}, \citetads{2006A&A...446..643V}, and \citetads{2011MNRAS.411..162G}.
From a radial velocity analysis using the broadening function technique (see \citeads{2017ApJ...842..110P} for details), we confirm a periodic variability of the radial velocity from our sample of spectra, with a period of 1.29\,day (0.78 cycle/day) and a peak-to-peak amplitude of $\approx 30$\,km\,s$^{-1}$ (Fig.~\ref{fig:rv_oscillation}).
This variability is driven primarily by changes in the asymmetry of the spectral lines, and exhibits a high phase stability over many years, as already noticed by \citetads{2011MNRAS.411..162G}.
It brings a significant contribution to the scatter of the determined radial velocities with respect to the expected orbital velocity modulation of Achernar A, whose amplitude is on the order of 10\,km\,s$^{-1}$.
For the determination of the barycentric radial velocity of the system (Sect.~\ref{sect:massA}), we therefore subtracted the 1.29\,day periodic variation. Other frequencies are present in the radial velocity signal, as shown in the Lomb-Scargle periodogram presented in Fig.~\ref{fig:RVperiodogram}.

\section{Analysis}\label{analysis}

\subsection{Mass of Achernar B\label{sect:massB}}

In absence of accurate radial velocities of both stars A and B, the fit of the astrometric orbit provides only a total mass of the system (from Kepler's third law).
To recover the mass of the Be component from this total mass, we need to estimate the mass of Achernar B and subtract its contribution from the total mass of the system.
In this Section, we estimate the luminosity and effective temperature of Achernar B based on its multiband photometry, with the goal to determine its mass from models.

\begin{figure}
     \centering
         \includegraphics[width=\hsize]{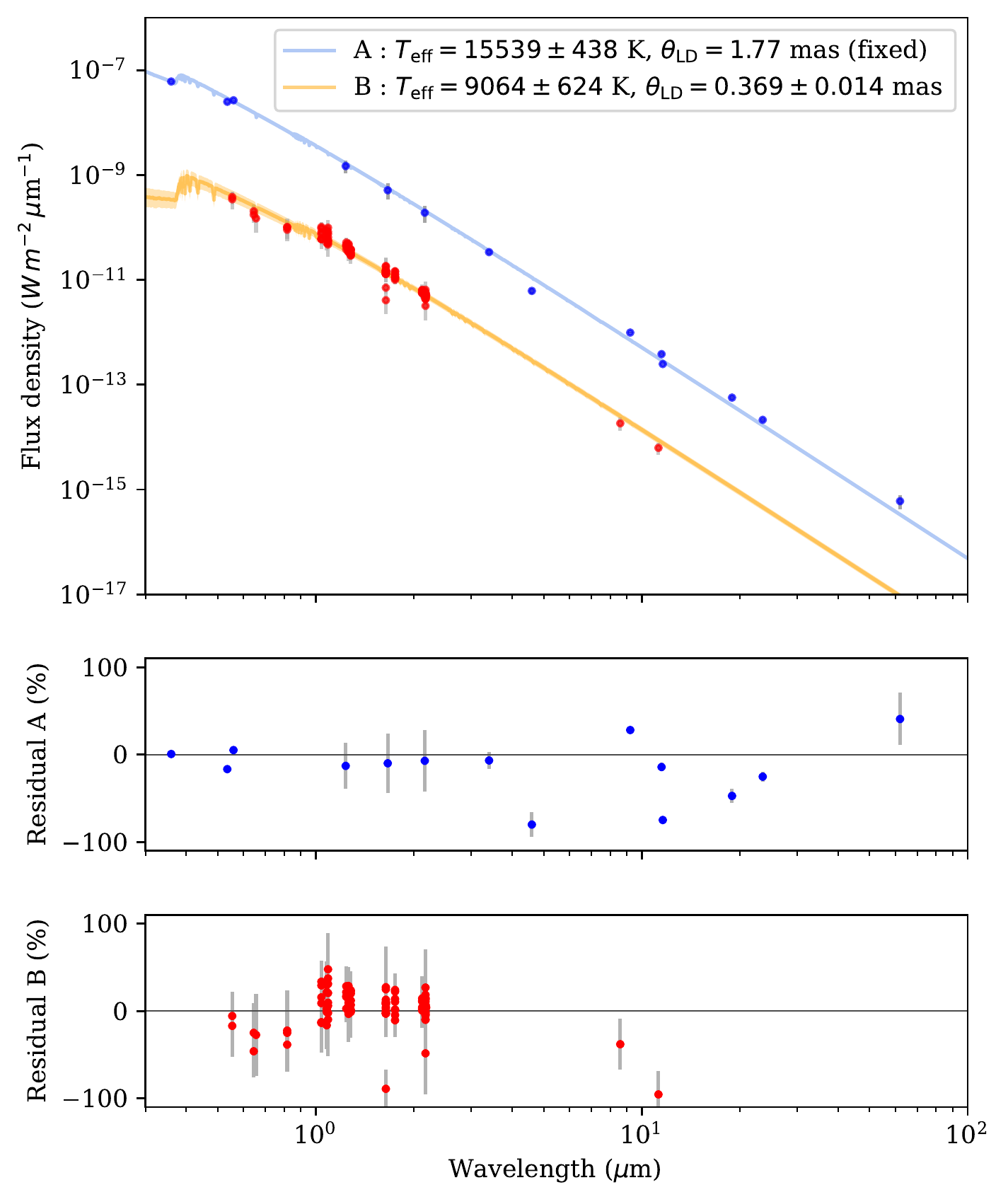}
     \caption{Spectral energy distributions of Achernar A and B.
     The photometric measurements for Achernar A are taken from the literature (Table~\ref{table:photA}) and from our measurements for Achernar B (Table~\ref{table:photB}).
     \label{fig:SED-Achernar-AB}}
\end{figure}

We first collected from the literature the available photometric data for Achernar AB (Table~\ref{table:photA}).
Due to the very tight angular separation of the two stars, the photometric measurements include the flux contributions from both components A and B. 
The $\approx 2\%$ flux contribution of the secondary is neglected for the adjustment of the spectral energy distribution of Achernar A ($f_A \approx f_{AB}$).
Relying on these measurements and the mean angular diameter of the star (including limb and gravity darkening) determined by \citetads{desouza2014} from interferometry ($\overline{\theta_\mathrm{LD}} = 1.77$\,mas, with an uncertainty of $\pm 2\%$) we determined the disk averaged effective temperature $T_\mathrm{eff}(A)$ (single parameter fit).
We based this determination on an interpolation of the set of Castelli-Kurucz model atmospheres \citepads{2003IAUS..210P.A20C} for solar metallicity.
We checked a posteriori that B's flux affects the determined stellar parameters of A by only $0.1\,\sigma$, and is therefore negligible.
This provided us with the spectral energy distribution (SED) of component A.
We subsequently used the SED of Achernar A to determine the flux densities of component B (Table~\ref{table:photB}) from the flux ratio $f_B/f_A$ that we measured in 19 filters (Table~\ref{tab:allastrometry1}).
Finally, the angular diameter $\theta_\mathrm{LD}(B)$ and effective temperature $T_\mathrm{eff}(B)$ of Achernar B were derived from fitting a Castelli-Kurucz SED model to these flux densities (two-parameter fit).
For the two SED fits (A and B), the determination of the best-fit parameters was obtained using a multiparameter Levenberg-Marquardt least-squares fitting algorithm on the complete data set. It is based on the \texttt{scipy.optimize.leastsq} routine of the SciPy\footnote{\url{https://scipy.org}} library.
The top panel of Fig.~\ref{fig:SED-Achernar-AB} shows the resulting best-fit SEDs of Achernar A and B together with the measurements.
The residual dispersion of the data points (bottom panels in Fig.~\ref{fig:SED-Achernar-AB}) with respect to the model are taken into account in the parameter uncertainties.
In the redder parts of the SED of Achernar A, there is an increasing uncertainty due to its Be nature, that leads to an excess.
This excess is variable and reflects the history of the disk over the last, typically, decades, and likely is at the origin of the higher scatter of the infrared photometry of Achernar A with respect to the best-fit SED model.
The bolometric luminosities of the two stars were determined from the integrated flux of these best-fit atmosphere models.
The best-fit parameters of the two stars are summarized in the lower part of Table~\ref{table:orbitalParams}.

\begin{table}
 \caption{Flux density $f_{AB}$ of Achernar AB.
 \label{table:photA}}
 \centering
  \begin{tabular}{lcccc}
  \hline \hline
  \noalign{\smallskip}
Ref. & Band & $\lambda$ & $f_{AB}$ & $\sigma(f_{AB})$ \\
	& & ($\mu$m) & (W\,m$^{-2}$\,$\mu$m$^{-1}$) & (W\,m$^{-2}$\,$\mu$m$^{-1}$) \\
 \hline \noalign{\smallskip} 

 1 & U & 0.360 & $6.06 \times 10^{-8}$ & $5.6\times 10^{-10}$  \\ 
 2 & Vt & 0.535 & $2.49 \times 10^{-8}$ & $2.3 \times 10^{-10}$  \\ 
 2 & Hp & 0.559 & $2.66 \times 10^{-8}$ & $5.9 \times 10^{-11}$  \\ 
 3 & J & 1.235 & $1.48 \times 10^{-9}$ & $3.9 \times 10^{-10}$  \\ 
 3 & H & 1.662 & $5.11 \times 10^{-10}$ & $1.8 \times 10^{-10}$  \\ 
 3 & K & 2.159 & $1.90 \times 10^{-10}$ & $6.8 \times 10^{-11}$  \\ 
 4 & L & 3.40 & $3.38 \times 10^{-11}$ & $3.3 \times 10^{-12}$  \\ 
 5 & 4.6\,$\mu$m & 4.60 & $6.16 \times 10^{-12}$ & $8.7 \times 10^{-13}$  \\ 
 6 & S9W & 9.218 & $9.86 \times 10^{-13}$ & $2.9 \times 10^{-14}$  \\ 
 7 & 12\,$\mu$m & 11.50 & $3.83 \times 10^{-13}$ & $1.9 \times 10^{-14}$  \\ 
 5 & 11.6\,$\mu$m & 11.60 & $2.48 \times 10^{-13}$ & $2.3 \times 10^{-15}$  \\ 
 6 & L18W & 18.92 & $5.67 \times 10^{-14}$ & $4.7 \times 10^{-15}$  \\ 
 7 & 25\,$\mu$m & 23.50 & $2.13 \times 10^{-14}$ & $1.3 \times 10^{-15}$  \\ 
 7 & 60\,$\mu$m & 62.00 & $6.01 \times 10^{-16}$ & $1.8 \times 10^{-16}$  \\ 
 
 \hline
\end{tabular}
\tablebib{
(1)~\citetads{2012AstL...38..331A}; (2)~\citetads{1997ESASP1200.....E};
(3)~\citetads{2006AJ....131.1163S}; (4)~\citetads{1978A&AS...34..477M};
(5)~\citetads{2012yCat.2311....0C}; (6)~\citetads{2010A&A...514A...1I};
(7)~\citetads{1988iras....1.....B}.
}
\end{table}

\begin{table}
 \caption{Flux density $f_B$ of Achernar B.
 \label{table:photB}}
 \centering
\begin{tabular}{lccc}
\hline \hline \noalign{\smallskip} 
Filter & $\lambda$ & $f_B$ & $\sigma(f_B)$ \\
	& ($\mu$m) & (W\,m$^{-2}$\,$\mu$m$^{-1}$) & (W\,m$^{-2}$\,$\mu$m$^{-1}$) \\
\hline \noalign{\smallskip} 
V & 0.554 & $3.64\times 10^{-10}$ & $8.0\times 10^{-11}$ \\
CntHa & 0.645 & $1.87\times 10^{-10}$ & $4.2\times 10^{-11}$ \\
B\_Ha & 0.656 & $1.48\times 10^{-10}$ & $6.9\times 10^{-11}$ \\
N\_I & 0.817 & $9.63\times 10^{-11}$ & $2.1\times 10^{-11}$ \\
NB\_1.04 & 1.04 & $7.23\times 10^{-11}$ & $6.3\times 10^{-12}$ \\
NB\_1.08 & 1.08 & $6.60\times 10^{-11}$ & $5.2\times 10^{-12}$ \\
NB\_1.09 & 1.09 & $6.20\times 10^{-11}$ & $5.2\times 10^{-12}$ \\
NB\_1.24 & 1.24 & $4.40\times 10^{-11}$ & $3.6\times 10^{-12}$ \\
NB\_1.26 & 1.26 & $3.91\times 10^{-11}$ & $3.1\times 10^{-12}$ \\
NB\_1.28 & 1.28 & $3.34\times 10^{-11}$ & $2.7\times 10^{-12}$ \\
NB\_1.64 & 1.64 & $1.46\times 10^{-11}$ & $7.2\times 10^{-13}$ \\
H & 1.64 & $1.13\times 10^{-11}$ & $4.7\times 10^{-13}$ \\
NB\_1.75 & 1.75 & $1.19\times 10^{-11}$ & $8.1\times 10^{-13}$ \\
NB\_2.12 & 2.12 & $5.84\times 10^{-12}$ & $2.1\times 10^{-13}$ \\
K & 2.17 & $4.39\times 10^{-12}$ & $1.8\times 10^{-13}$ \\
NB\_2.17 & 2.17 & $4.98\times 10^{-12}$ & $2.6\times 10^{-13}$ \\
NB\_2.18 & 2.18 & $5.11\times 10^{-12}$ & $2.7\times 10^{-13}$ \\
PAH1 & 8.59 & $1.83\times 10^{-14}$ & $5.3\times 10^{-15}$ \\
PAH2 & 11.25 & $6.23\times 10^{-15}$ & $1.6\times 10^{-15}$ \\
\hline
\end{tabular}
\end{table}

\begin{figure}
     \centering
         \includegraphics[width=8cm]{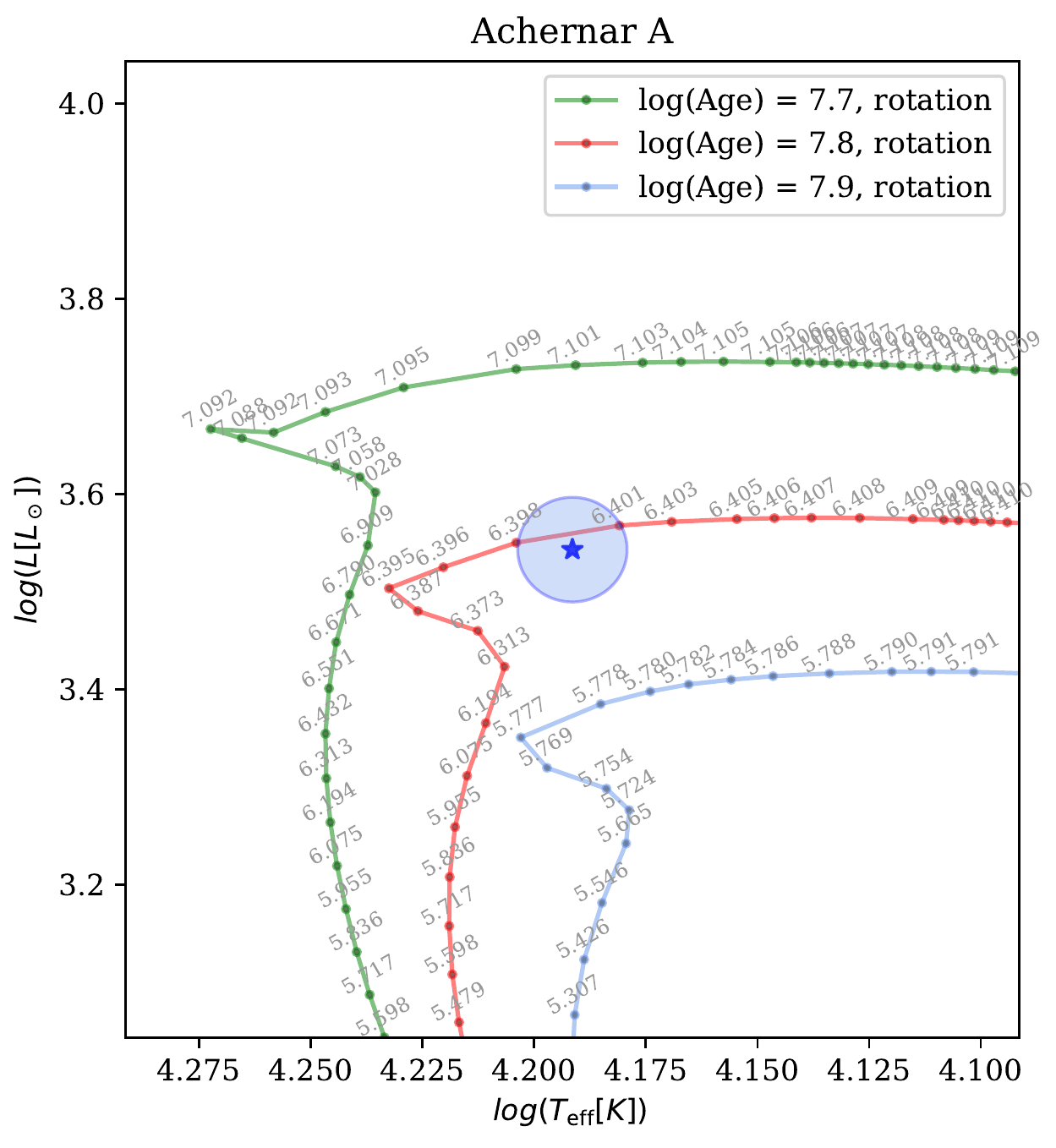}
     \caption{Position of Achernar A in the Herzsprung-Russell diagram, with isochrones from \citetads{2012A&A...537A.146E}.
     The age of each isochrone is in years, expressed in logarithmic scale.
     The labels on the isochrones indicate the mass values ($M_\odot$).
     \label{fig:Isochrones-Achernar-A}}
\end{figure}

\begin{figure}
     \centering
         \includegraphics[width=8cm]{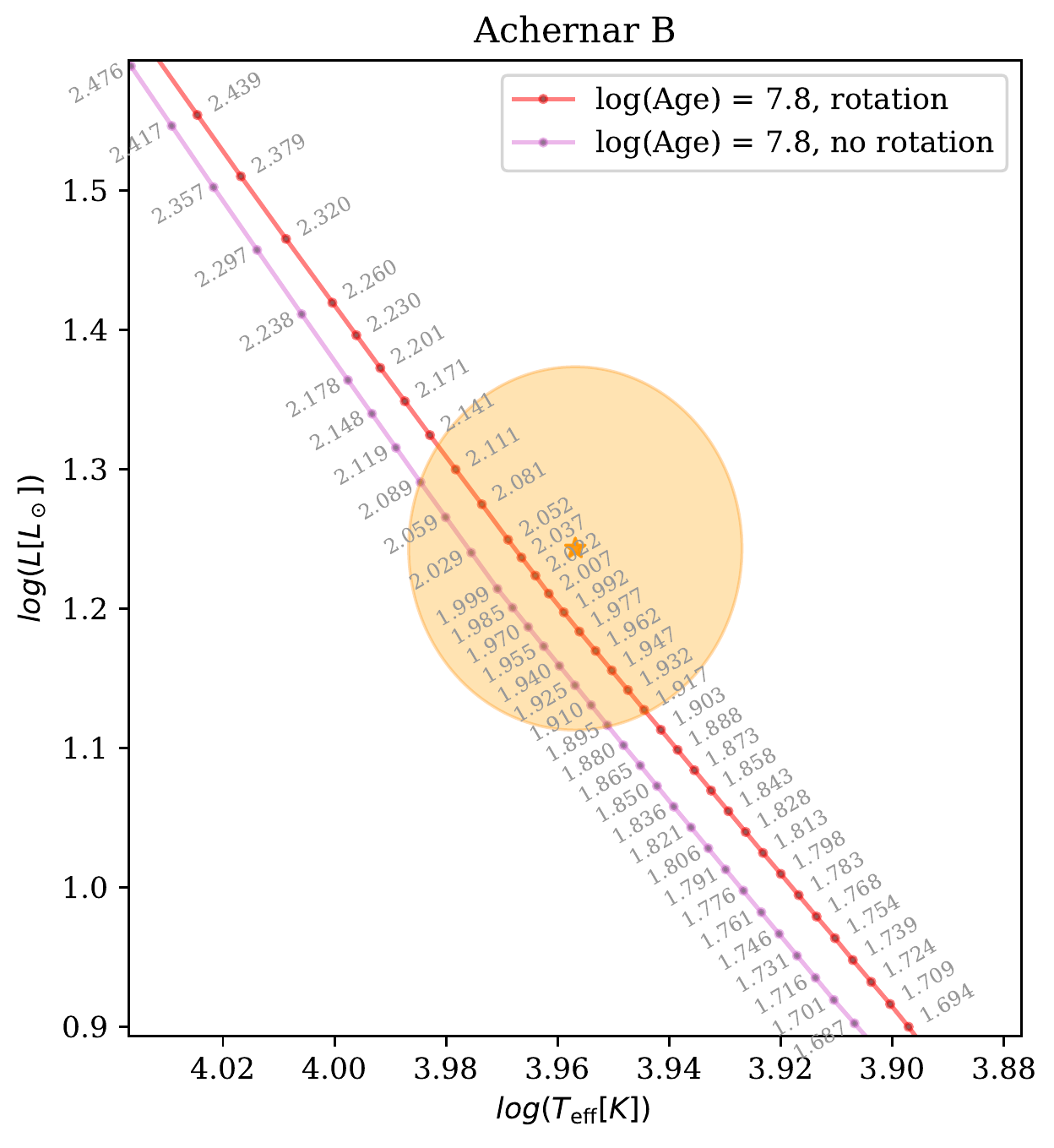}
     \caption{Position of Achernar B in the Herzsprung-Russell diagram, with isochrones from 
     \citetads{2012A&A...537A.146E}. The labels on the isochrones are the mass values ($M_\odot$).
     \label{fig:Isochrones-Achernar-B}}
\end{figure}

Placing Achernar A in the Herzsprung-Russell (HR) diagram with respect to the stellar evolution model isochrones from \citetads{2012A&A...537A.146E} points at an approximate age of $\log \mathrm{Age} = 7.8$ (Fig.~\ref{fig:Isochrones-Achernar-A}).
Based on this young age, we estimate the mass of Achernar B from its position of Achernar B in the HR diagram as shown in Fig.~\ref{fig:Isochrones-Achernar-B}.
As this star is on the main sequence, its mass is well constrained at $m_B = 2.02 \pm 0.11\,M_\sun$.
This estimate is largely independent of the assumed age of the system (between half and twice the adopted age from Achernar~A). 
The rotation status of Achernar B also has a negligible effect on its mass, as the isochrones with and without rotation are very similar (red and pink curves in Fig.~\ref{fig:Isochrones-Achernar-B}, respectively).
The mass of B that we obtain is consistent with a spectral type A2V-A3V from the grid by \citetads{2013ApJS..208....9P}\footnote{\url{http://www.pas.rochester.edu/~emamajek/EEM_dwarf_UBVIJHK_colors_Teff.txt}} (see also \citeads{2012ApJ...746..154P}).
It is in very good agreement with the A1V-A3V spectral type predicted by \citetads{kervella2008} based on a NACO spectrum of Achernar B. 

\subsection{Orbital parameters and mass of Achernar A\label{sect:massA}}

\begin{figure}
     \centering
         \includegraphics[width=\hsize]{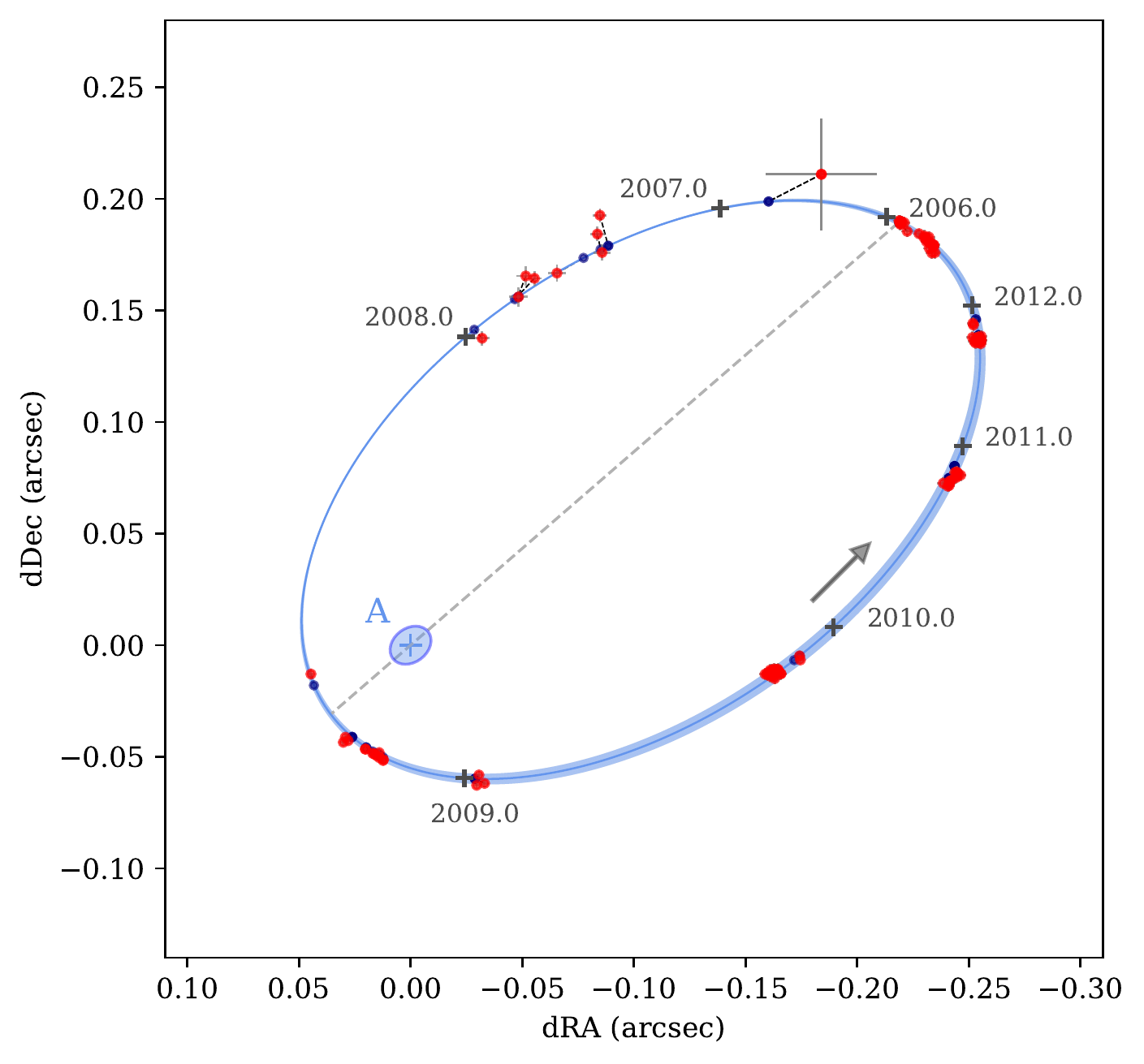}
     \caption{Best-fit relative orbit of Achernar B around A.
     The measurements are shown with red dots, and the corresponding model positions as dark gray dots (linked with dashed lines).
     The line of nodes is represented with a gray dashed segment.
     The thicker part of the orbital trajectory represents the section when Achernar B is closer to us than Achernar A.
     The rotationally flattened profile of the photosphere of Achernar A is represented enlarged by a factor ten. \label{fig:ABorbit}}
\end{figure}

\begin{figure}
     \centering
         \includegraphics[width=\hsize]{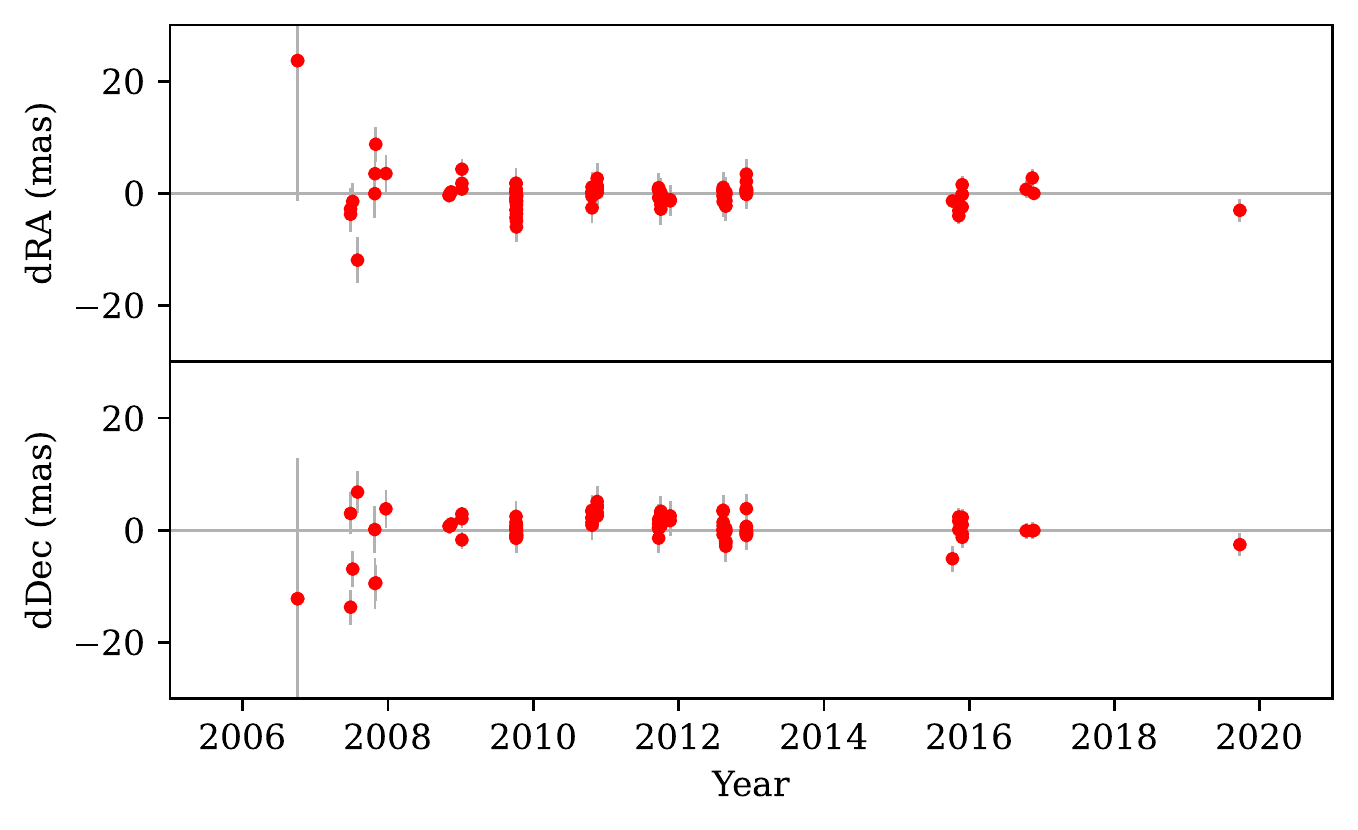}
     \caption{Residuals of the astrometry of Achernar B relative to A for the best-fit orbit as a function of time. \label{fig:ABastromResiduals}}
\end{figure}

\begin{figure}
     \centering
         \includegraphics[width=\hsize]{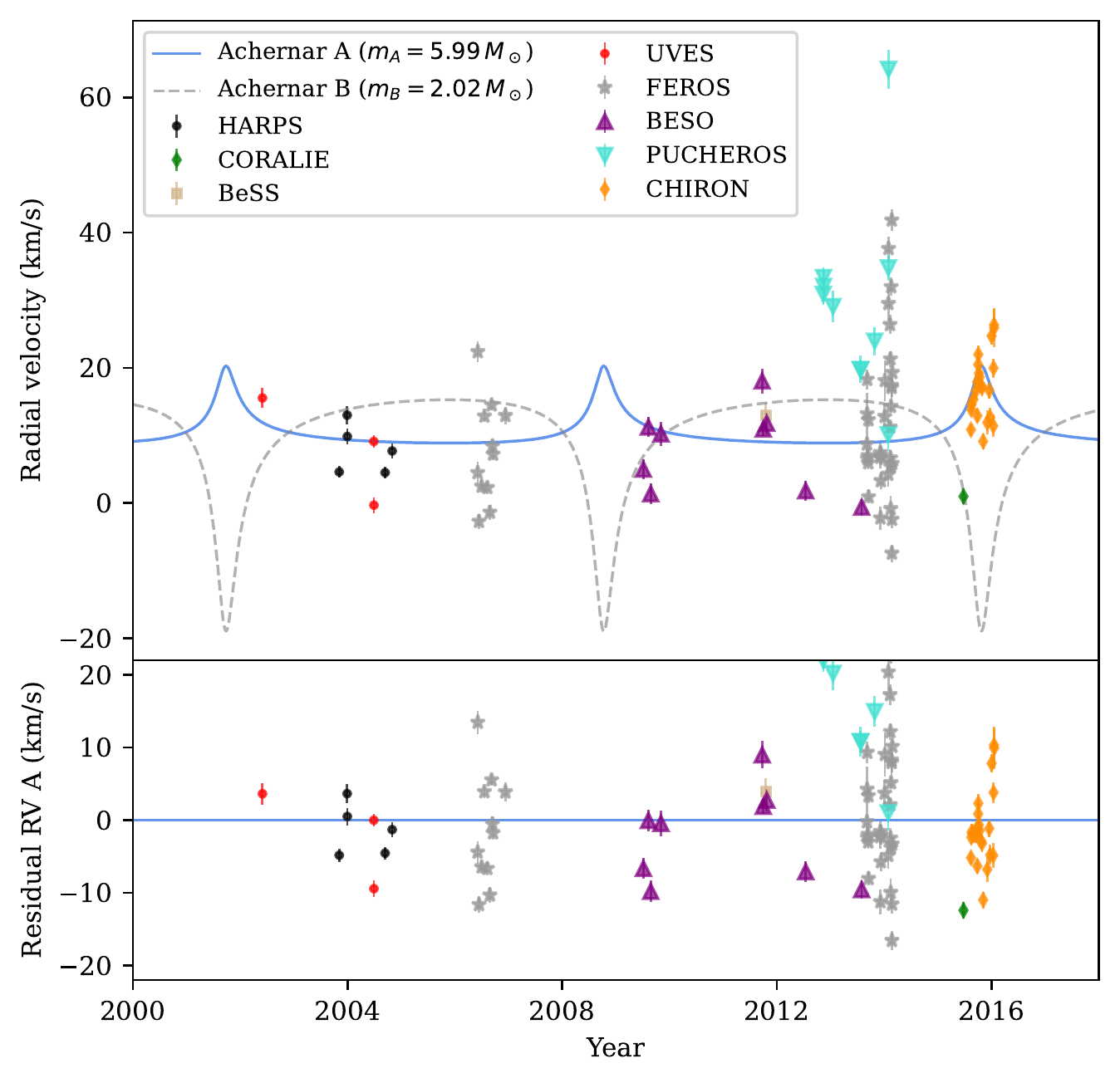}
     \caption{Radial velocity measurements of Achernar A compared to the best-fit orbital model of the AB pair (top panel) and the corresponding residuals (bottom panel). The short-period oscillations (Sect.~\ref{sect:RV}) with a period of 1.29\,day have been subtracted from the measurements. The dashed curve represents the predicted radial velocity curve of Achernar~B. \label{fig:ABRV}}
\end{figure}

To determine the orbital parameters of the system, we considered all the nonflagged relative astrometry listed in Table~\ref{tab:allastrometry1}. We complemented this data set with the radial velocity measurements of Achernar A summarized in Table~\ref{tab:allRV1}.
However, due to the low accuracy of the radial velocities, we derive the orbital elements, total mass and radial velocity of the barycenter of A and B in two steps:
\begin{enumerate}
\item Computation of all the orbital elements and the total mass of the system based exclusively on the differential astrometry and the Hipparcos parallax ($\varpi = 23.39 \pm 0.57$\,mas; \citeads{2007ASSL..350.....V}).
\item Determination of the radial velocity of the AB barycenter from a second fit including the radial velocities from which the 1.29\,day period oscillations have been subtracted (Fig.~\ref{fig:ABRV}), while fixing the other parameters to the best-fit values obtained at step 1.
\end{enumerate}
The orbital elements of Achernar B's orbit are summarized in Table~\ref{table:orbitalParams}, together with the total mass of the system and the barycentric radial velocity.
The orbital trajectory of B with respect to A on the sky is shown in Fig.~\ref{fig:ABorbit}.
We do not identify the signature of a third component in the inner Achernar system from the residuals of the astrometric measurements with respect to the best fit orbit (Fig.~\ref{fig:ABastromResiduals}).

The error budget of the total mass of the system ($\pm 7.4\%$) is heavily dominated by the uncertainty on the Hipparcos parallax of the system, as assuming a zero uncertainty on the parallax would result in a total mass uncertainty of only $\pm 2\%$.
Subtracting the mass $m_B = 2.02 \pm 0.11\,M_\sun$ determined in Sect.~\ref{sect:massB} from the total mass, we estimate the mass of Achernar A to $m_A = 5.99 \pm 0.60\,M_\sun$ ($\pm 10\%$).
This value is consistent within $0.7\sigma$ with the mass of $6.4\,M_\sun$ at an age of $\approx 60$\,Ma predicted for Achernar A from its position in the HR diagram (Fig.~\ref{fig:Isochrones-Achernar-A}) using the model isochrones by \citetads{2012A&A...537A.146E}.

As a side note, adopting the 3\% lower Hipparcos parallax value from the original reduction ($\varpi = 22.68 \pm 0.57$\,mas; \citeads{1997A&A...323L..49P}) instead of the reprocessed parallax by \citetads{2007ASSL..350.....V} results in a 10\% higher total mass of $m_\mathrm{tot} = 8.8 \pm 0.7\,M_\sun$ for the Achernar system (the total mass scales with $\varpi^{-3}$).
In the present work, we choose to adopt the parallax from \citetads{2007ASSL..350.....V}, that is well validated \citepads{2007A&A...474..653V}.
Achernar heavily saturates the Gaia detectors, and is absent for this reason from the recent Gaia DR3 catalog \citepads{2022arXiv220800211G}. However, as discussed by \citetads{2018IAUS..330..343S}, a significant improvement by up to a factor 10 of the Gaia astrometric parameters of naked eye stars is expected in the coming years, using dedicated data acquisition and processing techniques.
Such a high precision on the parallax will result in a much better constrained total mass for the Achernar system. In addition, the measurement by Gaia of the orbital astrometric wobble of Achernar~A combined with the orbital parameters from the present work will result in a strong constraint on the mass ratio $m_B/m_A$, potentially providing the mass of Achernar~A to an accuracy on the order of 1\%.

\begin{table}
 \caption{Orbital elements and parameters of Achernar A and B.
 \label{table:orbitalParams}}
 \centering
  \begin{tabular}{lrr}
\hline \hline \noalign{\smallskip} 
Parameter &  & \multicolumn{1}{c}{Value} \\
\hline \noalign{\smallskip} 

Hip. parallax\tablefootmark{a} (mas) & $\varpi_\mathrm{Hip}$ & $23.39 \pm 0.57$ (2.4\%) \\ 
 \hline \noalign{\smallskip} 
Semi-major axis (mas) & $a$ & $171.90 \pm 0.25$ \\ 
Semi-major axis (au) & $a$ & $7.35 \pm 0.18$ \\ 
Inclination (deg) & $i$ &   $30.32 \pm 0.35$ \\ 
Eccentricity & $e$ &     $0.7258 \pm 0.0015$ \\ 
Arg.\,of periastron (deg) & $\omega$ & $172.05 \pm 0.87$ \\ 
Long.\,of asc.\,node (deg) & $\Omega$   &    $310.91 \pm 0.80$ \\ 
Period (days) & $P$   &     $2570.94 \pm 0.53$ \\ 
Period (a) & $P$   &  $7.0389 \pm 0.0015$ \\ 
Ref.\,epoch (MJD) & $T_0$ &  $54743.44 \pm 0.53$ \\ 
Ref.\,epoch (Jyear) & $T_0$ &  $2008.7582 \pm 0.0014$ \\ 
Barycentric RV (km\,s$^{-1}$) & $v_R$ & $+10.49 \pm 0.51$ \\ 
Total mass ($M_\odot$) &  $m_\mathrm{tot}$ & $8.01 \pm 0.59$ (7.4\%) \\ 
 \hline \noalign{\smallskip} 
Ang. diam. of A\tablefootmark{b} (mas)  & $\overline{\theta_\mathrm{LD,A}}$ & $1.770 \pm 0.035$ (2.0\%) \\ 
Ang. diam. of B (mas)  & $\theta_\mathrm{LD,B}$ & $0.370 \pm 0.014$ (3.8\%) \\ 
Radius of A ($R_\odot$)  & $R_A$ & $8.14 \pm 0.26$ (3.2\%) \\ 
Radius of B ($R_\odot$)  & $R_B$ & $1.70 \pm 0.08$ (4.5\%) \\ 
$T_\mathrm{eff}$ of A (K)  & $T_\mathrm{eff, A}$ & $15539 \pm 438$ (2.8\%) \\ 
$T_\mathrm{eff}$ of B (K)  & $T_\mathrm{eff, B}$ & $9064 \pm 624$ (6.9\%) \\ 
Luminosity of A ($L_\odot$)  & $L_A$ & $3493 \pm 429$ (12\%) \\ 
Luminosity of B ($L_\odot$)  & $L_B$ & $17.5 \pm 5.1$ (29\%) \\ 
 \hline \noalign{\smallskip} 
Model mass of B ($M_\odot$)  & $m_B$ & $2.02 \pm 0.11$ (5\%) \\ 
Mass of A ($M_\odot$)  & $m_A$ & $5.99 \pm 0.60$ (10\%) \\ 

 \hline
\end{tabular}
\tablefoot{\tablefoottext{a}{Hipparcos parallax from \citetads{2007ASSL..350.....V}.
\tablefoottext{b}{The angular diameter $\overline{\theta_\mathrm{LD,A}}$ of Achernar A is the mean angular diameter determined by \citetads{desouza2014}.}
}}
\end{table}

\section{Discussion \label{discussion}}

\subsection{Spin-orbit relative orientation}

The inclination of the polar axis of Achernar A on the line of sight and its position angle with respect to the north have been determined by \citetads{desouza2014} using optical interferometry:
$i_\mathrm{rot,A}=60.6^{+7.1}_{-3.9}\,\deg$ and $\mathrm{PA}_\mathrm{rot,A} = 216.9 \pm 0.4\,\deg$.
These geometrical parameters differ from those of the orbital plane of Achernar B's orbit:
$i_\mathrm{AB} = 30.3 \pm 0.4\,\deg$ and $\Omega_\mathrm{AB} = 310.9 \pm 0.8\,\deg$ (Table~\ref{table:orbitalParams}).
The relative inclination between the equatorial plane of Achernar A and the orbital plane of the AB pair is
$\Delta i =  i_\mathrm{rot,A} - i_\mathrm{AB} = 30^{+7}_{-4}\,\deg$.
Such a spin-orbit misalignment is observed in other stellar systems (see, e.g., \citeads{2018MNRAS.478.1942S, 2007A&A...474..565A, 2011ApJ...726...68A, 2013ApJ...767...32A}).
The direction of the rotation of Achernar A and of the revolution of B is the same (counterclockwise in projection on the plane of the sky).
The difference in position angle on the sky of the equator of Achernar A with respect to the line of nodes of the orbit of B is small at
$\Delta \mathrm{PA} = (\mathrm{PA}_\mathrm{rot,A}+90^\circ)  - \Omega_\mathrm{AB} = -4 \pm 1\,\deg$ (Fig.~\ref{fig:ABorbit}).
A progressive alignment of the orbital plane of B with the spin of A may occur in the future either through tidal interactions \citepads{2017MNRAS.468.1387L} or possibly mass transfer when Achernar A will evolve into a supergiant (Sect.~\ref{futureevolution}).

\subsection{Achernar B and the Be decretion disk}

\citetads{2022MNRAS.509..931S} modeled the effect of the presence of a misaligned orbiting companion on the Be star disk. In the case of the Achernar system, the high eccentricity $e=0.726$ brings component B between $12.7$\,au (apastron) and $2.0$\,au (periastron).
At its closest approach from Achernar A, the secondary is therefore located at $\approx 47\,R_\mathrm{eq}$ from the primary, with $R_\mathrm{eq} = 9.2\,R_\sun$ the equatorial radius of Achernar A \citepads{desouza2014}.
The decretion disk of Achernar during the Be phases was measured by \citetads{vedova2017} using optical interferometry to extend in the near-infrared up to $\approx 2\,R_\mathrm{eq}$ (see also \citeads{carciofi2007}).
This is consistent with the results of a modeling with the SIMECA code by \citetads{2008A&A...486..785K}, who found that an equatorial disk with a radius $\leq 5\,R_\mathrm{eq}$ reproduces the observations.
These observations were collected during a phase of relatively moderate \half~emission of Achernar A (see also Sect.~\ref{sect:halpha} and Fig.~\ref{fig:halpha}), and the disk was likely less extended than during typical emission phases.
\citetads{2019ApJ...885..147K} observed the disk emission of Achernar in the millimeter domain around the 2015 periastron passage of Achernar B, but did not detect a significant variation caused by the approach of the companion.
The approximate formula by \citetads{1983ApJ...268..368E} gives a Roche radius of $\approx 1.0$\,au for Achernar~A at periastron, or half of the periastron separation. Mass transfer from the Be star to Achernar B therefore appears unlikely considering the limited extension of the decretion disk (on the order of 0.2\,au, considering an extension of $5\,R_\mathrm{eq}$), except possibly during particularly intense Be phases.

\subsection{Present evolutionary state of Achernar}

\begin{figure*}
     \centering
         \includegraphics[height=6cm]{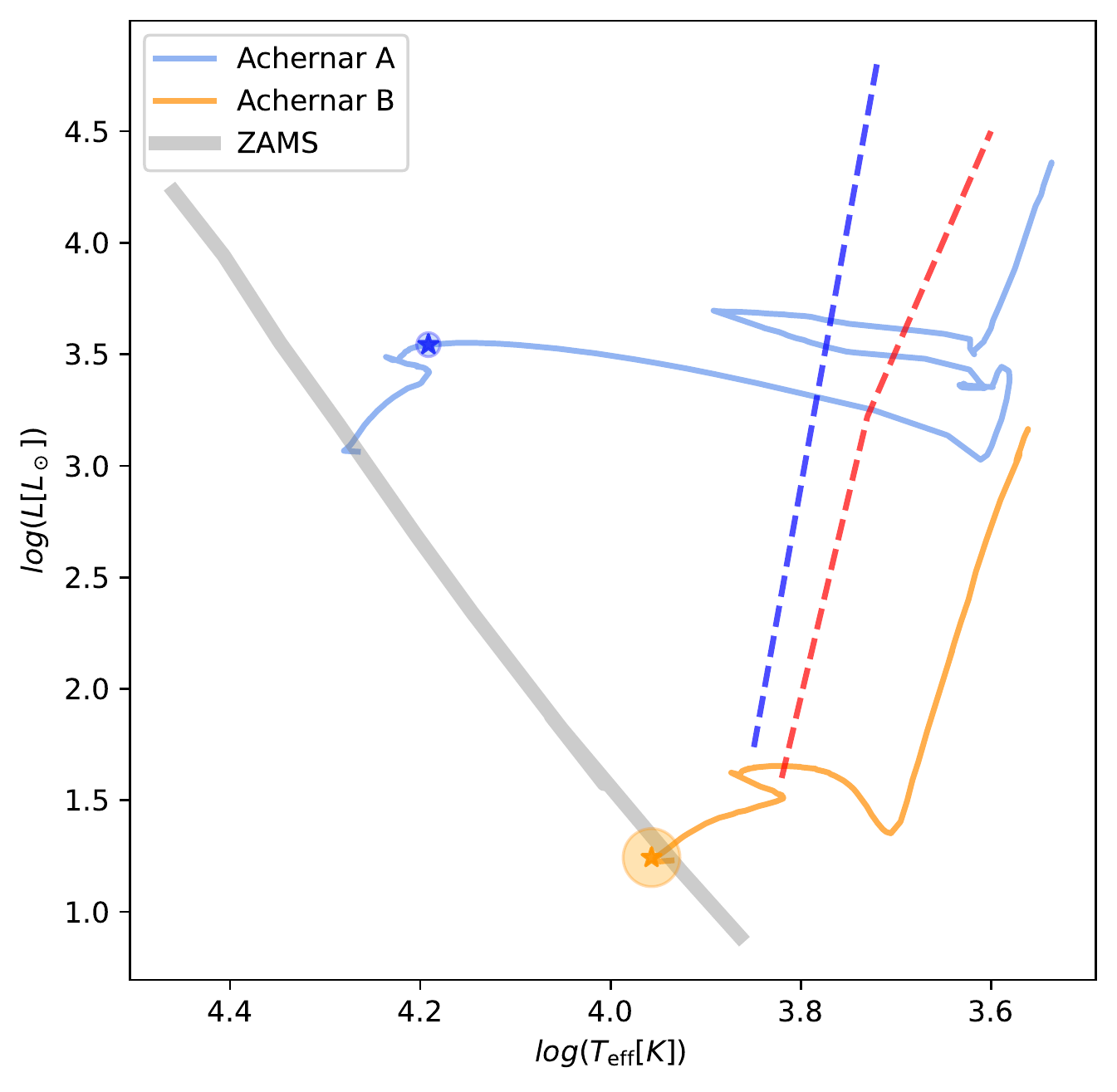}
         \includegraphics[height=6cm]{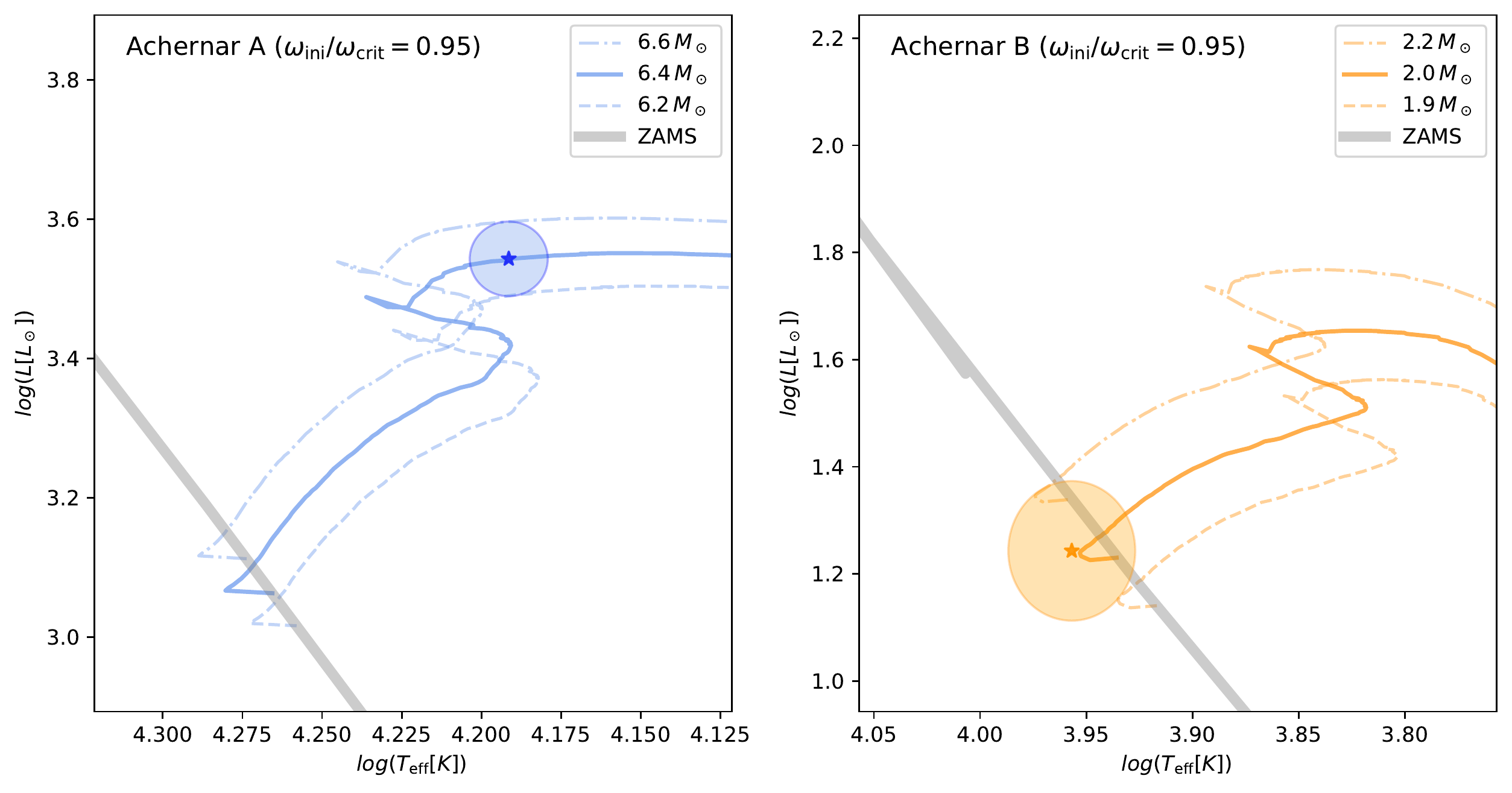}
     \caption{Evolutionary tracks for Achernar A and B from \citetads{2013A&A...553A..24G} for Z=0.014, $m_A = 6.4\,M_\sun$, $m_B = 2.0\,M_\sun$ and $\omega_\mathrm{ini}/\omega_\mathrm{crit}=0.95$ for both stars. The approximate boundaries of the instability strip from \citetads{2016A&A...591A...8A} are shown with blue and red dashed curves.
     \label{fig:Georgy-models}}
\end{figure*}

To evaluate the age and evolutionary state of Achernar, we adopt the stellar evolution models with rotation by \citetads{2013A&A...553A..24G}. 
Here we assume that Achernar A is not the result of a past stellar merger. This possibility appears unlikely considering the orbital configuration of the AB system and its relatively young age.
The metallicity is taken to be solar ($Z=0.014$) and the initial rotation velocity as 95\% of the critical velocity for both stars.
The latter hypothesis is justified by the present rotation velocity of Achernar A, which was determined to be very close to critical by \citetads{desouza2014} at $\omega/\omega_\mathrm{crit} = 0.98$.
Figure~\ref{fig:Georgy-models} shows the best-fit evolutionary tracks of both stars.
For Achernar B, the choice of rotation velocity hypothesis has mostly no effect on the estimation of its mass (see also Sect. \ref{sect:massB}).
The mass range for Achernar A is found to be $m_A = 6.4 \pm 0.2\,M_\sun$ at an age of $63 \pm 4$\,Ma, in agreement with the $6.0 \pm 0.6\,M_\sun$ mass determined in Sect.~\ref{sect:massA}.
The best-fit model is overall in good agreement with the observed properties of Achernar A, that passed the turn-off $\approx 70\,000$\,years ago.
With a present composition of its core of $X \approx 6\%$ and $Y\approx 93\%$, it is at the beginning of the hydrogen-shell burning phase, rapidly crossing the HR diagram toward the helium-burning phase.

\begin{figure}
     \centering
         \includegraphics[width=\hsize]{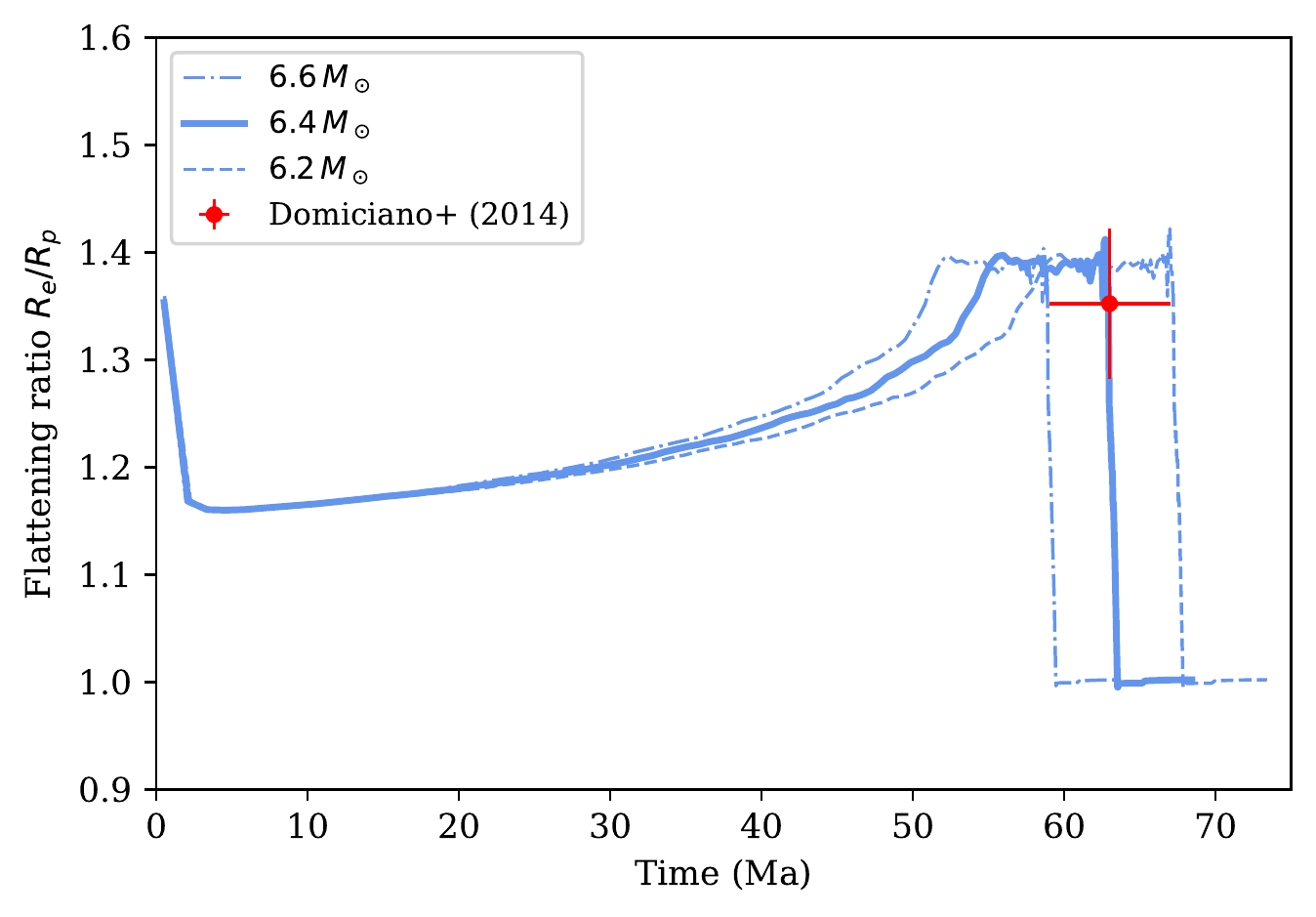}
     \caption{Evolution of the flattening ratio (equatorial radius divided by the polar radius) for three models of Achernar A, compared with the estimate from \citetads{desouza2014} shown as a red dot.
     \label{fig:Georgy-flattening}}
\end{figure}

As shown in Fig.~\ref{fig:Georgy-flattening}, the expected geometric flattening ratio of its photosphere $R_\mathrm{eq}/R_\mathrm{pol} = 1.39$ is consistent with the value of $1.35 \pm 0.07$ determined by \citetads{desouza2014} from interferometric observations.
The present state of Achernar corresponds to its strongest flattening over its complete evolution. This occurs shortly before the rapid inflation toward the supergiant class, when the flattening will promptly disappear.
The curve of the rotational velocity evolution depends on the adopted initial rotation rate. However, as soon as the star reaches critical rotation, the initial rotation velocity conditions become inaccessible, because an unspecified amount of angular momentum has been lost through the disk.
The evolution of Achernar during its first 50\,Ma is therefore uncertain as it depends significantly on the adopted initial rotation velocity.

The secondary component Achernar B is a regular main sequence star, whose early A spectral type implies that it has statistically a high probability of being also a fast rotator \citepads{2007A&A...463..671R,2012A&A...537A.120Z}.
Following the recent Be star survey by \citetads{2020A&A...641A..42B}, Achernar is thus presently the only known Be star with a main sequence companion.
Achernar B does not show evidence of a past mass-transfer with Achernar A.
This is therefore a different configuration from that reported recently by \citetads{2022arXiv220105614E} in the binary Be star \object{HD 15124}.
Considering the present evolutionary state of Achernar A, Achernar B will not evolve into a subdwarf such as the companions detected by \citetads{2022ApJ...926..213K} orbiting three classical Be stars (28 Cyg, V2119 Cyg, and 60 Cyg).
The longer orbital period of Achernar B (7 years) compared to these three binary systems (a few months) prevented mass transfer between the two components.
This absence of interactions with Achernar B implies that Achernar A is presently the first and only identified case of a single-star evolution track Be star. The proximity of the two stars at periastron also prevents the presence of a third component that could have spun up A in the past through mass transfer.
The single-star evolutionary models predict that the rotation of Achernar A became progressively closer to critical (Fig.~\ref{fig:Georgy-flattening}) and drove a formerly normal B star into the Be star regime without requiring mass-transfer.
This direct mechanism to form a Be star therefore appears to be a plausible alternative to the binary mass-transfer scenario discussed by \citetads{2020A&A...641A..42B} (see also \citeads{2021ApJ...908...67S}).
The properties of the Achernar system also strengthen the hypothesis by \citetads{2021A&A...653A.144H} that a significant fraction of Be stars were formed through single-star evolution.

As it is a main sequence star, the high contrast between Achernar A and B makes it a difficult detection. Additionally, as the shape of Achernar B's SED is generally comparable to that of Achernar A (Fig.~\ref{fig:SED-Achernar-AB}), its presence does not cause a significant color-dependent excess that could be employed to detect it.
Finally, the fact that Achernar A and probably also B are fast rotators prevents the spectroscopic detection of such main sequence companions from their signature in radial velocity (Sect.~\ref{sect:radvelocityA}).
As a consequence, direct imaging and long-baseline interferometry appear as the most efficient techniques to search for main sequence companions of Be stars.
To estimate the distance up to which Achernar B would be detectable using SPHERE adaptive optics, we observe that the detection of B is difficult close to periastron (Table~\ref{tab:allastrometry1}), that is, at a separation of $\approx 50$\,mas.
Considering the apastron separation at discovery, this implies in practice that the detection of companions similar to B using adaptive optics is limited to a radius of $\approx 6\times$ its actual distance, that is, 250\,pc (this could be improved using, e.g., coronagraphic techniques).
Thanks to its high angular resolution ($\approx 1$\,mas) and contrast ($\approx 200$;  \citeads{2019A&A...622A.164G}), optical long-baseline interferometry would have the capacity to detect a twin of Achernar B up to $\approx 3$\,kpc. At this distance, and neglecting interstellar reddening, the apparent K band magnitude of Achernar ($m_K \approx 10$) would be accessible to the GRAVITY fringe tracker with the VLTI Unit telescopes.

\subsection{Future evolution of Achernar \label{futureevolution}}
According to the position of the blue edge of the classical instability strip (IS) estimated by \citetads{2016A&A...591A...8A}, the best-fit evolution model of Achernar A ($m_A = 6.4\,M_\sun$) implies that it will enter the IS in $\approx 400\,000$\,years (Fig.~\ref{fig:Georgy-models}, left panel). The first crossing of the IS will last only $\approx 10\,000$\,years, during which Achernar will be an intermediate mass classical Cepheid, with a radius $R \approx 50\,R_\sun$ and a fundamental mode pulsation period $P \approx 5$\,days (adopting the period-mass-radius relation by \citeads{2018ApJ...862...43P}; see also \citeads{2021A&A...656A.102T}).
The Achernar system therefore appears to be a promising progenitor of the numerous Cepheid binary or multiple systems that are observed in the Milky Way \citepads{2013A&A...552A..21G, 2014A&A...561L...3G, 2019A&A...622A.164G, 2019A&A...623A.117K, 2019A&A...623A.116K} and in the Magellanic Clouds \citepads{2015ApJ...815...28G,2010Natur.468..542P,2013MNRAS.436..953P,2015ApJ...806...29P,2018ApJ...862...43P}.
After its first crossing of the IS, the radius of Achernar A will continue to increase, reaching a maximum of $120\,R_\sun$ in $\approx 600\,000$\,years. At this stage, due to the relatively high orbital eccentricity, interactions may occur between the supergiant and Achernar B (periastron at $\approx 430\,R_\sun$). The approach of the secondary star (mostly unevolved compared to its present stage) within $< 4\,R_\star$ from the primary during this phase may result in a modification of their evolutionary path, that is not reflected in the present model.

\subsection{Comoving star and the Tuc-Hor association \label{comoving_star}}

\begin{figure*}
     \centering
         \includegraphics[width=13cm, page=1]{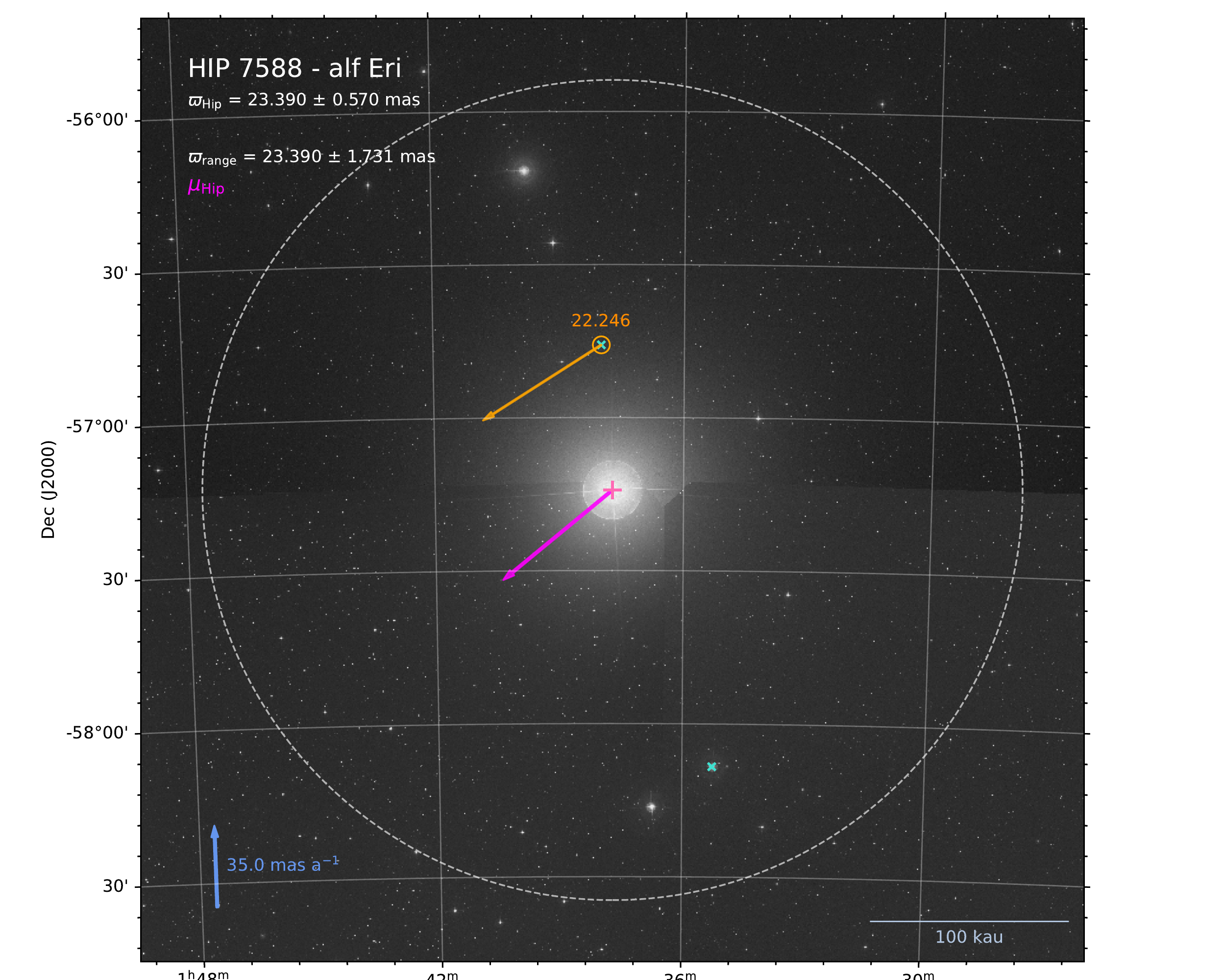}
     \caption{Achernar (center) and the comoving star 2MASS J01375879-5645447. The dashed circle has a radius of 1\,pc at the distance of Achernar. \label{fig:AchernarCPM}}
\end{figure*}

From their survey of nearby common proper motion (hereafter PM) companions in the solar neighborhood, \citetads{2022A&A...657A...7K} identified a star comoving with Achernar, \object{2MASS J01375879-5645447} (\object{TIC 230982053}, \object{Gaia EDR3 4911252913514571648}).
For convenience, we refer to this star in the following as 2M.
It is a relatively faint red dwarf ($G=13.48$, $K=9.53$) with an estimated mass of $0.41\,M_\sun$ and a radius of $0.40\,R_\sun$ (\citeads{2022A&A...657A...7K}, using the relations by \citeads{2015ApJ...804...64M}).
These parameters correspond to a red dwarf of approximate spectral type M2.5V \citepads{2013ApJS..208....9P}. This is slightly warmer and more massive than the M3.8V($\pm 0.3$) spectral type reported by \citetads{2014AJ....147..146K}, who classified 2M as a candidate member of the young Tucana-Horologium (Tuc-Hor) moving group.
These authors also determined its radial velocity to be $v_\mathrm{rad}(\mathrm{2M}) = 8.5 \pm 0.6$\,km\,s$^{-1}$, and its projected rotation velocity $v_\mathrm{rot}\,\sin i = 9.8 \pm 0.7$\,km\,s$^{-1}$.
2M is listed as a single star in the catalog by \citetads{2017A&A...599A..70J}, but the relatively high value of its Gaia EDR3 RUWE (1.5) points at the possible presence of an additional close-in object (see Sect.~4.2 of \citeads{2022A&A...657A...7K}).
\citetads{2015MNRAS.454..593B} and \citetads{2015ApJ...798...73G} both find that 2M is a member of the Tuc-Hor moving group, whose age is estimated to $45 \pm 4$\,Ma, and propose M4.0V and M3.9V spectral types, respectively.

The astrometric parameters of Achernar and 2M are listed in Table~\ref{table:comoving}. The Gaia EDR3 parallax of 2M is within $2\sigma$ of the Hipparcos parallax of Achernar.
2M is presently located at a projected separation of 73\,kau from Achernar, and an extrapolation of their projected separation in the past shows that it passed through a minimum of $40$\,kau approximately 150,000 years ago.
Their PM vectors are tightly aligned to less than 5\,degrees, and their relative space velocity is only $\Delta v = 2.55 \pm 0.88$\,km\,s$^{-1}$. This value is significantly larger than the present escape velocity of the three stars, that is at most $v_\mathrm{esc} = \sqrt{2 G M_\mathrm{tot}/R} = 0.45$\,km\,s$^{-1}$ for the total mass of $M_\mathrm{tot} = 8.4\,M_\sun$ and the projected separation $R=73$\,kau. We conclude that Achernar AB and 2M are not gravitationally bound.
However, their very low relative velocity points at a probable membership of Achernar in the Tuc-Hor association.
This possibility is reinforced by the similarity of the ages of Achernar ($\approx 60$\,Ma; Figs.~\ref{fig:Isochrones-Achernar-A} and \ref{fig:Georgy-models}) and of Tuc-Hor ($45 \pm 4$\,Ma; \citeads{2015MNRAS.454..593B}).
If confirmed, this membership would make of Achernar the most massive object of this young stellar association, and therefore a valuable age indicator for the moving group.

\begin{table*}
 \caption{Astrometry and radial velocity of Achernar and 2MASS J01375879-5645447 (Gaia EDR3 4911252913514571648).
 \label{table:comoving}}
 \centering
  \begin{tabular}{lccccc}
  \hline \hline
  \noalign{\smallskip}
Parameter (ICRS) &  & Value & Epoch & Reference \\
 \hline \noalign{\smallskip} 
Achernar position & $(\alpha,\delta)_\mathrm{AB}$ & (01:37:43.01696, $-57$:14:12.9219) $\pm (5.1,4.4)$\,mas & 1991.25 & 1 \\
Achernar parallax & $\varpi_\mathrm{AB}$ & $23.39 \pm 0.57$\,mas & 1991.25 & 1\\
Achernar PM & $\mu_\mathrm{AB}$ & $(+87.00 \pm 0.58, -38.24 \pm 0.50)$\,mas\,a$^{-1}$ & 1991.25 & 1\\
Achernar RV & $v_{R,\mathrm{AB}}$ & $+10.5 \pm 0.5$\,km\,s$^{-1}$ & 2013 & 2 \\
 \hline \noalign{\smallskip} 
2M position & $(\alpha,\delta)_\mathrm{2M}$ & (01:37:58.97272, $-56$:45:45.3427) $\pm (0.014,0.014)$\,mas & 2016.0 & 3\\
2M parallax  & $\varpi_\mathrm{AB}$ & $22.246 \pm 0.020$\,mas & 2016.0 & 3 \\
2M PM & $\mu_\mathrm{2M}$ & $(+92.992 \pm 0.017, -32.009 \pm 0.018)$,mas\,a$^{-1}$ & 2016.0 & 3 \\
2M RV & $v_{R,\mathrm{2M}}$ & $+8.5 \pm 0.6$\,km\,s$^{-1}$ & 2012.5 & 4 \\
 \hline
\end{tabular}
\tablebib{
(1)~\citetads{2007ASSL..350.....V};
(2)~Present work;
(3)~\citetads{GaiaEDR3content};
(4)~\citetads{2014AJ....147..146K}.}
\end{table*}

\section{Conclusion \label{conclusion}}

Using a combination of high angular resolution astrometric measurements of Achernar A and B collected over 13\,years with the VLT(I), we established the elements of the 7-year orbit of the system.
We also collected archival and new radial velocity measurements from spectroscopy spread over more than 12\,years.
Due to the presence of a significant scatter in these velocities induced in particular by the oscillations of Achernar A, we could not clearly detect the signature of the orbital motion in the radial velocity series.
However, we used these measurements to determine the barycentric radial velocity of the system.
Adopting the Hipparcos parallax, we determined the total mass of Achernar A and B, and deduced the mass of the main Be component ($m_A = 5.99 \pm 0.60\,M_\sun$) by subtracting a model-based estimate of the mass of the A2V-A3V secondary ($m_B = 2.02 \pm 0.11\,M_\sun$).
We do not observe a significant correlation between the orbital motion of Achernar B and the Be emission phases of Achernar A, although a moderate emission enhancement is observed around the periastron from late 2015.
As no significant mass-transfer occurred with Achernar B in the past, we argue that Achernar A is presently the first and only identified example of a single-star evolution track Be star. It is also the only known Be star with a regular main sequence companion.
Achernar A therefore demonstrates the existence of a direct mechanism to form a Be star, without the need for binary mass-transfer \citepads{2020A&A...641A..42B} as found, for example, in the $\phi$\,Per system \citepads{2015A&A...577A..51M}.
Achernar A will enter in the classical instability strip in $\approx 400\,000$\,years, potentially becoming an intermediate-mass Cepheid pulsating star.
Due to the eccentricity of the orbit of Achernar B, there is however a possibility that Achernar A and B interact during the transition of the primary to the supergiant phase, possibly affecting the evolutionary paths of the two stars.
The close similarity between the space motion of Achernar and the red dwarf 2MASS J01375879-5645447 leads us to propose that Achernar is a member of the  young Tucana-Horologium moving group. The age of this association ($45 \pm 4$\,Ma; \citeads{2015MNRAS.454..593B}) appears comparable to the value that we determine for Achernar A ($\approx 60$\,Ma) using the evolutionary tracks by \citetads{2013A&A...553A..24G}.

\begin{acknowledgements}
We dedicate this work to the memory of Dr Dimitri Pourbaix (1969-2021).
We thank Dr Richard I. Anderson for discussions on the evolution of Achernar that lead to improvements of this article.
Based on observations collected at the European Organisation for Astronomical Research in the Southern Hemisphere under ESO programmes:
082.D-0227(A), 082.D-0227(B), 082.D-0227(C), 082.D-0227(D) for AMBER (we thank the CNRS-France for the GTO time allocated to this VISA-CNRS program);
266.D-5655(A), and 073.D-0547(A) for UVES; 098.D-0522(B) for PIONIER;
098.D-0522(A) and 0102.D-0303(C) for GRAVITY;
096.D-0353(A) for SPHERE;
078.D-0295(A), 078.D-0295(B) for VISIR;
079.C-0036(B), 279.D-5064(A), 082.C-0577(A), 382.D-0065(A), 384.D-0504(A), 386.D-0706(E), 087.D-0150(B), 088.D-0145(A), 089.D-0800(A), 090.D-0755(A), 60.A-9026(A) for NACO;
072.C-0513(B), 60.A-9036(A), 073.C-0784(B), and 074.C-0012(A) for HARPS;
077.D-0390(A), 077.D-0605(C), 60.A-9120(B), 089.A-9032(B), and 091.A-9032(D) for FEROS.
Based on observations made through the Chilean Telescope Time under program ID CN-15B-4 for CHIRON data (plan ID 348). Additional observations were collected with the CORALIE spectrograph on the Euler telescope (Swiss Observatory) at La Silla, Chile, under program number CN2015A-6 (observer: A. Gallenne); with the BESO spectrograph at Cerro Armazones, Chile; with the PUCHEROS spectrograph at PUC Observatory, Santiago, Chile; and from the BeSS database. 
This work has made use of the BeSS database, operated at LESIA, Observatoire de Meudon, France (\url{http://basebe.obspm.fr}). We thank all the observers who acquired and provided the BeSS Achernar spectra used in this work.
This work has made use of the BeSOS Catalogue, operated by the Instituto de F\'{i}sica y Astronom\'{i}a, Universidad de Valpara\'{i}so, Chile (\url{http://besos.ifa.uv.cl}) and funded by Fondecyt iniciaci\'{o}n N\degr~11130702. The page is maintained thanks to FONDECYT N\degr~11190945.
This work has made use of data from the European Space Agency (ESA) mission {\it Gaia} (\url{http://www.cosmos.esa.int/gaia}), processed by the {\it Gaia} Data Processing and Analysis Consortium (DPAC, \url{http://www.cosmos.esa.int/web/gaia/dpac/consortium}).
Funding for the DPAC has been provided by national institutions, in particular the institutions participating in the {\it Gaia} Multilateral Agreement.
The authors acknowledge the support of the French Agence Nationale de la Recherche (ANR), under grants ANR-13-JS05-0005 (SAM), ANR-15-CE31-0012-01 (UnlockCepheids), ANR-21-CE31-0018-01 (MASSIF).
The research leading to these results  has received funding from the European Research Council (ERC) under the European Union's Horizon 2020 research and innovation program (projects CepBin, grant agreement 695099, and UniverScale, grant agreement 951549).
A.\,G. aknowledges support from the CONICYT/FONDECYT grant No. 3130361 and ANID-ALMA fund No. ASTRO20-0059.
A.\,C.\,C.~acknowledges support from CNPq (grant 311446/2019-1) and FAPESP (grants 2018/04055-8 and 2019/13354-1). 
This research has made use of Astropy\footnote{Available at \url{http://www.astropy.org/}}, a community-developed core Python package for Astronomy \citepads{2013A&A...558A..33A,2018AJ....156..123A}, the Numpy library \citepads{Harris20}, the Scipy library \citepads{scipy}, the Astroquery library \citepads{2019AJ....157...98G} and the Matplotlib graphics environment \citepads{Hunter:2007}.
This research has also made use of the Jean-Marie Mariotti Center \texttt{LITpro} service co-developed by CRAL, IPAG and LAGRANGE.
We used the SIMBAD and VizieR databases and catalogue access tool at the CDS, Strasbourg (France), and NASA's Astrophysics Data System Bibliographic Services.
The original description of the VizieR service was published in \citetads{2000A&AS..143...23O}.
The Digitized Sky Surveys were produced at the Space Telescope Science Institute under U.S. Government grant NAG W-2166. The images of these surveys are based on photographic data obtained using the Oschin Schmidt Telescope on Palomar Mountain and the UK Schmidt Telescope. 
The UK Schmidt Telescope was operated by the Royal Observatory Edinburgh, with funding from the UK Science and Engineering Research Council, until 1988 June, and thereafter by the Anglo-Australian Observatory. Original plate material is copyright (c) of the Royal Observatory Edinburgh and the Anglo-Australian Observatory. The plates were processed into the present compressed digital form with the permission of these institutions.
This publication makes use of data products from the Two Micron All Sky Survey, which is a joint project of the University of Massachusetts and the Infrared Processing and Analysis Center/California Institute of Technology, funded by the National Aeronautics and Space Administration and the National Science Foundation.
\end{acknowledgements}

\bibliographystyle{aa} 
\bibliography{Bibliography-Kervella}

\begin{appendix}

\onecolumn
\section{Astrometric and photometric data}

\input{Tables/all-obs-table}

\twocolumn
\FloatBarrier
\section{GRAVITY observations and modeling \label{GRAVITYplots}}

The individual GRAVITY observations of Achernar, together with the best-fit model for the complete data set, are presented in Fig.~\ref{fig:GRAVITYfits1}.

\begin{figure}[h!]
     \centering
         \includegraphics[width=\hsize]{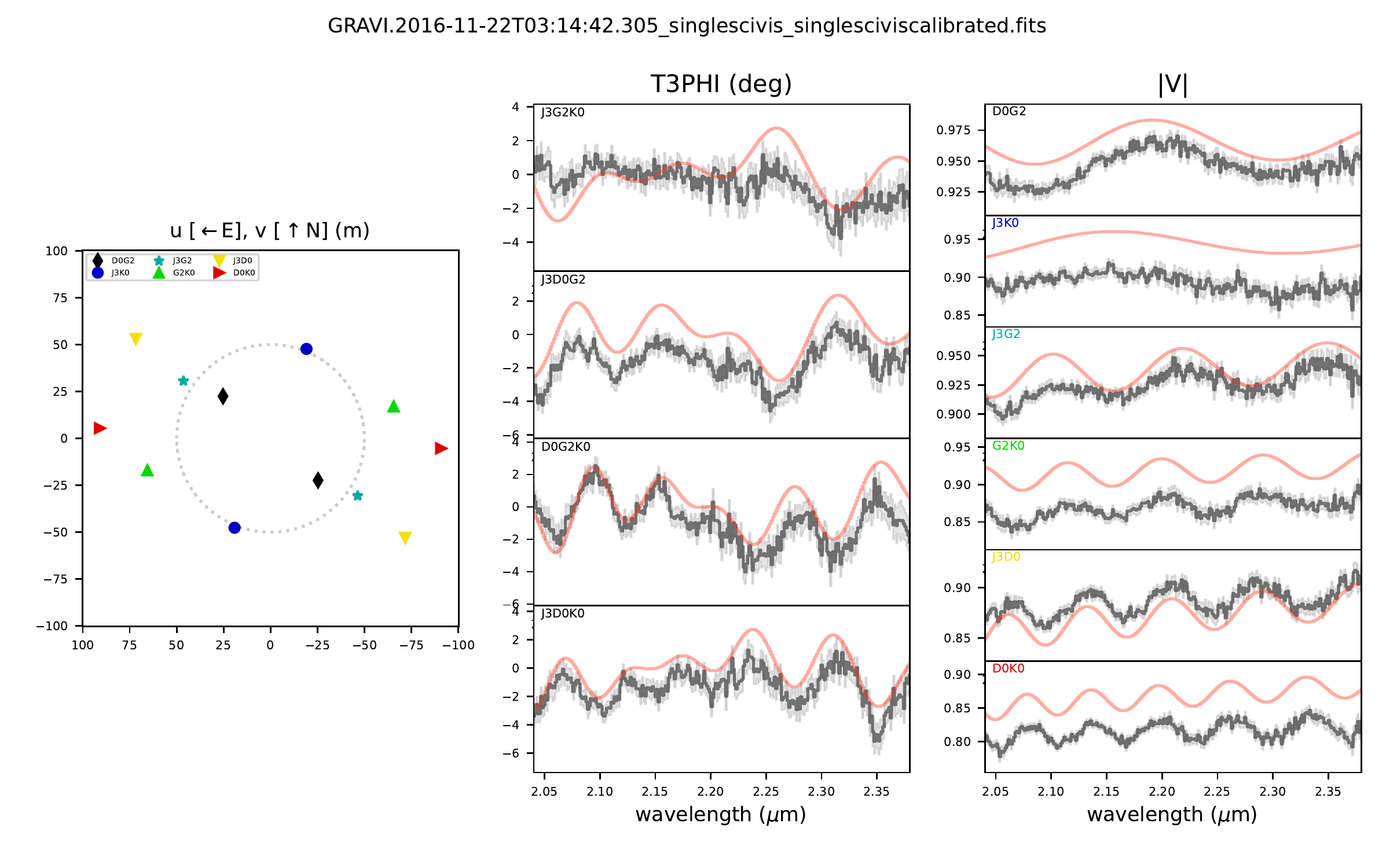}
         \includegraphics[width=\hsize]{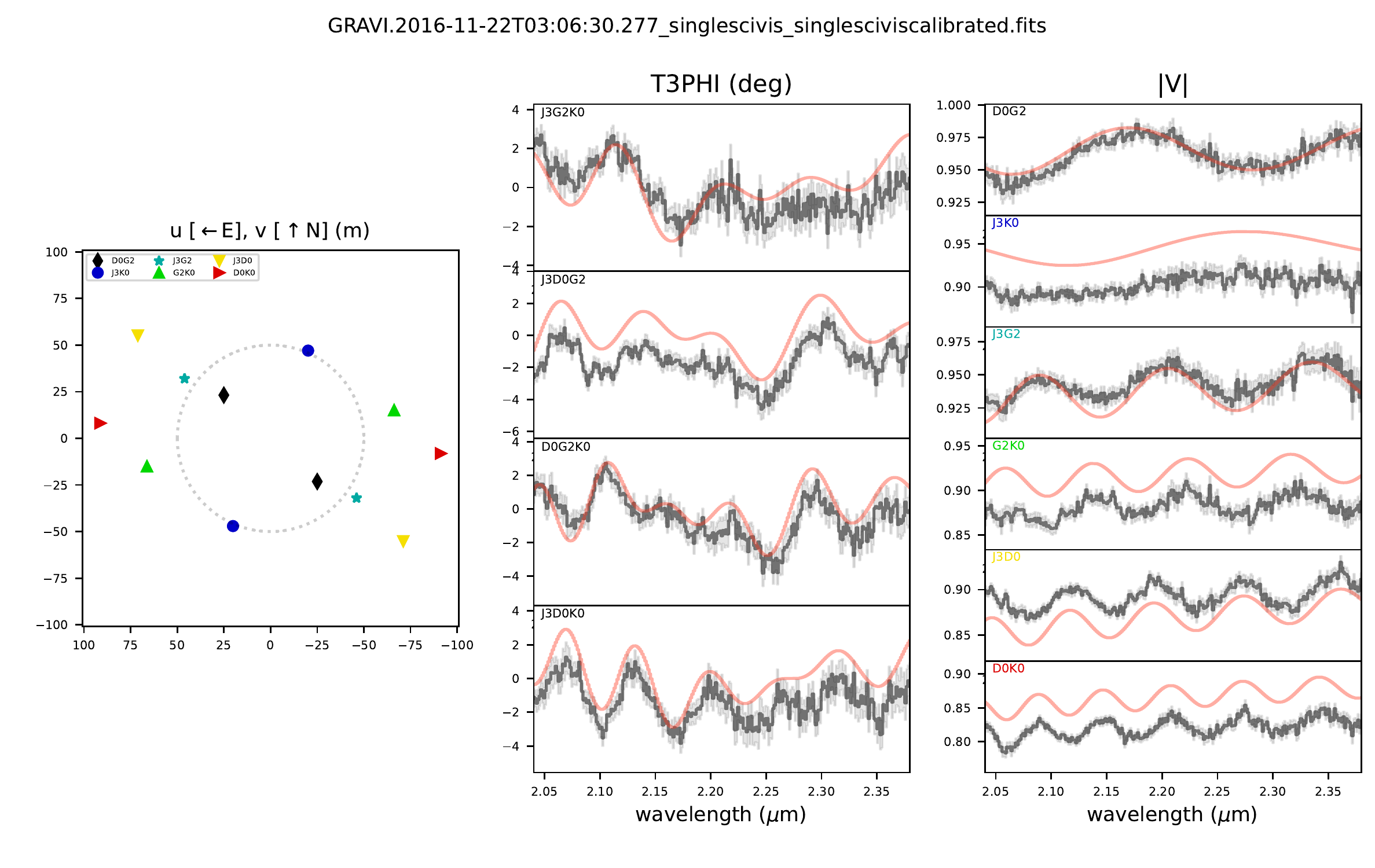}
         \includegraphics[width=\hsize]{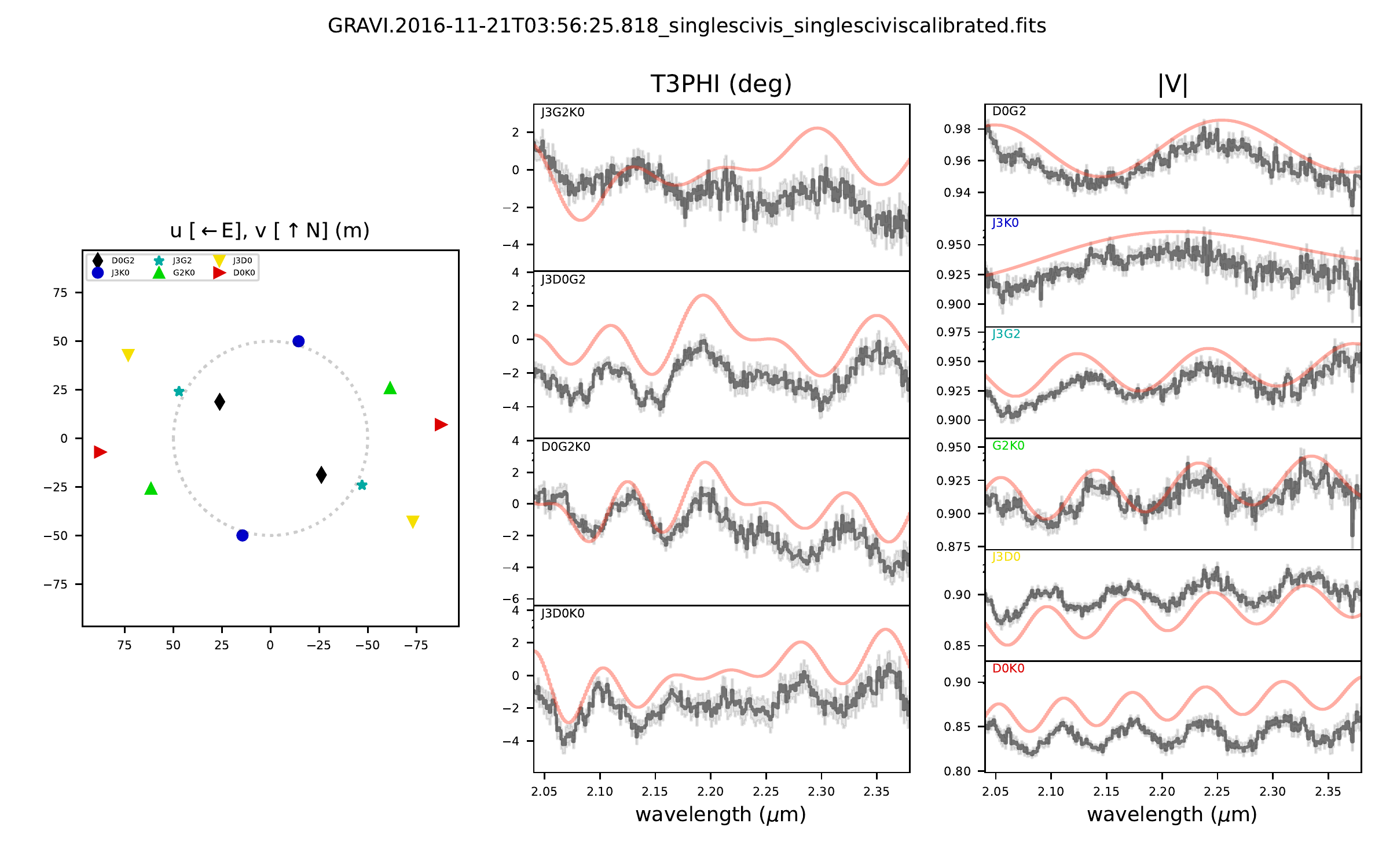}
     \caption{Detection of Achernar~B from the GRAVITY individual observations in the closure phase signal $T3\phi$ for each of the four telescope triplets, and amplitude of the fringe visibility $|V|$ for the six baselines. The observations are shown as black curves, and the best-fit model as red curves. The observing dates are given above each panel (see also Fig.~\ref{fig:pmoired_bestfit} and Sect.~\ref{sect:gravity} for further details).
     \label{fig:GRAVITYfits1}}
\end{figure}
         
\begin{figure}[h!]
    \renewcommand\thefigure{B.1}
     \centering
         \includegraphics[width=\hsize]{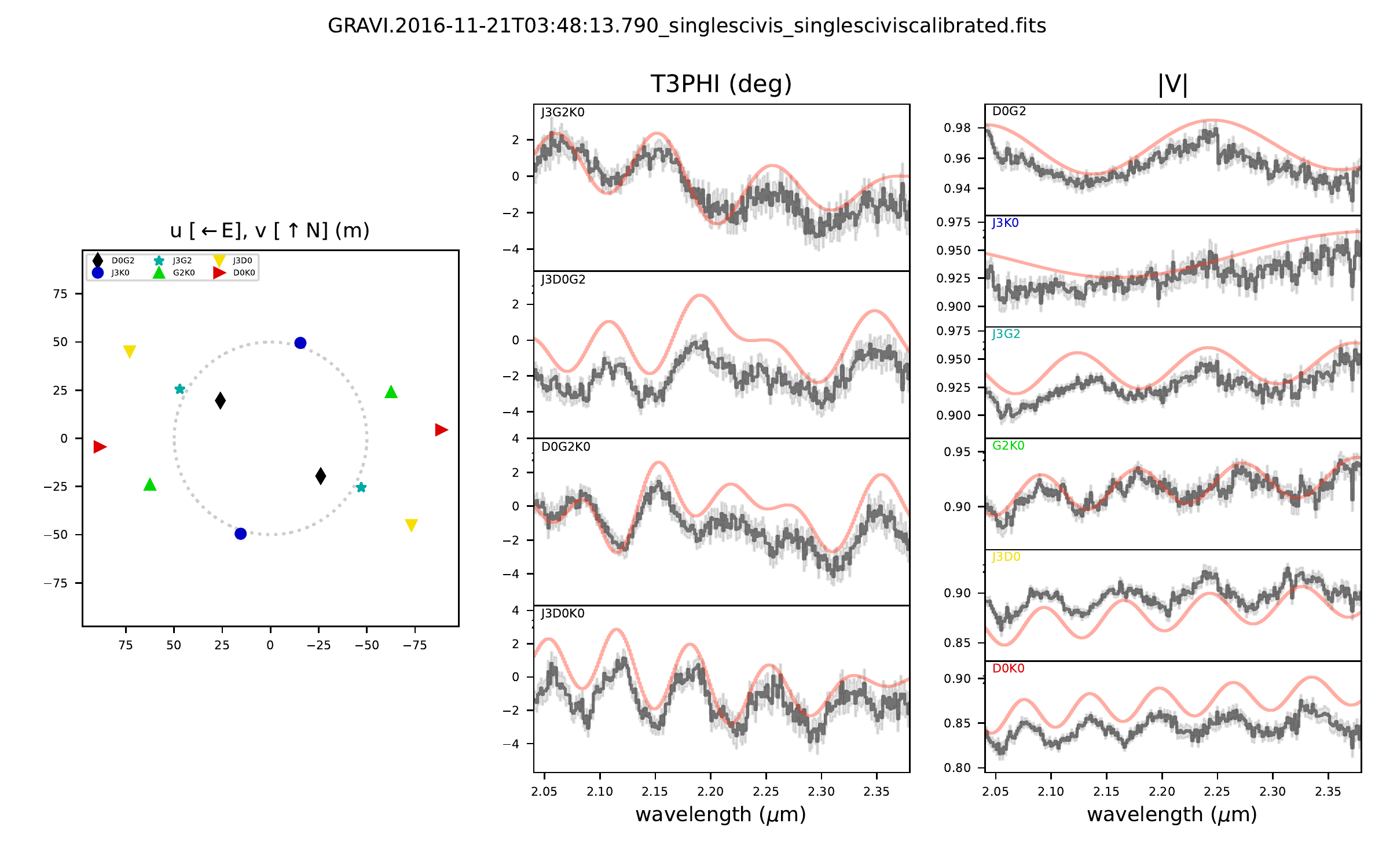}
         \includegraphics[width=\hsize]{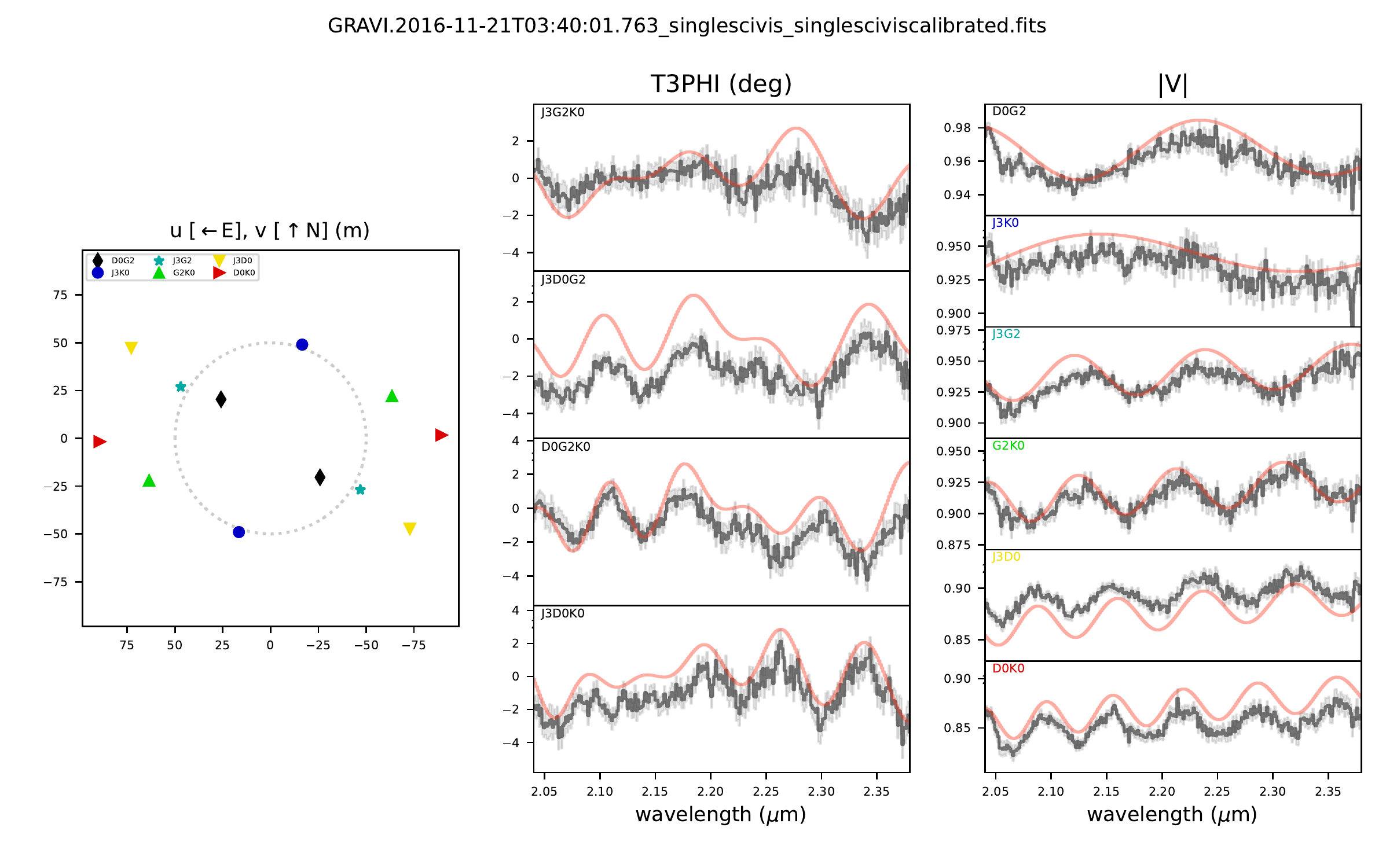}
         \includegraphics[width=\hsize]{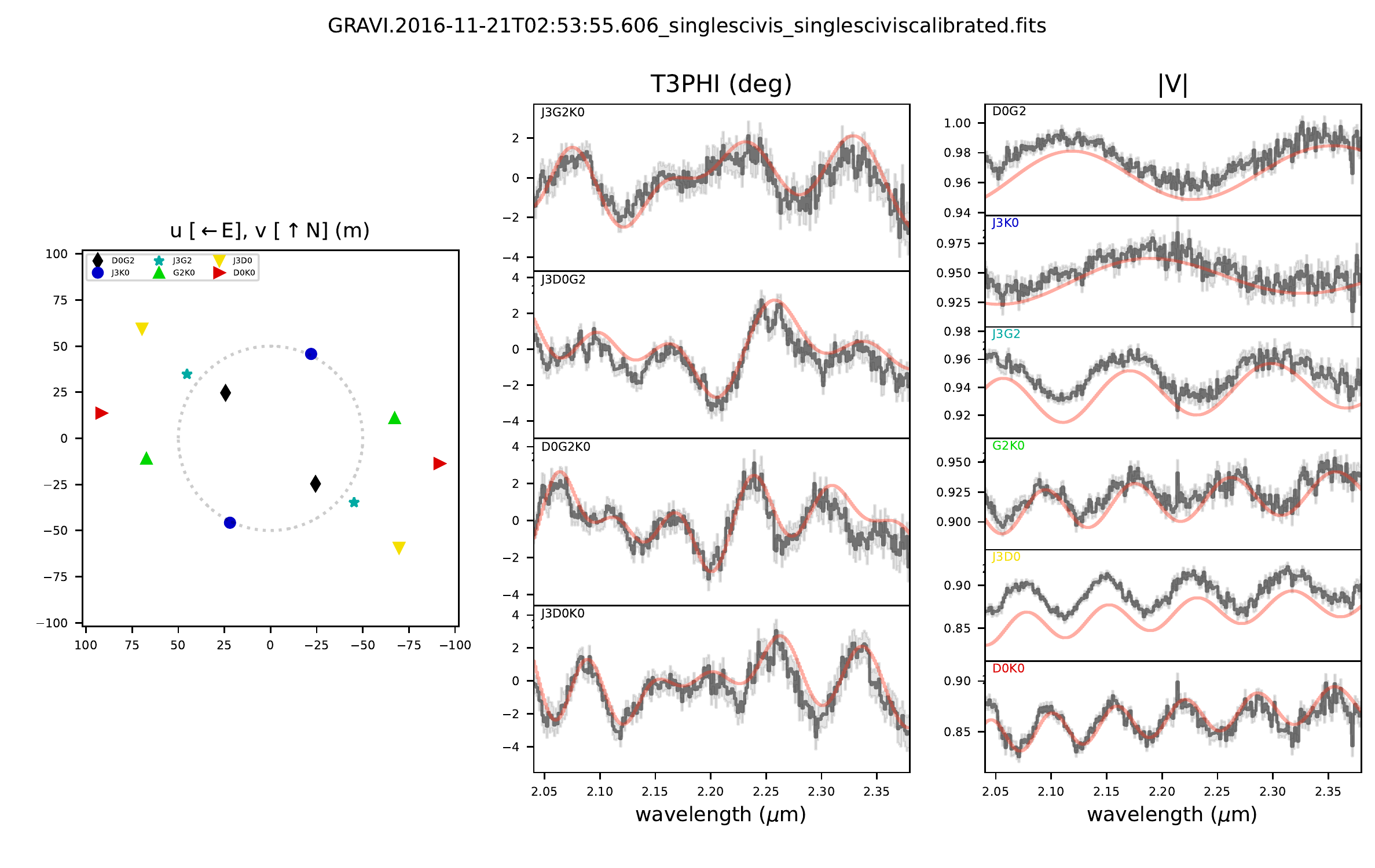}
         \includegraphics[width=\hsize]{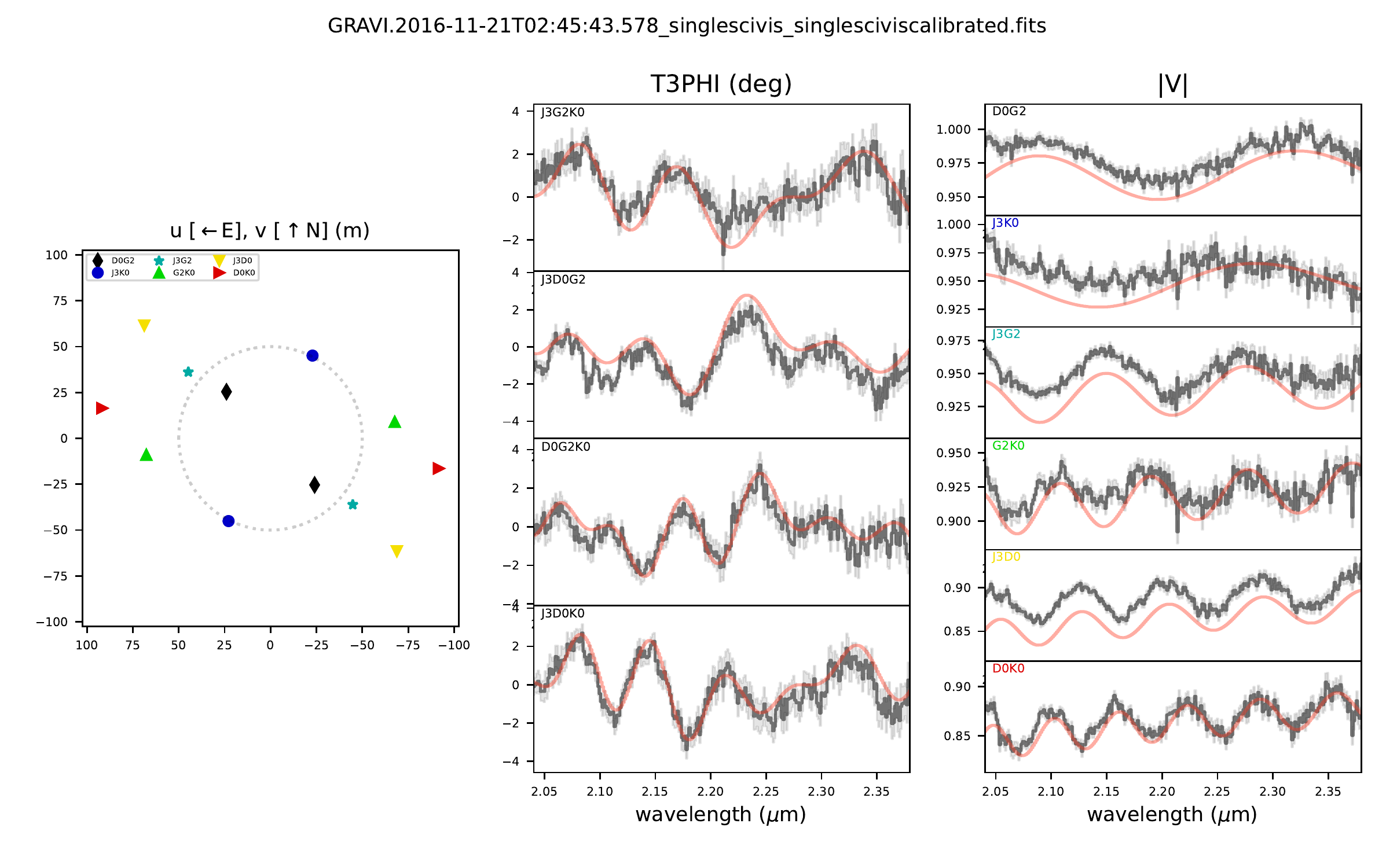}
     \caption{continued.}
\end{figure}

\begin{figure}[h!]
    \renewcommand\thefigure{B.1}
     \centering
         \includegraphics[width=\hsize]{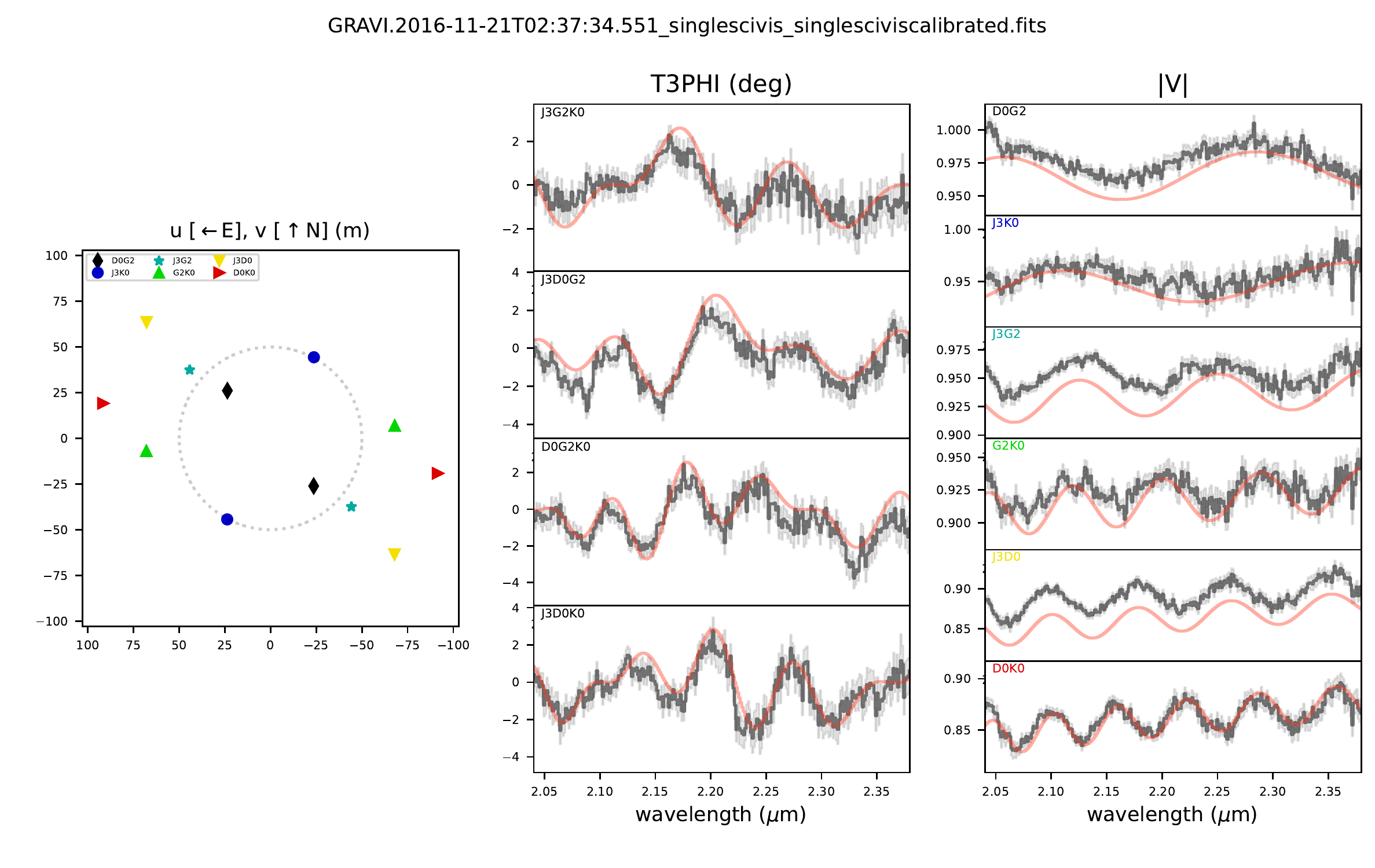}
         \includegraphics[width=\hsize]{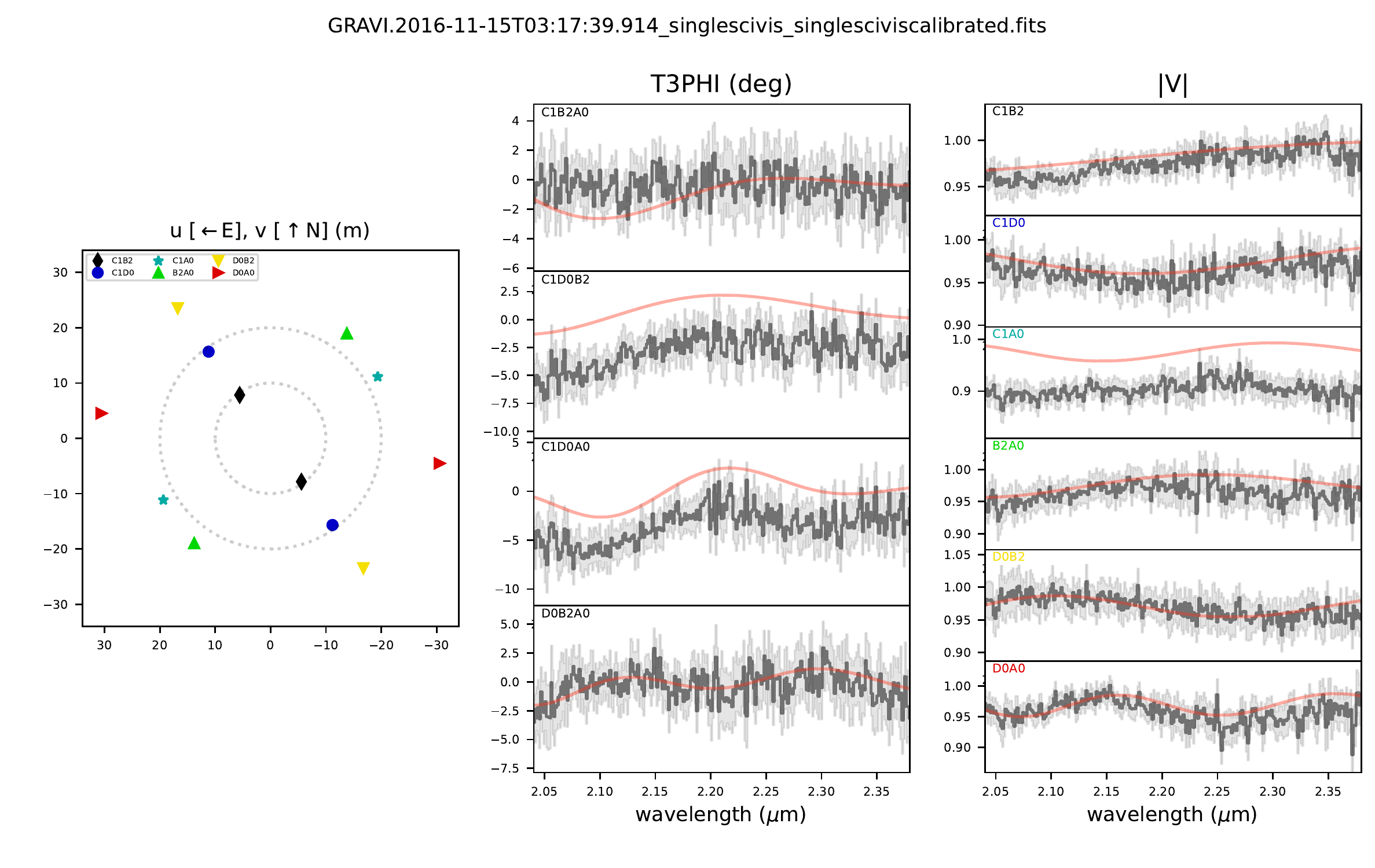}
         \includegraphics[width=\hsize]{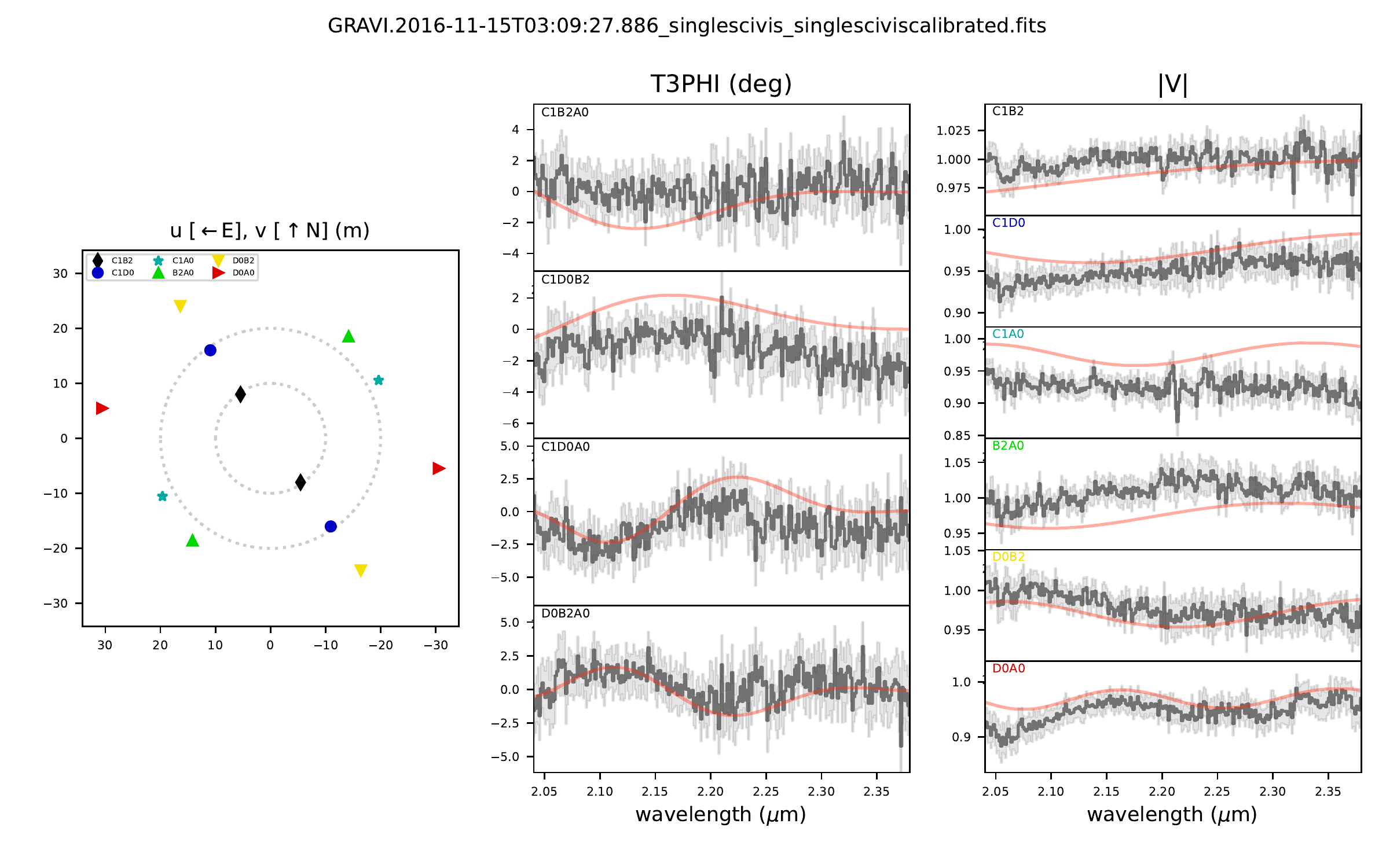}
         \includegraphics[width=\hsize]{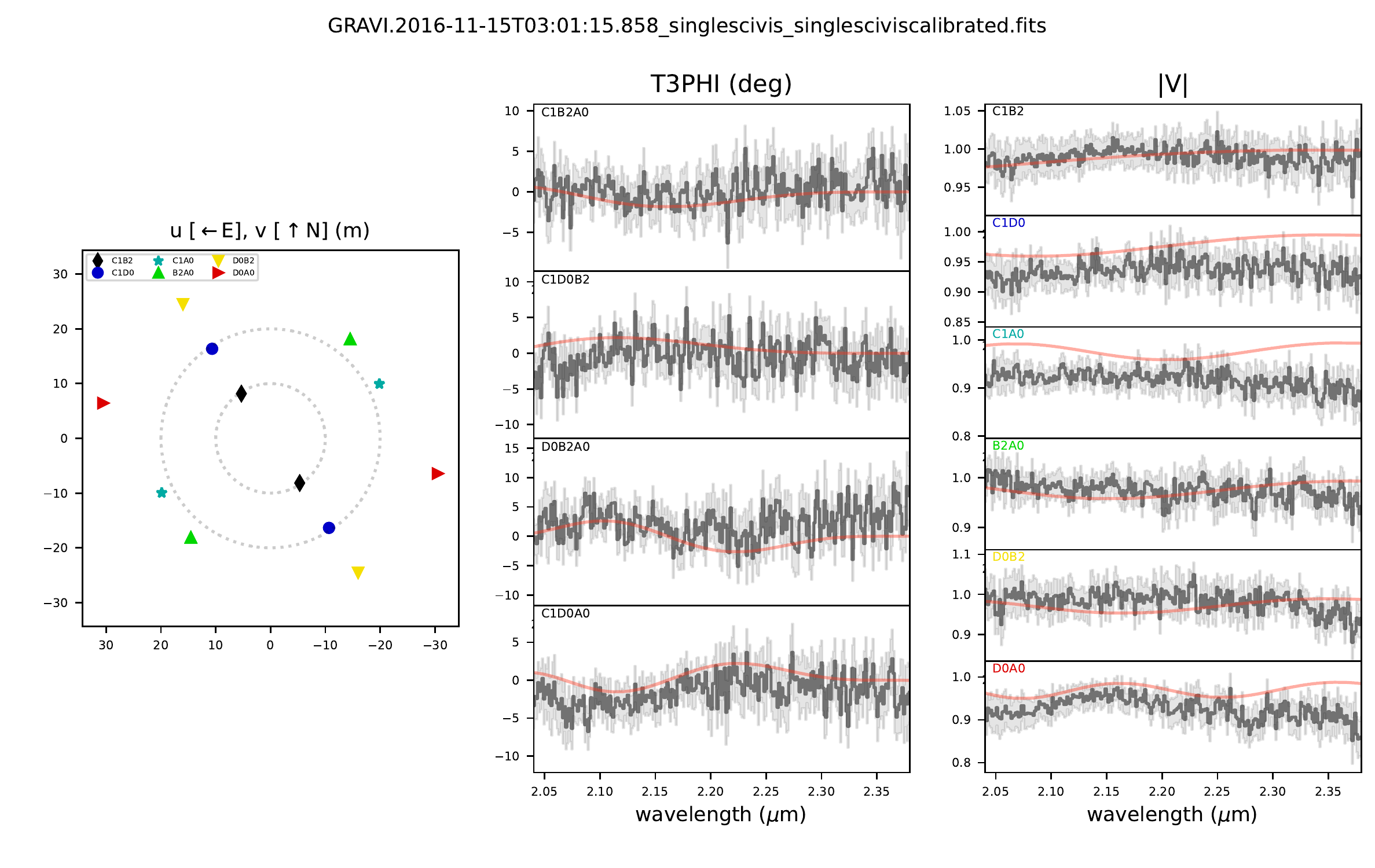}
     \caption{continued.}
\end{figure}

\begin{figure}[h!]
    \renewcommand\thefigure{B.1}
     \centering
         \includegraphics[width=\hsize]{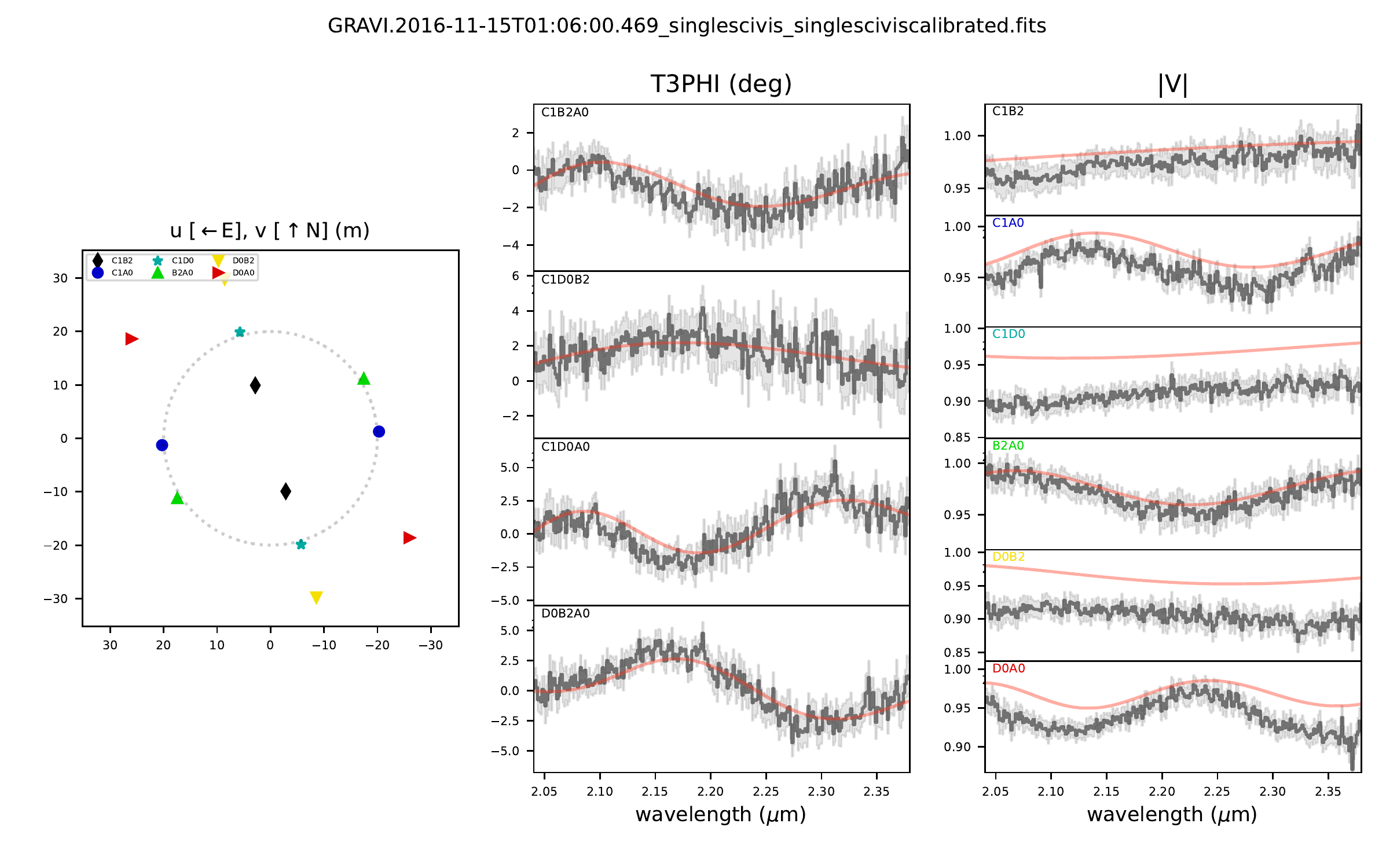}
         \includegraphics[width=\hsize]{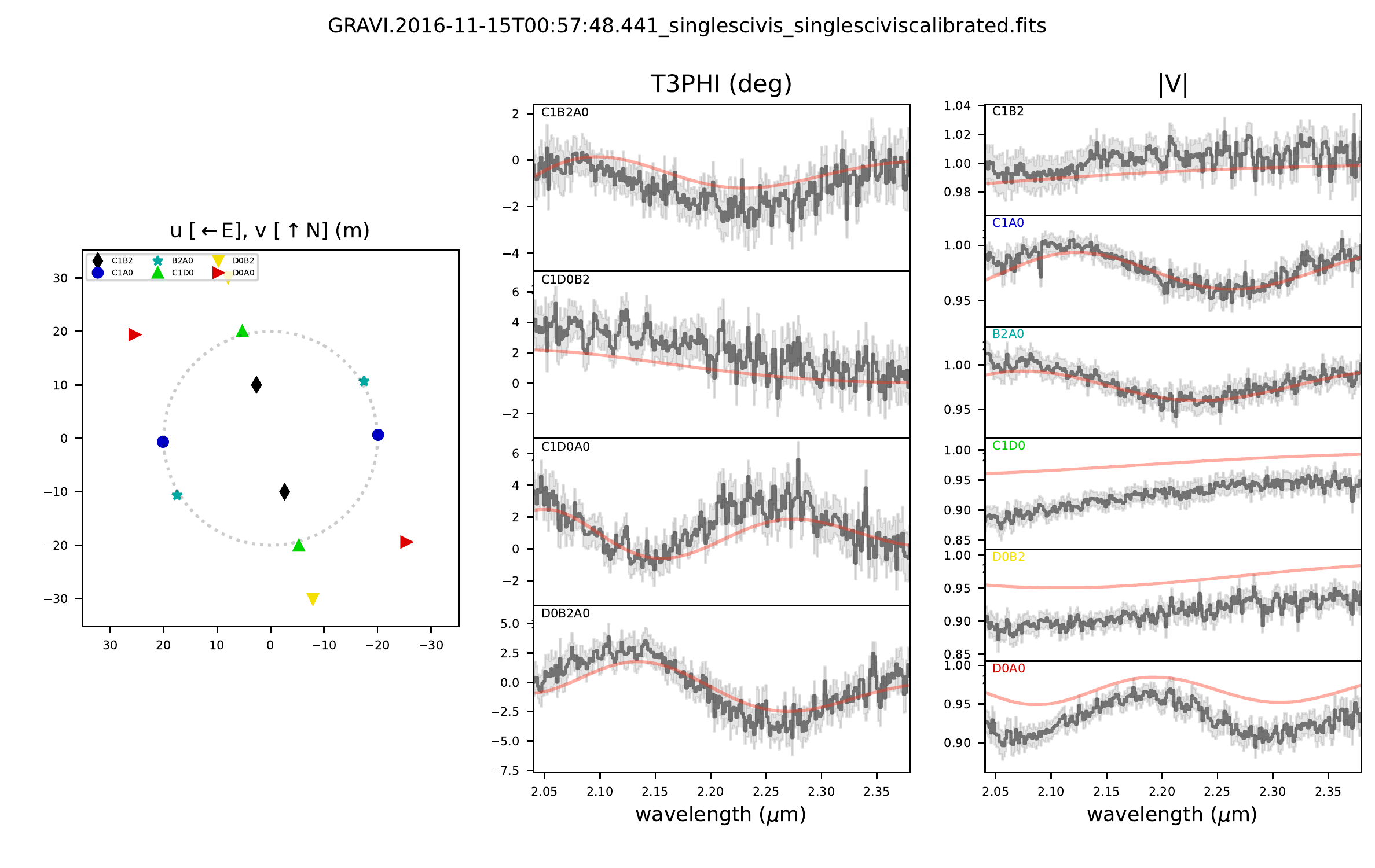}
         \includegraphics[width=\hsize]{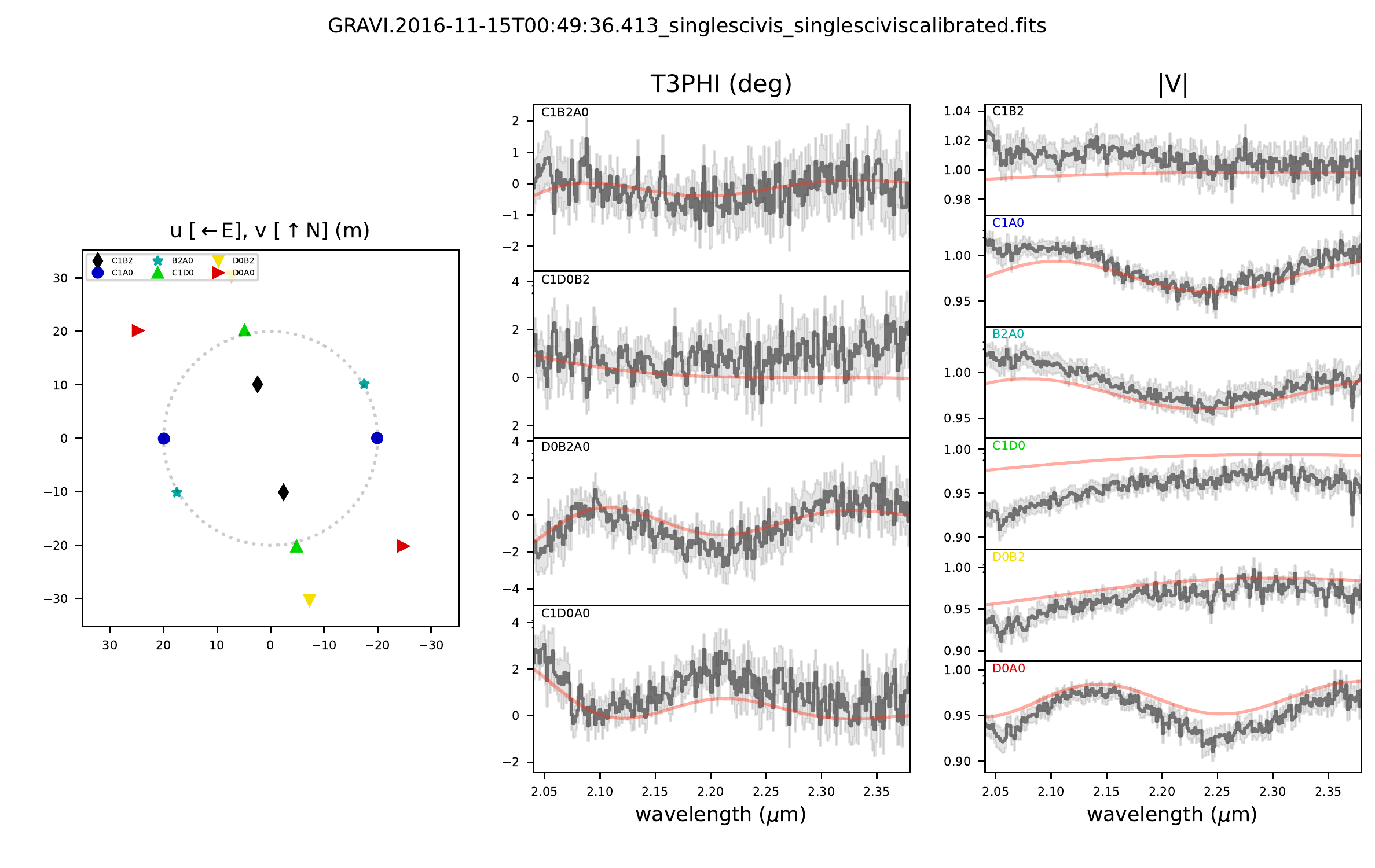}
     \caption{continued.}
\end{figure}

\FloatBarrier
\section{Radial velocities of Achernar\label{sect:radvelA}}

\input{Tables/spectroscopy-radvel}

\FloatBarrier
\end{appendix}

\end{document}

%% file: Tables/all-obs-table.tex
The complete set of relative astrometric and photometric measurements of Achernar B with respect to A is presented in Table~\ref{tab:allastrometry1}. 

\begin{table*}[h!]
\caption{Astrometry and photometry of Achernar B relative to Achernar A. The data points marked with a ``$\bullet$'' flag were excluded from the orbital and spectral energy distribution fits.
\label{tab:allastrometry1}}
\centering \small
  \begin{tabular}{lclrrcrc}
    \hline
    \hline
Date & MJD & Instrument & $\Delta \alpha$ & $\Delta \delta$ & $\lambda$ & $f_B/f_A$ & Flag \\
         &     &             & (mas)           & (mas)        & ($\mu$m) & (\%)   \\
    \hline  \noalign{\smallskip}
2006-10-04 & 54012.1328 & VISIR &  &  & 11.25 & $2.14 \pm 0.57$ &  \\ 
2006-10-04 & 54012.1328 & VISIR & $-184.00 \pm 25.0$ & $211.00 \pm 25.0$ & 8.59 & $1.95 \pm 0.57$ &  \\ 
2007-06-27 & 54278.3673 & NACO & $-85.77 \pm 3.69$ & $175.95 \pm 3.74$ & 2.17 & $2.17 \pm 0.63$ &  \\ 
2007-06-27 & 54278.3677 & NACO & $-84.82 \pm 3.10$ & $192.64 \pm 3.09$ & 2.17 & $2.94 \pm 0.77$ &  \\ 
2007-07-08 & 54289.3243 & NACO & $-78.48 \pm 6.87$ & $178.79 \pm 7.08$ & 2.17 & $2.02 \pm 0.72$ & $\bullet$ \\ 
2007-07-08 & 54289.3246 & NACO & $-83.69 \pm 3.30$ & $184.24 \pm 3.29$ & 2.17 & $2.28 \pm 0.71$ &  \\ 
2007-08-01 & 54313.4218 & NACO & $-61.87 \pm 5.91$ & $157.48 \pm 5.63$ & 2.17 & $3.38 \pm 0.84$ & $\bullet$ \\ 
2007-08-01 & 54313.4221 & NACO & $-65.57 \pm 4.06$ & $166.72 \pm 3.75$ & 2.17 & $3.27 \pm 1.44$ &  \\ 
2007-08-09 & 54321.3553 & NACO & $-80.18 \pm 5.18$ & $184.03 \pm 5.61$ & 2.17 & $2.61 \pm 0.77$ & $\bullet$ \\ 
2007-08-09 & 54321.3558 & NACO & $-86.02 \pm 6.55$ & $179.24 \pm 6.59$ & 2.17 & $2.21 \pm 0.68$ & $\bullet$ \\ 
2007-08-09 & 54321.4110 & NACO & $-55.21 \pm 4.33$ & $160.10 \pm 3.81$ & 2.17 & $4.55 \pm 1.28$ & $\bullet$ \\ 
2007-08-09 & 54321.4115 & NACO & $-54.51 \pm 5.70$ & $161.14 \pm 4.56$ & 2.17 & $2.66 \pm 1.11$ & $\bullet$ \\ 
2007-10-27 & 54400.1953 & NACO & $-18.75 \pm 19.2$ & $141.43 \pm 13.0$ & 2.17 & $3.52 \pm 1.69$ & $\bullet$ \\ 
2007-10-27 & 54400.1962 & NACO & $-17.47 \pm 17.1$ & $133.81 \pm 13.5$ & 2.17 & $4.27 \pm 1.74$ & $\bullet$ \\ 
2007-10-27 & 54400.1989 & NACO & $-48.36 \pm 4.31$ & $156.05 \pm 4.26$ & 1.64 & $3.09 \pm 1.52$ &  \\ 
2007-10-27 & 54400.1996 & NACO & $-35.32 \pm 7.17$ & $133.61 \pm 8.40$ & 1.64 & $4.00 \pm 1.52$ & $\bullet$ \\ 
2007-10-27 & 54400.2015 & NACO & $-54.25 \pm 3.38$ & $147.05 \pm 3.58$ & 1.09 & $4.06 \pm 2.17$ & $\bullet$ \\ 
2007-10-27 & 54400.2021 & NACO & $-64.08 \pm 4.60$ & $146.15 \pm 4.72$ & 1.09 & $3.52 \pm 2.52$ & $\bullet$ \\ 
2007-10-28 & 54401.1258 & NACO & $-49.93 \pm 4.18$ & $176.24 \pm 4.91$ & 2.17 & $1.62 \pm 0.98$ & $\bullet$ \\ 
2007-10-28 & 54401.1267 & NACO & $-48.02 \pm 4.40$ & $174.89 \pm 4.94$ & 2.17 & $1.89 \pm 0.79$ & $\bullet$ \\ 
2007-10-28 & 54401.1280 & NACO & $-50.59 \pm 5.20$ & $169.46 \pm 5.60$ & 1.64 & $2.03 \pm 0.93$ & $\bullet$ \\ 
2007-10-28 & 54401.1287 & NACO & $-51.59 \pm 4.36$ & $165.43 \pm 4.45$ & 1.64 & $2.60 \pm 1.06$ &  \\ 
2007-10-28 & 54401.1301 & NACO & $-52.22 \pm 5.41$ & $173.39 \pm 6.63$ & 1.09 & $1.59 \pm 1.44$ & $\bullet$ \\ 
2007-10-28 & 54401.1307 & NACO & $-41.49 \pm 5.05$ & $166.23 \pm 4.82$ & 1.09 & $2.54 \pm 1.73$ & $\bullet$ \\ 
2007-11-01 & 54405.0067 & NACO & $-52.22 \pm 3.52$ & $174.34 \pm 4.00$ & 2.17 & $1.24 \pm 0.93$ & $\bullet$ \\ 
2007-11-01 & 54405.0076 & NACO & $-53.44 \pm 3.50$ & $164.66 \pm 3.54$ & 2.17 & $1.70 \pm 6.05$ & $\bullet$ \\ 
2007-11-01 & 54405.0088 & NACO & $-63.94 \pm 3.81$ & $169.62 \pm 3.79$ & 1.64 & $2.20 \pm 1.12$ & $\bullet$ \\ 
2007-11-01 & 54405.0095 & NACO & $-55.48 \pm 3.14$ & $164.39 \pm 3.17$ & 1.64 & $3.20 \pm 1.27$ &  \\ 
2007-11-01 & 54405.0108 & NACO & $-80.13 \pm 14.3$ & $162.74 \pm 6.57$ & 1.09 & $4.09 \pm 2.17$ & $\bullet$ \\ 
2007-11-01 & 54405.0114 & NACO & $-79.43 \pm 10.1$ & $160.13 \pm 5.72$ & 1.09 & $4.19 \pm 2.18$ & $\bullet$ \\ 
2007-11-07 & 54411.0984 & NACO & $-36.77 \pm 4.83$ & $168.43 \pm 5.30$ & 2.17 & $1.17 \pm 1.53$ & $\bullet$ \\ 
2007-11-07 & 54411.0992 & NACO & $-43.58 \pm 3.68$ & $167.54 \pm 3.92$ & 2.17 & $1.69 \pm 1.97$ & $\bullet$ \\ 
2007-11-07 & 54411.1008 & NACO & $-37.98 \pm 4.57$ & $166.69 \pm 4.75$ & 1.64 & $2.54 \pm 1.54$ & $\bullet$ \\ 
2007-11-07 & 54411.1015 & NACO & $-47.98 \pm 5.69$ & $165.64 \pm 5.60$ & 1.64 & $2.98 \pm 1.78$ & $\bullet$ \\ 
2007-11-07 & 54411.1028 & NACO & $-40.58 \pm 5.99$ & $174.98 \pm 7.47$ & 1.09 & $1.01 \pm 1.66$ & $\bullet$ \\ 
2007-11-07 & 54411.1034 & NACO & $-46.33 \pm 4.02$ & $180.09 \pm 5.30$ & 1.09 & $0.92 \pm 3.07$ & $\bullet$ \\ 
2007-12-22 & 54456.0693 & NACO & $-29.90 \pm 3.88$ & $156.23 \pm 4.12$ & 2.17 & $1.38 \pm 1.27$ & $\bullet$ \\ 
2007-12-22 & 54456.0701 & NACO & $-25.76 \pm 3.56$ & $152.19 \pm 3.65$ & 2.17 & $1.90 \pm 1.14$ & $\bullet$ \\ 
2007-12-22 & 54456.0725 & NACO & $-27.71 \pm 4.43$ & $156.61 \pm 4.90$ & 1.64 & $1.88 \pm 2.05$ & $\bullet$ \\ 
2007-12-22 & 54456.0731 & NACO & $-28.40 \pm 3.66$ & $154.56 \pm 3.80$ & 1.64 & $2.48 \pm 2.04$ & $\bullet$ \\ 
2007-12-22 & 54456.0742 & NACO & $-34.64 \pm 6.67$ & $144.28 \pm 6.21$ & 1.09 & $3.00 \pm 1.93$ & $\bullet$ \\ 
2007-12-22 & 54456.0748 & NACO & $-32.06 \pm 3.30$ & $137.53 \pm 3.34$ & 1.09 & $3.90 \pm 1.63$ &  \\ 
2008-11-04 & 54779.0000 & AMBER & & & 1.64 & $2.26 \pm 0.09$ &  \\ 
2008-11-04 & 54779.0000 & AMBER &  &  & 2.17 & $2.29 \pm 0.09$ &  \\ 
2008-11-04 & 54774.1213 & AMBER & $20.23 \pm 0.50$ & $-46.54 \pm 0.95$ & HK &  &  \\ 
2008-11-08 & 54778.1213 & AMBER & $16.99 \pm 0.55$ & $-48.42 \pm 0.97$ & HK &  &  \\ 
2008-11-10 & 54780.1213 & AMBER & $15.37 \pm 0.61$ & $-49.36 \pm 1.00$ & HK &  &  \\ 
2008-11-14 & 54784.1213 & AMBER & $12.13 \pm 0.76$ & $-51.24 \pm 1.08$ & HK &  &  \\ 
2009-01-04 & 54835.0214 & NACO & $-44.45 \pm 3.51$ & $-56.62 \pm 3.10$ & 1.75 & $1.36 \pm 1.14$ & $\bullet$ \\ 
2009-01-04 & 54835.0254 & NACO & $-38.65 \pm 3.59$ & $-58.26 \pm 3.31$ & 1.64 & $1.79 \pm 0.79$ & $\bullet$ \\ 
2009-01-04 & 54835.0290 & NACO & $-44.27 \pm 3.33$ & $-81.38 \pm 3.51$ & 2.12 & $2.12 \pm 0.96$ & $\bullet$ \\ 
2009-01-04 & 54835.0319 & NACO & $-46.44 \pm 3.81$ & $-86.91 \pm 4.36$ & 2.17 & $1.35 \pm 0.80$ & $\bullet$ \\ 
2009-01-07 & 54838.0356 & NACO & $-34.91 \pm 2.99$ & $-44.26 \pm 3.19$ & 2.17 & $2.12 \pm 1.46$ & $\bullet$ \\ 
2009-01-07 & 54838.0381 & NACO & $-38.88 \pm 2.94$ & $-43.46 \pm 3.05$ & 2.12 & $1.68 \pm 1.37$ & $\bullet$ \\ 
2009-01-07 & 54838.0451 & NACO & $-33.65 \pm 12.8$ & $-87.41 \pm 30.0$ & 1.28 & $-0.80 \pm 2.30$ & $\bullet$ \\ 
2009-01-07 & 54838.0475 & NACO & $-37.68 \pm 2.87$ & $-69.06 \pm 2.87$ & 1.26 & $-0.57 \pm 2.50$ & $\bullet$ \\ 
2009-01-07 & 54838.0500 & NACO & $-37.00 \pm 11.1$ & $-67.25 \pm 11.3$ & 1.09 & $-1.69 \pm 2.00$ & $\bullet$ \\ 
    \hline
    \end{tabular}
\end{table*}

\begin{table*}
\renewcommand\thetable{A.1}
\caption{continued.}
\centering \small
  \begin{tabular}{lclrrcrc}
    \hline
    \hline
Date & MJD & Instrument & $\Delta \alpha$ & $\Delta \delta$ & $\lambda$ & $f_B/f_A$ & Flag \\
         &     &             & (mas)           & (mas)        & ($\mu$m) & (\%)   \\
    \hline  \noalign{\smallskip}
2009-01-07 & 54838.0945 & NACO SAM & $-33.14 \pm 1.74$ & $-61.91 \pm 1.66$ & 1.64 & $2.67 \pm 0.19$ &  \\  
2009-01-07 & 54838.0945 & NACO SAM & $-30.63 \pm 1.54$ & $-58.14 \pm 1.68$ & 2.18 & $2.55 \pm 0.13$ &  \\  
2009-01-07 & 54838.0945 & NACO SAM & $-29.60 \pm 1.06$ & $-62.76 \pm 1.16$ & 2.12 & $2.60 \pm 0.10$ &  \\  
2009-10-06 & 55110.2083 & NACO & $-159.79 \pm 2.69$ & $-13.17 \pm 2.68$ & 2.12 & $2.91 \pm 0.73$ &  \\ 
2009-10-06 & 55110.2095 & NACO & $-165.91 \pm 2.68$ & $-12.74 \pm 2.68$ & 1.75 & $2.77 \pm 0.61$ &  \\ 
2009-10-06 & 55110.2105 & NACO & $-164.29 \pm 2.66$ & $-11.06 \pm 2.66$ & 1.64 & $2.53 \pm 0.58$ &  \\ 
2009-10-06 & 55110.2117 & NACO & $-164.48 \pm 2.68$ & $-11.50 \pm 2.68$ & 1.28 & $2.75 \pm 0.60$ &  \\ 
2009-10-06 & 55110.2128 & NACO & $-163.19 \pm 2.67$ & $-12.63 \pm 2.67$ & 1.26 & $3.14 \pm 0.69$ &  \\ 
2009-10-06 & 55110.2140 & NACO & $-165.83 \pm 2.66$ & $-12.80 \pm 2.66$ & 1.24 & $2.86 \pm 0.58$ &  \\ 
2009-10-06 & 55110.2151 & NACO & $-163.10 \pm 2.67$ & $-14.96 \pm 2.67$ & 1.09 & $2.95 \pm 0.74$ &  \\ 
2009-10-06 & 55110.2163 & NACO & $-164.89 \pm 2.69$ & $-11.66 \pm 2.70$ & 1.08 & $3.09 \pm 0.74$ &  \\ 
2009-10-06 & 55110.2174 & NACO & $-162.66 \pm 2.68$ & $-13.22 \pm 2.68$ & 1.04 & $3.05 \pm 0.78$ &  \\ 
2009-10-06 & 55110.2189 & NACO & $-161.11 \pm 2.69$ & $-13.91 \pm 2.69$ & 2.17 & $2.81 \pm 0.74$ &  \\ 
2009-10-08 & 55112.2239 & NACO & $-159.84 \pm 2.69$ & $-12.44 \pm 2.68$ & 2.17 & $2.78 \pm 0.66$ &  \\ 
2009-10-08 & 55112.2252 & NACO & $-158.77 \pm 2.69$ & $-12.97 \pm 2.69$ & 2.12 & $2.81 \pm 0.65$ &  \\ 
2009-10-08 & 55112.2264 & NACO & $-163.43 \pm 2.72$ & $-11.67 \pm 2.72$ & 1.75 & $3.08 \pm 0.63$ &  \\ 
2009-10-08 & 55112.2275 & NACO & $-164.53 \pm 2.66$ & $-10.73 \pm 2.66$ & 1.64 & $2.43 \pm 0.51$ &  \\ 
2009-10-08 & 55112.2287 & NACO & $-164.23 \pm 2.67$ & $-11.16 \pm 2.67$ & 1.28 & $2.62 \pm 0.60$ &  \\ 
2009-10-08 & 55112.2299 & NACO & $-163.31 \pm 2.67$ & $-10.94 \pm 2.67$ & 1.26 & $3.01 \pm 0.69$ &  \\ 
2009-10-08 & 55112.2311 & NACO & $-162.73 \pm 2.69$ & $-10.67 \pm 2.69$ & 1.24 & $3.13 \pm 0.71$ &  \\ 
2009-10-08 & 55112.2323 & NACO & $-161.88 \pm 2.75$ & $-11.26 \pm 2.75$ & 1.09 & $3.25 \pm 0.75$ &  \\ 
2009-10-08 & 55112.2335 & NACO & $-161.11 \pm 2.74$ & $-11.14 \pm 2.74$ & 1.08 & $3.28 \pm 0.76$ &  \\ 
2009-10-08 & 55112.2346 & NACO & $-164.22 \pm 2.70$ & $-11.93 \pm 2.70$ & 1.04 & $3.25 \pm 0.79$ &  \\ 
2010-10-23 & 55492.1995 & NACO & $-240.63 \pm 2.67$ & $71.28 \pm 2.67$ & 2.17 & $2.35 \pm 0.43$ &  \\ 
2010-10-23 & 55492.2007 & NACO & $-242.28 \pm 2.66$ & $73.93 \pm 2.66$ & 1.75 & $2.43 \pm 0.47$ &  \\ 
2010-10-23 & 55492.2018 & NACO & $-241.42 \pm 2.68$ & $72.63 \pm 2.68$ & 1.28 & $2.08 \pm 0.38$ &  \\ 
2010-10-23 & 55492.2030 & NACO & $-238.56 \pm 2.67$ & $72.60 \pm 2.67$ & 1.24 & $2.30 \pm 0.36$ &  \\ 
2010-10-23 & 55492.2041 & NACO & $-241.21 \pm 2.69$ & $71.52 \pm 2.69$ & 1.08 & $2.22 \pm 0.41$ &  \\ 
2010-10-23 & 55492.2053 & NACO & $-241.31 \pm 2.67$ & $73.44 \pm 2.67$ & 1.04 & $1.90 \pm 0.42$ &  \\ 
2010-11-18 & 55518.0295 & NACO & $-245.01 \pm 2.65$ & $77.03 \pm 2.65$ & 1.64 & $2.55 \pm 0.52$ &  \\ 
2010-11-18 & 55518.0307 & NACO & $-244.59 \pm 2.66$ & $77.67 \pm 2.66$ & 2.17 & $2.75 \pm 0.55$ &  \\ 
2010-11-18 & 55518.0318 & NACO & $-243.77 \pm 2.65$ & $77.39 \pm 2.65$ & 1.64 & $2.34 \pm 0.44$ &  \\ 
2010-11-18 & 55518.0330 & NACO & $-244.83 \pm 2.66$ & $75.80 \pm 2.66$ & 1.26 & $2.17 \pm 0.39$ &  \\ 
2010-11-18 & 55518.0341 & NACO & $-246.35 \pm 2.67$ & $76.18 \pm 2.67$ & 1.09 & $1.99 \pm 0.39$ &  \\ 
2010-11-18 & 55518.0353 & NACO & $-244.28 \pm 2.67$ & $75.08 \pm 2.67$ & 1.04 & $1.91 \pm 0.37$ &  \\ 
2011-09-23 & 55827.2460 & NACO & $-255.57 \pm 2.66$ & $138.45 \pm 2.66$ & 2.17 & $2.31 \pm 0.45$ &  \\ 
2011-09-23 & 55827.2487 & NACO & $-255.39 \pm 2.67$ & $135.12 \pm 2.67$ & 2.17 & $2.17 \pm 0.44$ &  \\ 
2011-09-23 & 55827.2501 & NACO & $-253.92 \pm 2.66$ & $136.54 \pm 2.66$ & 2.17 & $2.37 \pm 0.47$ &  \\ 
2011-09-23 & 55827.2513 & NACO & $-255.68 \pm 2.68$ & $136.75 \pm 2.68$ & 2.12 & $2.55 \pm 0.50$ &  \\ 
2011-09-23 & 55827.2524 & NACO & $-255.20 \pm 2.67$ & $135.85 \pm 2.67$ & 1.75 & $2.16 \pm 0.42$ &  \\ 
2011-09-23 & 55827.2535 & NACO & $-255.44 \pm 2.74$ & $136.65 \pm 2.74$ & 1.64 & $2.30 \pm 0.46$ &  \\ 
2011-10-04 & 55838.2645 & NACO & $-254.59 \pm 2.66$ & $137.88 \pm 2.66$ & 2.17 & $2.68 \pm 0.51$ &  \\ 
2011-10-04 & 55838.2660 & NACO & $-254.27 \pm 2.66$ & $138.04 \pm 2.66$ & 2.12 & $2.79 \pm 0.53$ &  \\ 
2011-10-04 & 55838.2673 & NACO & $-253.79 \pm 2.66$ & $136.67 \pm 2.66$ & 1.75 & $3.15 \pm 0.56$ &  \\ 
2011-10-04 & 55838.2684 & NACO & $-253.80 \pm 2.67$ & $137.26 \pm 2.67$ & 1.64 & $2.58 \pm 0.47$ &  \\ 
2011-10-04 & 55838.2697 & NACO & $-253.07 \pm 2.70$ & $135.40 \pm 2.70$ & 1.28 & $2.71 \pm 0.50$ &  \\ 
2011-10-04 & 55838.2709 & NACO & $-254.43 \pm 2.74$ & $135.72 \pm 2.75$ & 1.26 & $2.47 \pm 0.48$ &  \\ 
2011-10-04 & 55838.2721 & NACO & $-253.74 \pm 2.68$ & $136.47 \pm 2.68$ & 1.24 & $2.71 \pm 0.46$ &  \\ 
2011-10-04 & 55838.2733 & NACO & $-252.47 \pm 2.76$ & $136.15 \pm 2.76$ & 1.09 & $2.25 \pm 0.41$ &  \\ 
2011-10-04 & 55838.2746 & NACO & $-253.35 \pm 2.88$ & $137.56 \pm 2.88$ & 1.08 & $2.23 \pm 0.44$ &  \\ 
2011-10-04 & 55838.2758 & NACO & $-251.65 \pm 2.81$ & $137.89 \pm 2.81$ & 1.04 & $2.37 \pm 0.45$ &  \\ 
2011-11-20 & 55885.0603 & NACO & $-252.09 \pm 2.66$ & $143.42 \pm 2.66$ & 2.17 & $2.58 \pm 0.49$ &  \\ 
2011-11-20 & 55885.0631 & NACO & $-251.86 \pm 2.66$ & $144.27 \pm 2.66$ & 2.17 & $2.56 \pm 0.48$ &  \\ 
2012-08-12 & 56151.3938 & NACO & $-234.42 \pm 2.66$ & $179.37 \pm 2.66$ & 2.17 & $2.50 \pm 0.52$ &  \\ 
2012-08-12 & 56151.3958 & NACO & $-234.15 \pm 2.66$ & $179.13 \pm 2.66$ & 2.12 & $2.52 \pm 0.52$ &  \\ 
2012-08-12 & 56151.3971 & NACO & $-234.27 \pm 2.66$ & $179.21 \pm 2.66$ & 1.75 & $2.29 \pm 0.43$ &  \\ 
2012-08-12 & 56151.3991 & NACO & $-233.24 \pm 2.66$ & $180.07 \pm 2.66$ & 1.28 & $2.13 \pm 0.67$ &  \\ 
2012-08-12 & 56151.4013 & NACO & $-233.47 \pm 2.68$ & $175.71 \pm 2.68$ & 1.26 & $2.20 \pm 0.74$ &  \\ 
2012-08-12 & 56151.4033 & NACO & $-234.24 \pm 2.73$ & $178.45 \pm 2.73$ & 1.09 & $1.86 \pm 0.78$ &  \\ 
2012-08-12 & 56151.4045 & NACO & $-234.85 \pm 2.71$ & $175.90 \pm 2.72$ & 1.08 & $1.87 \pm 0.51$ &  \\ 
2012-08-12 & 56151.4056 & NACO & $-232.32 \pm 2.71$ & $177.83 \pm 2.71$ & 1.04 & $1.91 \pm 0.67$ &  \\ 
2012-08-25 & 56164.3359 & NACO & $-232.59 \pm 2.66$ & $180.17 \pm 2.66$ & 2.17 & $2.56 \pm 0.49$ &  \\ 

    \hline
    \end{tabular}
\end{table*}

\begin{table*}
\renewcommand\thetable{A.1}
\caption{continued.}
\centering \small
  \begin{tabular}{lclrrcrc}
    \hline
    \hline
Date & MJD & Instrument & $\Delta \alpha$ & $\Delta \delta$ & $\lambda$ & $f_B/f_A$ & Flag \\
         &     &             & (mas)           & (mas)        & ($\mu$m) & (\%)   \\
    \hline  \noalign{\smallskip}
2012-08-25 & 56164.3371 & NACO & $-235.44 \pm 2.70$ & $196.32 \pm 2.78$ & 2.12 & $2.95 \pm 0.76$ & $\bullet$ \\ 
2012-08-25 & 56164.3382 & NACO & $-232.24 \pm 2.65$ & $180.53 \pm 2.65$ & 1.75 & $2.68 \pm 0.50$ &  \\ 
2012-08-25 & 56164.3392 & NACO & $-231.06 \pm 2.67$ & $180.82 \pm 2.67$ & 1.64 & $2.52 \pm 0.42$ &  \\ 
2012-08-25 & 56164.3402 & NACO & $-230.96 \pm 2.68$ & $182.47 \pm 2.68$ & 1.28 & $2.26 \pm 0.45$ &  \\ 
2012-08-25 & 56164.3412 & NACO & $-230.13 \pm 2.72$ & $183.38 \pm 2.72$ & 1.26 & $2.39 \pm 0.47$ &  \\ 
2012-08-25 & 56164.3432 & NACO & $-232.20 \pm 2.74$ & $182.86 \pm 2.74$ & 1.09 & $2.56 \pm 0.59$ &  \\ 
2012-08-25 & 56164.3444 & NACO & $-230.06 \pm 2.71$ & $182.51 \pm 2.71$ & 1.08 & $2.78 \pm 0.51$ &  \\ 
2012-12-07 & 56268.0620 & NACO & $-219.09 \pm 2.65$ & $189.35 \pm 2.65$ & 2.17 & $2.44 \pm 0.45$ &  \\ 
2012-12-07 & 56268.0634 & NACO & $-219.87 \pm 2.65$ & $189.78 \pm 2.65$ & 2.12 & $2.50 \pm 0.47$ &  \\ 
2012-12-07 & 56268.0649 & NACO & $-219.55 \pm 2.65$ & $189.85 \pm 2.65$ & 1.75 & $2.43 \pm 0.43$ &  \\ 
2012-12-07 & 56268.0663 & NACO & $-219.50 \pm 2.66$ & $189.73 \pm 2.66$ & 1.64 & $2.39 \pm 0.45$ &  \\ 
2012-12-07 & 56268.0679 & NACO & $-218.80 \pm 2.67$ & $190.21 \pm 2.67$ & 1.28 & $2.39 \pm 0.53$ &  \\ 
2012-12-07 & 56268.0695 & NACO & $-219.19 \pm 2.69$ & $189.99 \pm 2.69$ & 1.26 & $2.59 \pm 0.54$ &  \\ 
2012-12-07 & 56268.0711 & NACO & $-219.34 \pm 2.67$ & $188.59 \pm 2.67$ & 1.24 & $2.68 \pm 0.55$ &  \\ 
2012-12-07 & 56268.0726 & NACO & $-221.12 \pm 2.76$ & $189.33 \pm 2.76$ & 1.09 & $2.17 \pm 0.57$ &  \\ 
2012-12-07 & 56268.0742 & NACO & $-219.69 \pm 2.69$ & $188.77 \pm 2.69$ & 1.08 & $2.15 \pm 0.43$ &  \\ 
2012-12-07 & 56268.0757 & NACO & $-222.43 \pm 2.69$ & $185.43 \pm 2.69$ & 1.04 & $2.56 \pm 0.90$ &  \\ 
2015-10-09 & 57304.1001 & SPHERE & $44.61 \pm 1.22$ & $-12.91 \pm 2.34$ & 0.8168 & $1.48 \pm 0.68$ &  \\ 
2015-10-09 & 57304.1001 & SPHERE & $48.01 \pm 2.26$ & $-15.39 \pm 2.15$ & 0.554 & $0.10 \pm 0.44$ & $\bullet$ \\ 
2015-10-09 & 57304.1090 & SPHERE & $47.90 \pm 4.28$ & $-12.41 \pm 4.12$ & 0.6556 & $0.33 \pm 0.50$ & $\bullet$ \\ 
2015-10-09 & 57304.1090 & SPHERE & $47.55 \pm 1.86$ & $-14.65 \pm 1.86$ & 0.6449 & $0.61 \pm 0.50$ & $\bullet$ \\ 
2015-11-10 & 57336.1727 & SPHERE & $27.94 \pm 0.92$ & $-42.77 \pm 1.06$ & 0.8168 & $1.31 \pm 0.41$ &  \\ 
2015-11-10 & 57336.1727 & SPHERE & $29.20 \pm 1.16$ & $-41.21 \pm 1.13$ & 0.554 & $1.47 \pm 0.41$ &  \\ 
2015-11-10 & 57336.1855 & SPHERE & $31.90 \pm 2.68$ & $-46.81 \pm 4.09$ & 0.6556 & $0.97 \pm 0.23$ & $\bullet$ \\ 
2015-11-10 & 57336.1855 & SPHERE & $30.12 \pm 1.53$ & $-43.52 \pm 1.63$ & 0.6449 & $1.12 \pm 0.34$ &  \\ 
2015-11-17 & 57343.0441 & SPHERE & $23.48 \pm 1.03$ & $-49.88 \pm 1.06$ & 0.8168 & $0.19 \pm 0.44$ & $\bullet$ \\ 
2015-11-17 & 57343.0441 & SPHERE & $26.20 \pm 1.31$ & $-45.97 \pm 1.17$ & 0.554 & $0.88 \pm 0.66$ & $\bullet$ \\ 
2015-11-17 & 57343.0591 & SPHERE & $24.42 \pm 1.59$ & $-47.29 \pm 1.22$ & 0.6449 & $0.68 \pm 0.42$ & $\bullet$ \\ 
2015-11-17 & 57343.0591 & SPHERE & $24.33 \pm 1.38$ & $-47.86 \pm 1.28$ & 0.6556 & $0.46 \pm 0.38$ & $\bullet$ \\ 
2015-11-18 & 57344.0226 & SPHERE & $19.83 \pm 0.98$ & $-50.16 \pm 0.84$ & 0.8168 & $0.52 \pm 0.43$ & $\bullet$ \\ 
2015-11-18 & 57344.0226 & SPHERE & $22.94 \pm 0.91$ & $-46.02 \pm 0.82$ & 0.554 & $0.47 \pm 0.42$ & $\bullet$ \\ 
2015-11-27 & 57353.1197 & SPHERE & $12.32 \pm 1.47$ & $-51.64 \pm 1.43$ & 0.8168 & $1.45 \pm 0.57$ &  \\ 
2015-11-27 & 57353.1197 & SPHERE & $16.30 \pm 1.70$ & $-48.72 \pm 1.20$ & 0.554 & $1.33 \pm 0.47$ &  \\ 
2015-11-27 & 57353.1330 & SPHERE & $14.04 \pm 1.65$ & $-48.10 \pm 1.92$ & 0.6556 & $1.09 \pm 0.51$ &  \\ 
2015-11-27 & 57353.1330 & SPHERE & $14.03 \pm 1.11$ & $-50.37 \pm 1.11$ & 0.6449 & $1.31 \pm 0.44$ &  \\ 
2016-10-12 & 57674.7517 & PIONIER & $-162.70 \pm 1.60$ & $-13.90 \pm 1.40$ & 1.64 & $1.23 \pm 0.27$ &  \\ 
2016-11-11 & 57705.5752 & PIONIER & $-174.60 \pm 1.60$ & $-6.60 \pm 1.50$ & 1.64 & $0.71 \pm 0.32$ &  \\ 
2016-11-21 & 57713.4683 & GRAVITY & $-174.24 \pm 0.02$ & $-4.79 \pm 0.05$ & 2.17 & $1.61 \pm 0.76$ &  \\
2019-11-22 & 58749.0833 & MATISSE & $-227.70 \pm 2.00$ & $184.39 \pm 2.00$ & 3.50 &  & \\
     \hline
    \end{tabular}
\end{table*}

%% file: Tables/spectroscopy-radvel.tex
\subsection{Spectroscopy data set}

\begin{table*}
\caption{{\bf Achernar A spectroscopic data set}. The table's left panel describe the instrument characteristics: $\Delta\lambda$ refers to the spectrograph wavelength coverage, while $R$ refers to the spectral resolution. On the middle panel, $N_{\rm sp}$ refers to the total number of spectra gathered from each spectrograph, while $N_{\rm epoch}$ refers to the number of distinct observation epochs (\ie~observation nights). The table's right panel details the instrument coverage of the 13 significant metallic absorption lines of Achernar A used in this study (see text).}
\label{tab:spectro}
 \centering
 \small
\renewcommand{\arraystretch}{1}
\setlength\tabcolsep{4.5pt}
  \begin{tabular}{l c c | c c | c c c c c}
    \hline
    \hline
    Instrument or database & $\Delta\lambda$ & $R = \lambda / \mathrm{d} \lambda$ & $N_{\rm sp}$   & $N_{\rm epoch}$  & Line set 1 & Line set 2 & Line set 3 & C II & O I \\
                           & (\AA)           &                                    &  ($\dagger$)   &                  & (3) & (3) & (5)  & (4267.26 \AA)    & (7774 \AA)  \\
    \hline  \noalign{\smallskip}
    ESO/HARPS$^{(a)}$      & 3900 -- 6800    & 115~000                            & 39 (19)    & 5  &   y    &   y    &  y     &  y   &  n      \\
    ESO/UVES$^{(b)}$       & 3900 -- 8800    & $\sim$55~000                       & 8          & 3  &   y    &   y    &  y     &  y   &  y      \\
    ESO/FEROS$^{(c)}$      & 3900 -- 8800    & 48~000                             & 175 (172)  & 44 &   y    &   y    &  y     &  y   &  y      \\
    OCA/BESO$^{(d)}$       & 3900 -- 8000    & 48~000                             & 20 (9)     & 9  &   y    &   y    &  y     &  y   &  y      \\
    CTIO/CHIRON$^{(e)}$    & 4500 -- 8000    & $\sim$130~000                      & 73 (68)    & 22 &   n    &   n    &  y     &  n   &  y      \\
    BeSOS/PUCHEROS$^{(f)}$ & 4250 -- 7300    & 18~000                             & 457 (401)  & 10 &   n    &   y    &  y     &  y   &  n      \\
    Euler/CORALIE$^{(g)}$  & 3900 -- 6800    & 55~000                             & 1          &  1 &   y    &   y    &  y     &  y   &  n      \\
    BeSS$^{(h)}$           & 4250 -- 7150    & 10~000                             & 1          &  1 &   n    &   y    &  y     &  n   &  n      \\
    \hline
    \end{tabular}
\tablefoot{{\it (a)} The {\it High-Accuracy Radial-velocity Planet Searcher} \citepads[HARPS,][]{pepe2002} echelle spectrograph is mounted on the 3.6m~telescope at La Silla Observatory (Chile). {\it (b)} The {\it Ultra Violet and Echelle Spectrograph} \citepads[UVES,][]{dekker2000} is mounted on the Unitary Telescope 2 (UT2, Kueyen) at the {\it Very Large Telescope} (VLT) ESO Observatory (Paranal, Chile). {\it (c)} FEROS \citepads{kaufer1998} is mounted on the 2.2m~telescope at La Silla.  {\it (d)} The {\it Bochum Echelle Spectrograph for OCA} \citepads[BESO,][]{steiner2006} is mounted on the 1.5m~Hexapod-Telescope at the Observatorio Cerro Armazones (OCA, Cerro Armazones, Chile). {\it (e)} The CHIRON echelle spectrograph \citepads{tokovinin2013} is mounted on the 1.5m~telescope at Cerro Tololo Interamerican Observatory (CTIO, Cerro Tololo, Chile). {\it (f)} PUCHEROS \citepads[the {\it Pontificia Universidad Catolica High Echelle Resolution Optical Spectrograph},][]{vanzi2012} is mounted on the ESO 50-cm telescope of the Observatory UC Santa Martina near Santiago, Chile. {\it (g)} CORALIE \citepads{queloz2001} is mounted on the Euler 1.2m~telescope of the Swiss Observatory at La Silla. {\it (h)} BeSS 2011-10-18 (observer: B. Heathcote).
{\it ($\dagger$)} The number in brackets refers to the number of spectra actually used for RV computation after prior spectrum selection, if some spectra are rejected (see text).}
\end{table*}
\renewcommand{\arraystretch}{1}

The different spectrographs from which we collected spectra of Achernar and the overall properties of the data are listed in Table~\ref{tab:spectro}.

\subsection{\half~and other Balmer lines}\label{sect:halpha}

\begin{figure*}
     \centering
         \includegraphics[width=0.47\hsize]{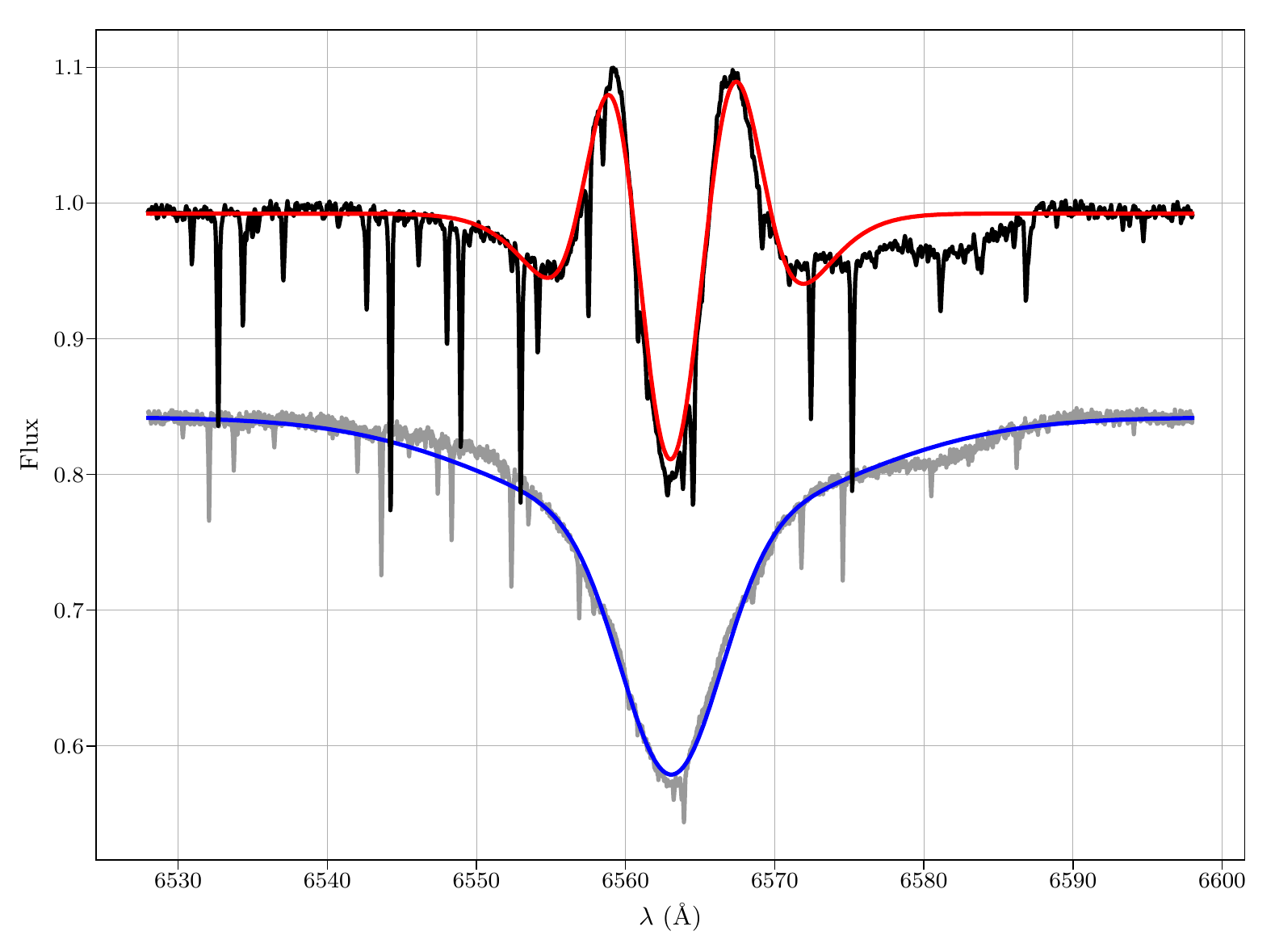}
         \includegraphics[width=0.47\hsize]{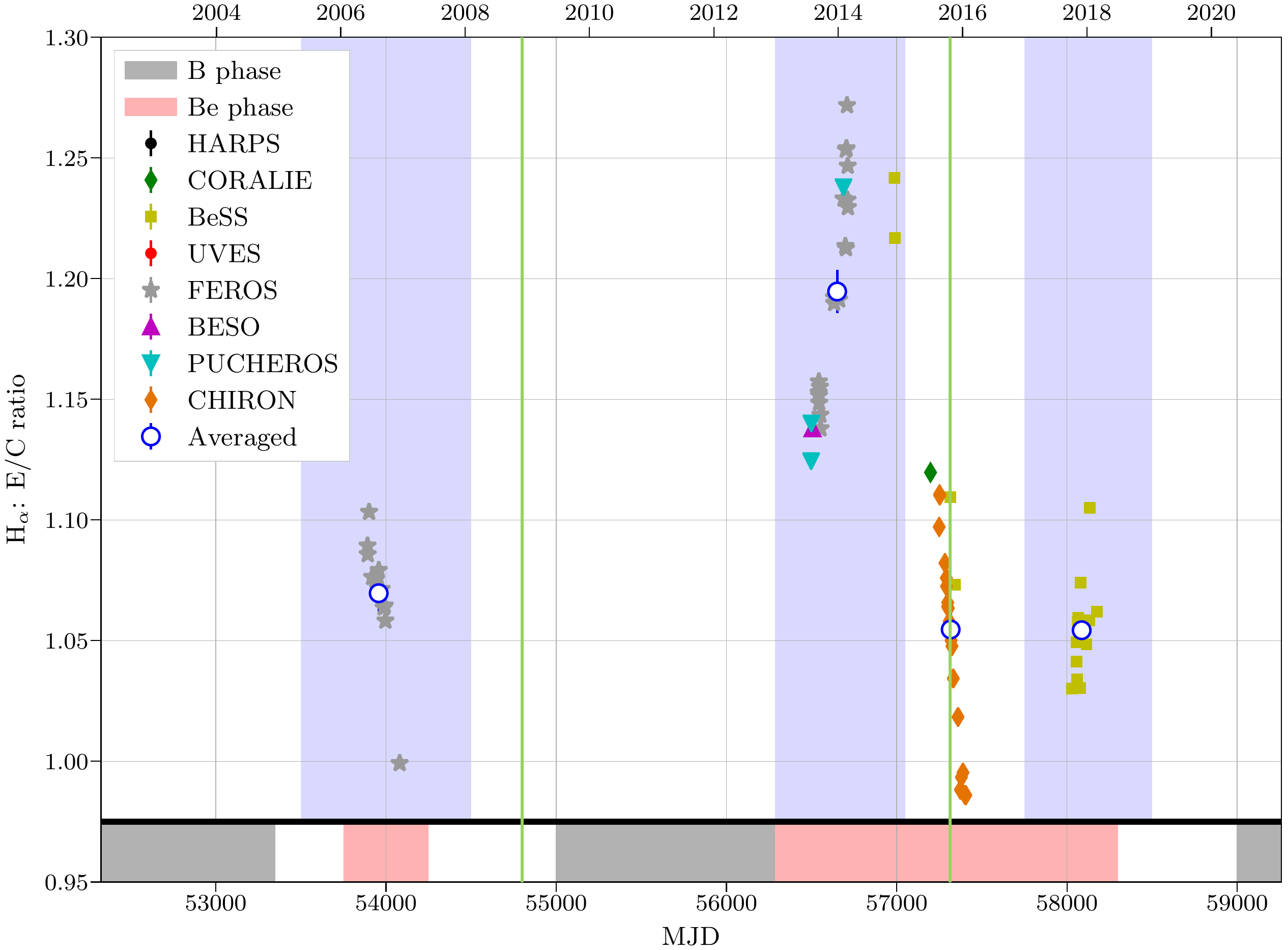}         
         \includegraphics[width=0.47\hsize]{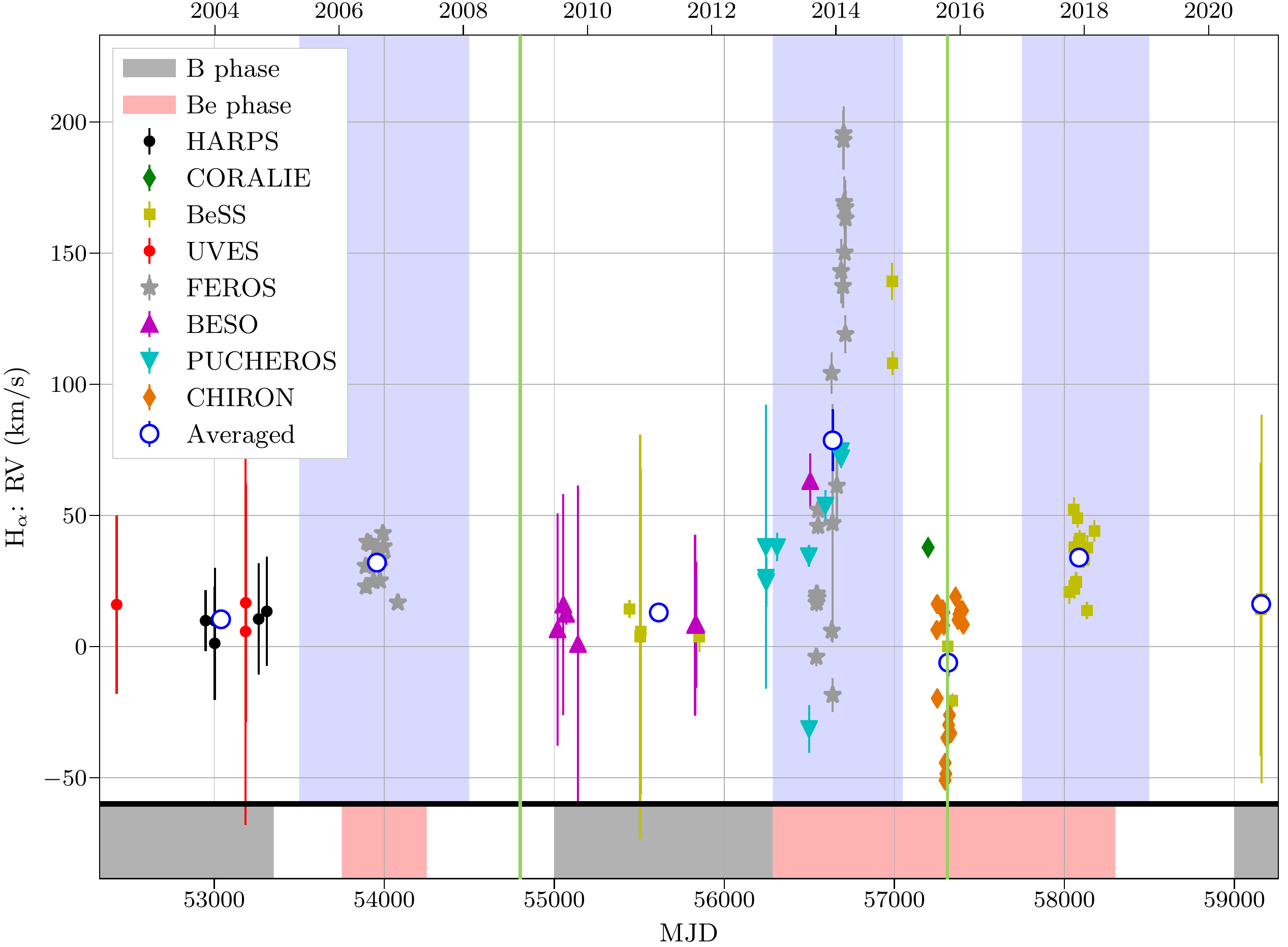}
         \includegraphics[width=0.47\hsize]{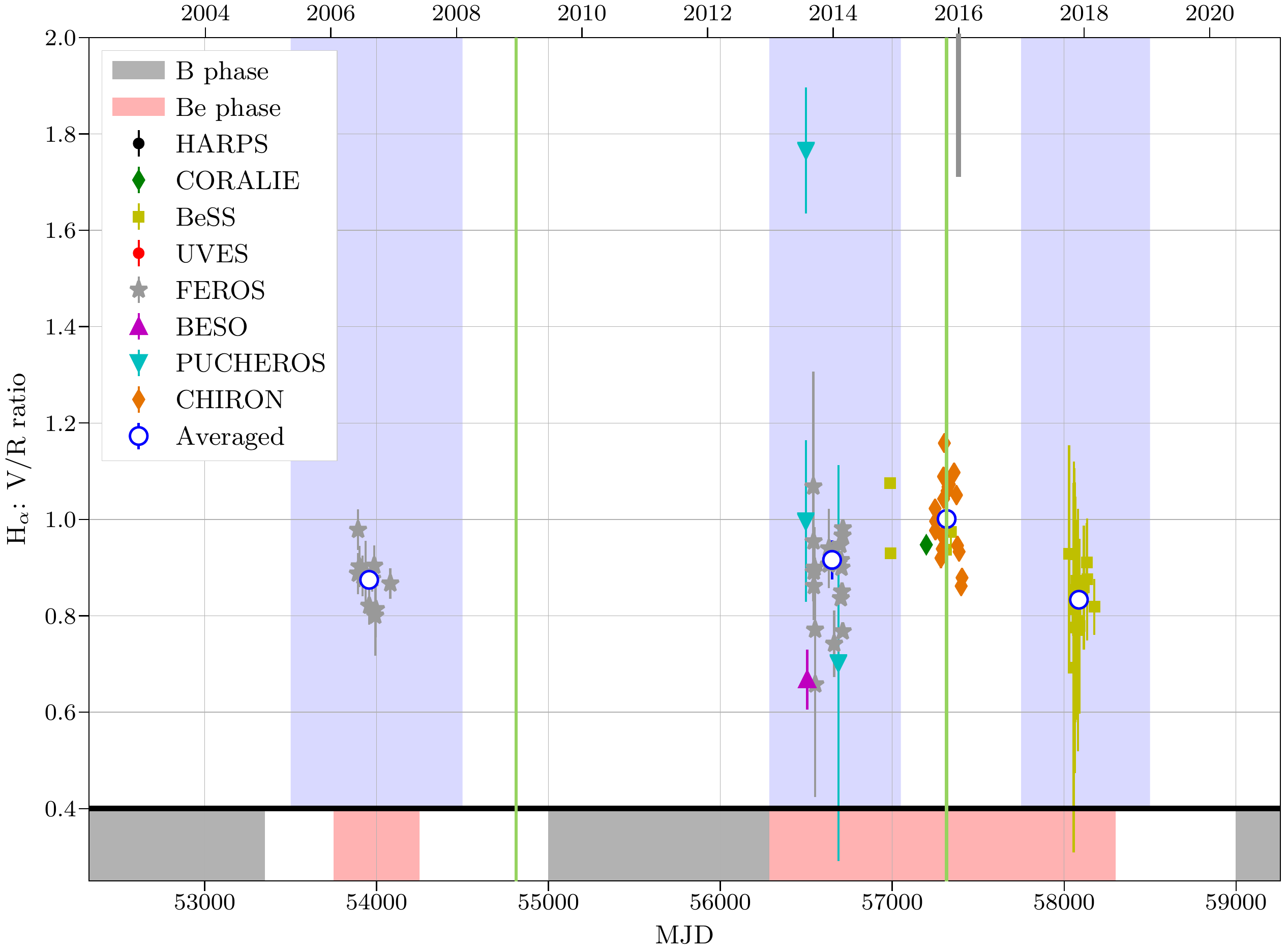}
     \caption{Variability of Achernar in the \half~line. The \half~line is displayed in the {\it top left panel} for the two cases: the emission, Be phase (2006 FEROS spectrum, upper black), and the quiescent, B phase (2003 HARPS spectrum, lower gray, flux-shifted for clarity). Empiric line models are displayed as the red and blue curves, respectively (see text). The three other panels display the time series of the observables derived from the \half~line modeling ({\it from left to right and top to bottom}): the E/C ratio, the $H_\alpha$ line RV, and the V/R ratio. At the bottom of each panel, observed emission and quiescent phases are highlighted in red and gray, respectively. The periastron passages at epochs 2008.78 and 2015.81 are marked with green vertical lines. Open blue circles represent data values averaged over large time intervals, highlighted in blue or white in the background.}\label{fig:halpha}
\end{figure*}

As can be expected for a B-type star, the spectrum of Achernar exhibits very few spectral lines. The Hydrogen Balmer and Paschen series are dominating and are all in absorption except for \half, which displays intermittent emission. This behavior is relatively well-explained by the existence of a circumstellar gaseous Keplerian disk fed by mass ejected from the star. In all the spectra considered here, the disk is weak (no emission in H$_{\beta}$ or in the other Balmer lines, only weak emission in \half) and transient, meaning that Achernar alternates between a quiescent, emission-free phase, and an \half-emission phase when the disk is present \citepads[B~$\rightleftharpoons$~Be,][]{2013A&A...559L...4R, 2013A&ARv..21...69R}. When in emission, the \half~line of Achernar displays a typical two-peak structure embedded into a wide absorption profile (Fig.~\ref{fig:halpha}, top left panel), that is used as a proxy for the disk behavior.\\

 For most of our spectroscopic data set, several consecutive spectra were acquired at each observation epoch (Table~\ref{tab:spectro}). We stacked these consecutive spectra together to increase the spectral S/N and to reduce the possible high-frequency jitter in the data. Then we fit \half~with two basic, empirical models to monitor Achernar B~$\rightleftharpoons$~Be phases, and investigate a possible link with the companion orbit. The emission-free model consisted in the sum of two Gaussians in absorption, one with a shallow, broad peak, and the other with a narrower deeper peak. For the emission model, we considered the sum of a shallow Gaussian in absorption plus two narrow Gaussians in emission. We emphasize that we did not try to model \half~perfectly but that we only tried to monitor its overall behavior. From the models, we derived three observables: the emission-to-continuum ratio (E/C; that is, the emission peak height above the continuum), the radial velocity (RV), and the ratio of the violet and red peaks \citepads[V/R,][]{2013A&ARv..21...69R}.\\

We are able to distinguish several different phases from our \half~modeling (Fig.~\ref{fig:halpha}). From mid-2002 to at least mid-2004, Achernar was in a quiescent B phase (no significant \half~emission), before entering an emission Be phase in 2006. This \half~behavior was already monitored by previous studies \citepads{carciofi2007,2013A&A...559L...4R}. The disk responsible for this emission must have vanished some time later as from 2009 to 2013 \half~was again in absorption. Then, a new disk started forming between November 2012 and January 2013, with \half~emission apparently lasting at least until 2018. In late 2020, this new disk has again vanished. The timing of the start of this second Be phase is confirmed by interferometric observations of the disk formation \citepads{desouza2014,vedova2017}. We note that both Be phases seem to occur around a passage at periastron of the binary companion, with the first observed Be phase happening roughly two years before the passage in late 2008, and with the second Be phase enclosing the next passage in late 2015 (CHIRON epochs).

We observe that Achernar \half~emission phases remain quite weak for a Be star, with an E/C ratio always below 1.3, confirming previous studies \citepads{2013A&A...559L...4R}. Interestingly, the strength of the \half~emission seems to vary only on long timescales: the E/C ratio is continuously decreasing over the 2006 FEROS data set, as well as over the 2015-2016 CHIRON data set, while it is almost continuously increasing over the 2014 FEROS data set as well as over the 2018 BeSS data set (Fig.~\ref{fig:halpha}). The \half~RV are highly variable during the Be phases. Due to the very large uncertainties derived from the emission-free model, it is difficult to assess the \half~RV during Achernar Be phases, though the RV values seem to be close to the V$_{\gamma}$~$\simeq$~19~\kms~from the litterature.

For several Be stars with short-period binary companions, the variability of the V/R ratio of several Be stars was found to be synchronized with the orbital motion \citepads{2013A&A...560A..30Z,saad2020}. For Achernar, the V/R ratio is quite stable over the two observed Be phases, with most data in the $0.75-1$ range over the 2006, 2013-2014, and 2018 data sets.
However, the V/R displays a small but significant average increase over the 2015-2016 (CHIRON) data set, with data in the 0.85-1.15 range ($\sim$1 in average). Interestingly, this corresponds to the expected passage at periastron from late 2015. Yet, given the smallness of this apparent V/R shift and the data dispersion, we consider it uncertain that this is caused by the proximity of the companion to the primary star.

\begin{figure*}
     \centering
         \includegraphics[width=\hsize]{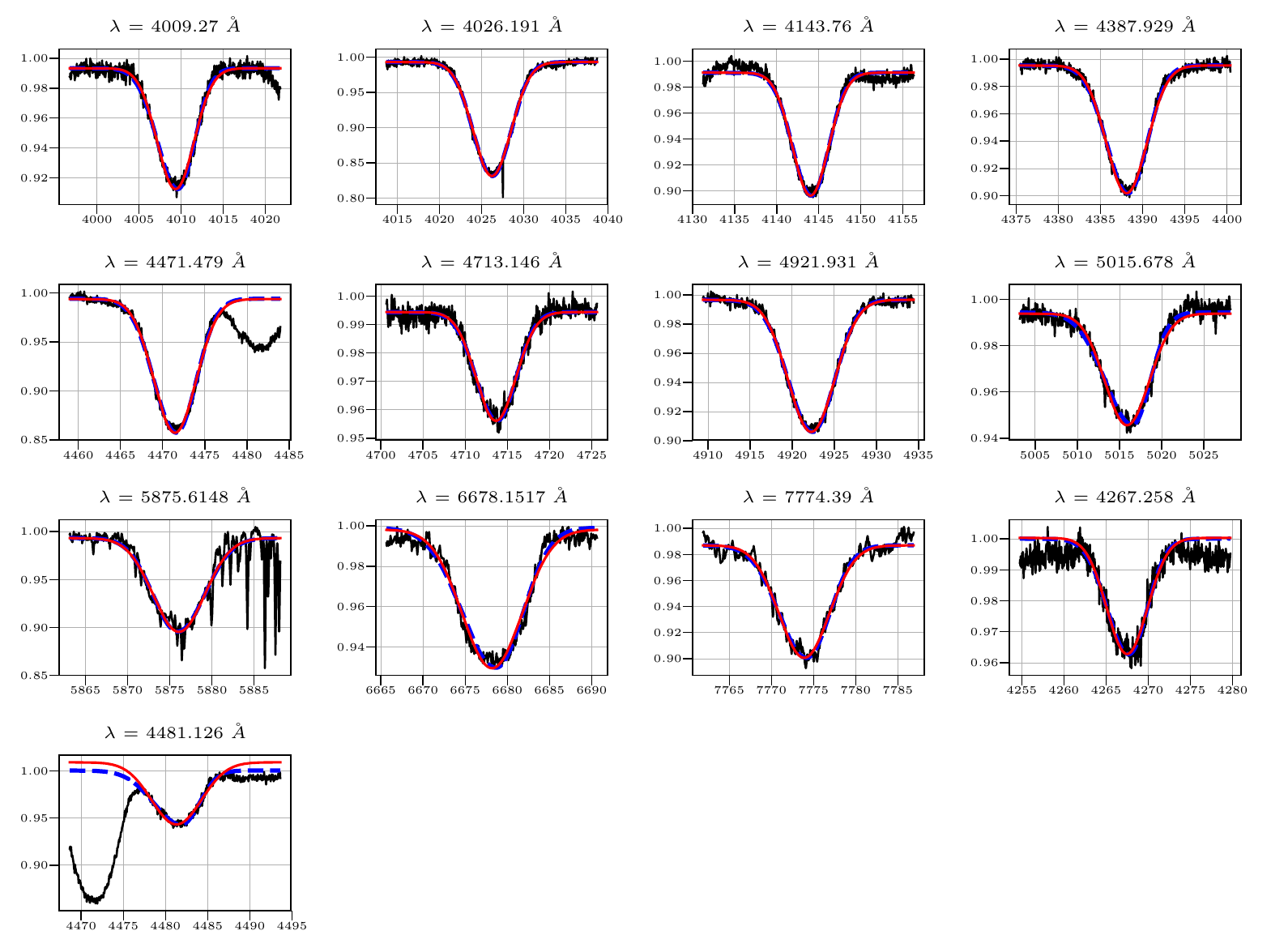}
     \caption{Significant photospheric absorption lines of Achernar A, taken from a 2006 FEROS stacked spectrum (see text). For each displayed line, the stacked spectrum is displayed in black, a Gaussian fit of the line is displayed as a red solid curve, and a biGaussian fit of the line is displayed as a blue dashed curve. The vertical axis is the ratio of the flux to the continuum, and the horizontal axes the wavelength in $\AA$}\label{fig:lines}
\end{figure*}

\begin{figure*}
     \centering
     \includegraphics[width=16cm]{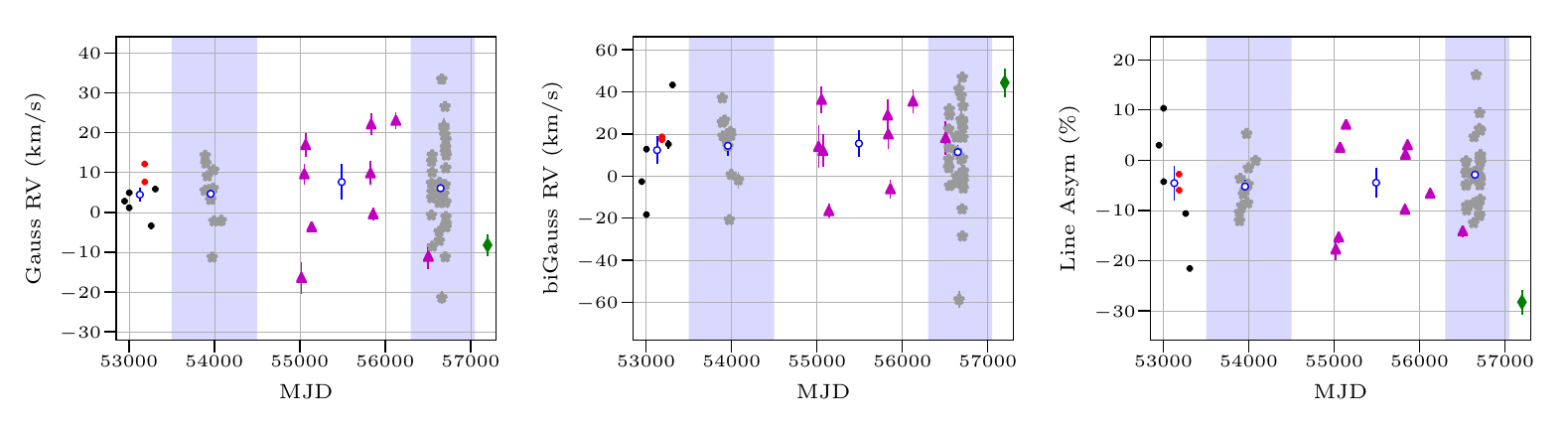}
     \includegraphics[width=16cm]{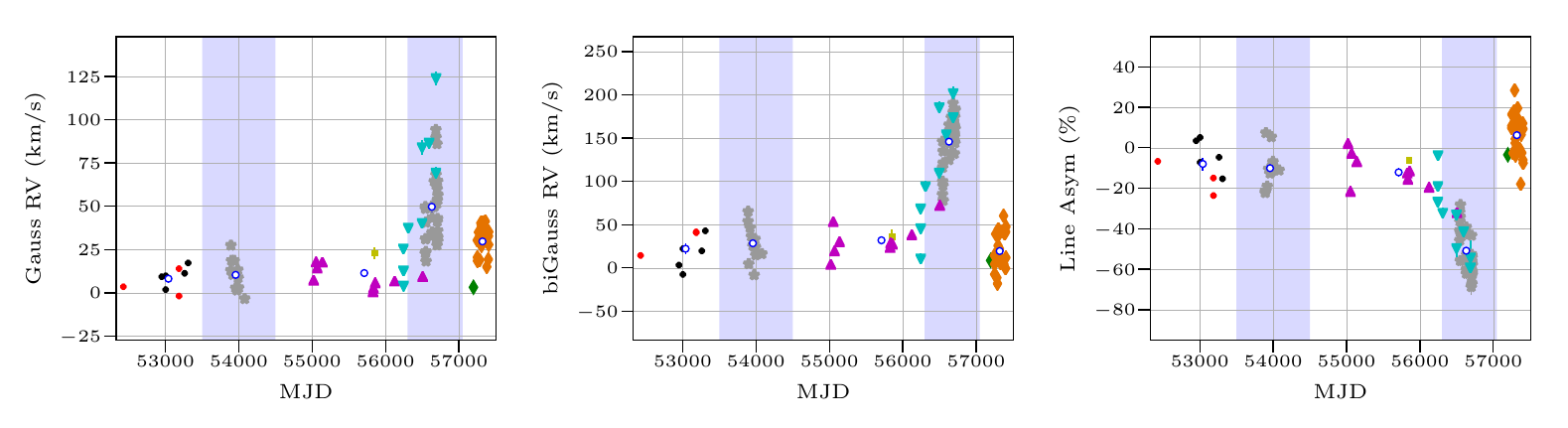}
     \caption{Single-line radial velocity data of Achernar A. The data are displayed for two lines: the He I $\lambda = $ 4009.26~\AA~(top row), and the He I $\lambda =$ 5015.68~\AA~(bottom row), respectively. On each row are displayed from left to right: the Gaussian RVs, the biGaussian RVs, and the line asymmetry as measured within the biGaussian line model (see text). Each spectrograph is color- and symbol-coded (\eg,~HARPS/UVES: black/red circles, FEROS: gray stars, BESO/PUCHEROS: purple/cyan triangles, CHIRON/CORALIE: orange/green diamonds, BeSS: yellow square).}\label{fig:line_rv}
\end{figure*}

\subsection{Helium and metallic lines}

\subsubsection{Significant metallic absorption lines:} 

Apart from the Hydrogen Balmer and Paschen series, we count only thirteen significant photospheric absorption lines that are deep enough for characterization and radial velocity (RV) computation (Fig.~\ref{fig:lines}). Ten of these lines are of neutral Helium (He~I), while the three others are of singly ionized metals. The list of observed lines is the following\footnote{We used both the study by \cite{wiese2009} and the Handbook of Basic Atomic Spectroscopic Data \citepads[][available online at: \url{https://www.nist.gov/pml/handbook-basic-atomic-spectroscopic-data}]{sansonetti2005} to identify the absorption lines.}:
\begin{itemize}
    \item three He~I lines at $\lambda = 4009.26$, 4026.21, 4143.76~\AA~(line set 1 in Table~\ref{tab:spectro}),
    \item the ionized Carbon (C~II) line at $\lambda = 4267.26$~\AA,
    \item two He~I lines at $\lambda = 4387.93$~and 4471.48~\AA, and the ionized Magnesium (Mg~II) line at $\lambda = 4481.13$~\AA~(line set 2),
    \item five He~I lines at $\lambda = 4713.17$, 4921.93, 5015.68, 5875.66, and 6678.15~\AA~(line set 3),
    \item the ionized Oxygen (O~I) line at $\lambda = 7774.39$~\AA.
\end{itemize}
These absorption lines are quite shallow (with a relative depth between 5 and 15\%~of the continuum) and strongly broadened, with a typical Full Width at Half Maximum (FWHM) as large as $\sim 5\,\AA$~($\approx 300$\,km\,s$^{-1}$; Fig.~\ref{fig:lines}).
This is evidently caused by the very fast rotation rate of Achernar A. These thirteen lines do not exhibit any significant emission feature over our data set. Many of them are commonly used to derive RV time series of B and Be stars that are spectroscopic binaries \citepads{wade2017, saad2020, 2022A&A...658A..69B}.
Other lines are also visible, such as He\,I $\lambda = 7065\,\AA$, C\,II $\lambda = 4267\,\AA$, Ca\,II $\lambda = 3934\,\AA$, Si\,II $\lambda = 6347/73\,\AA$, Si\,II $\lambda = 4128/30\,\AA$ and C\,II $\lambda=6578/82$ (visible in Fig.~\ref{fig:halpha}).

\subsubsection{Single-line RV data:}

We first computed single-line RV times series for the thirteen absorption lines detailed above.  Each line in each stacked spectrum was fit ({\it i}) with a four-parameter Gaussian model, and ({\it ii}) with a five-parameter biGaussian model, that takes into account the line asymmetry (Fig.~\ref{fig:lines}).
The star shows periodic line profile variability, with a pattern often described as ``swaying'' in the early days. That is (one of) the periods that is also seen in photometry. While \citetads{2022ApJ...924..117B} consider the photometric signature ambiguous, \citetads{2003A&A...411..229R} ascribe the accompanying spectroscopic signature to pulsation.
The fitting process and parameters for both models are detailed in \cite{borgniet2019}. From each model, we thus obtain a RV time series, as well as a time series of the (biGaussian) line asymmetry as defined by \cite{nardetto2006}. The RV data generally exhibit a large dispersion (from 10 to 15 \kms~for the Gaussian RVs, and from 15 to 60 \kms~for the biGaussian RVs, respectively).
The RV uncertainties computed within the line fits are typically on the order of 1~\kms. Significantly, the line asymmetry sometimes also exhibits very large values, thereby impacting the corresponding RV data (Fig.~\ref{fig:line_rv}, lower panel) and contributing to further hide the companion-induced Doppler shift. Moreover, some of the absorption lines are not covered by all instruments, depending on the spectrograph wavelength range (see Fig.~\ref{fig:line_rv}, upper panel, and Table~\ref{tab:spectro}).
Crucially, the five bluest He~I lines (line sets 1 and 2), that are generally the deepest lines and thus best-suited for a more precise RV computation (Fig.~\ref{fig:lines}), are not covered by CHIRON at the expected time of the companion passage at periastron. All these factors make it unrealistic to detect the companion-induced Doppler shift in single-line RV data.\\

\begin{figure*}
     \centering
     \includegraphics[width=0.4\hsize]{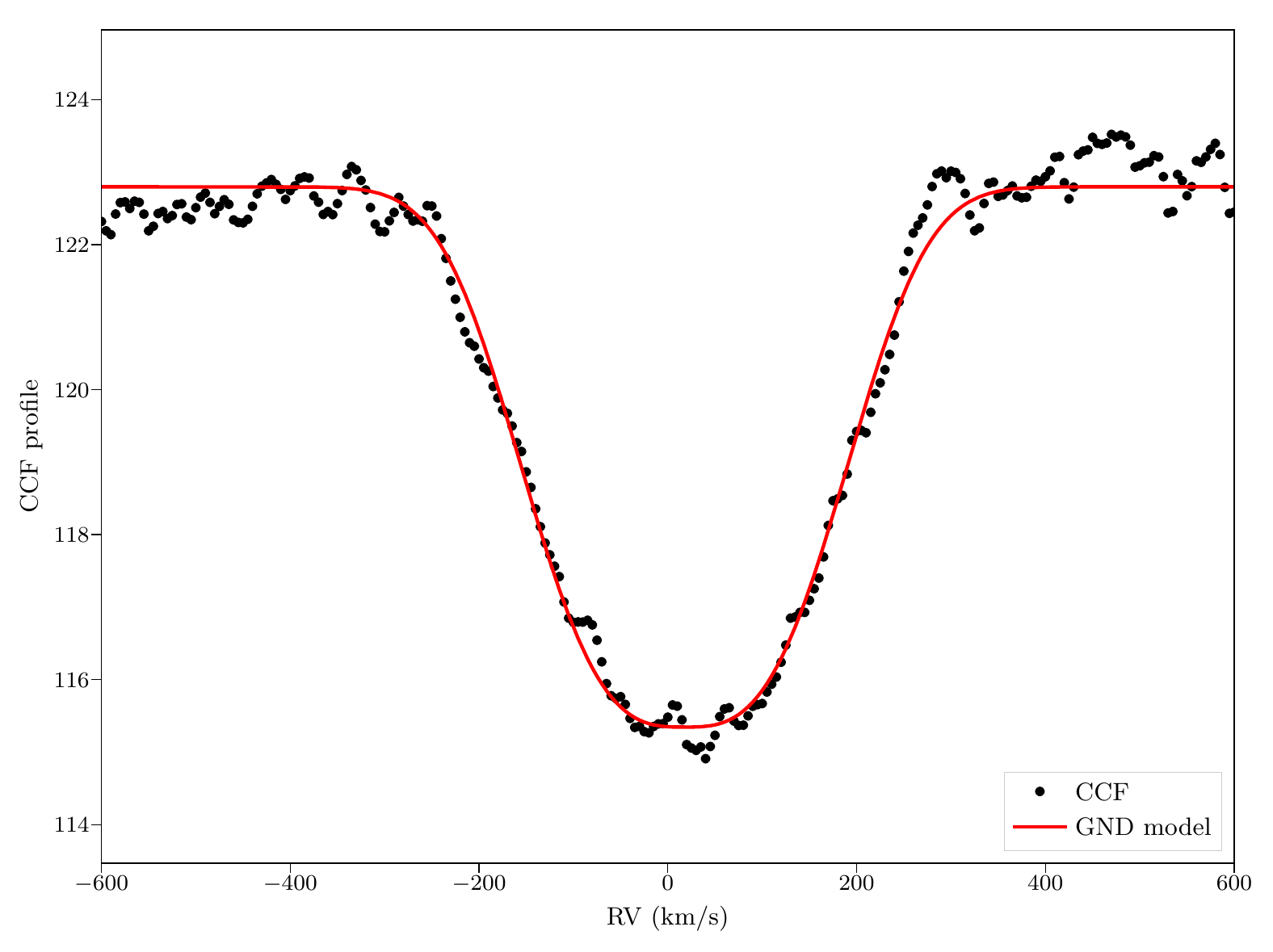} \ \ 
     \includegraphics[width=0.4\hsize]{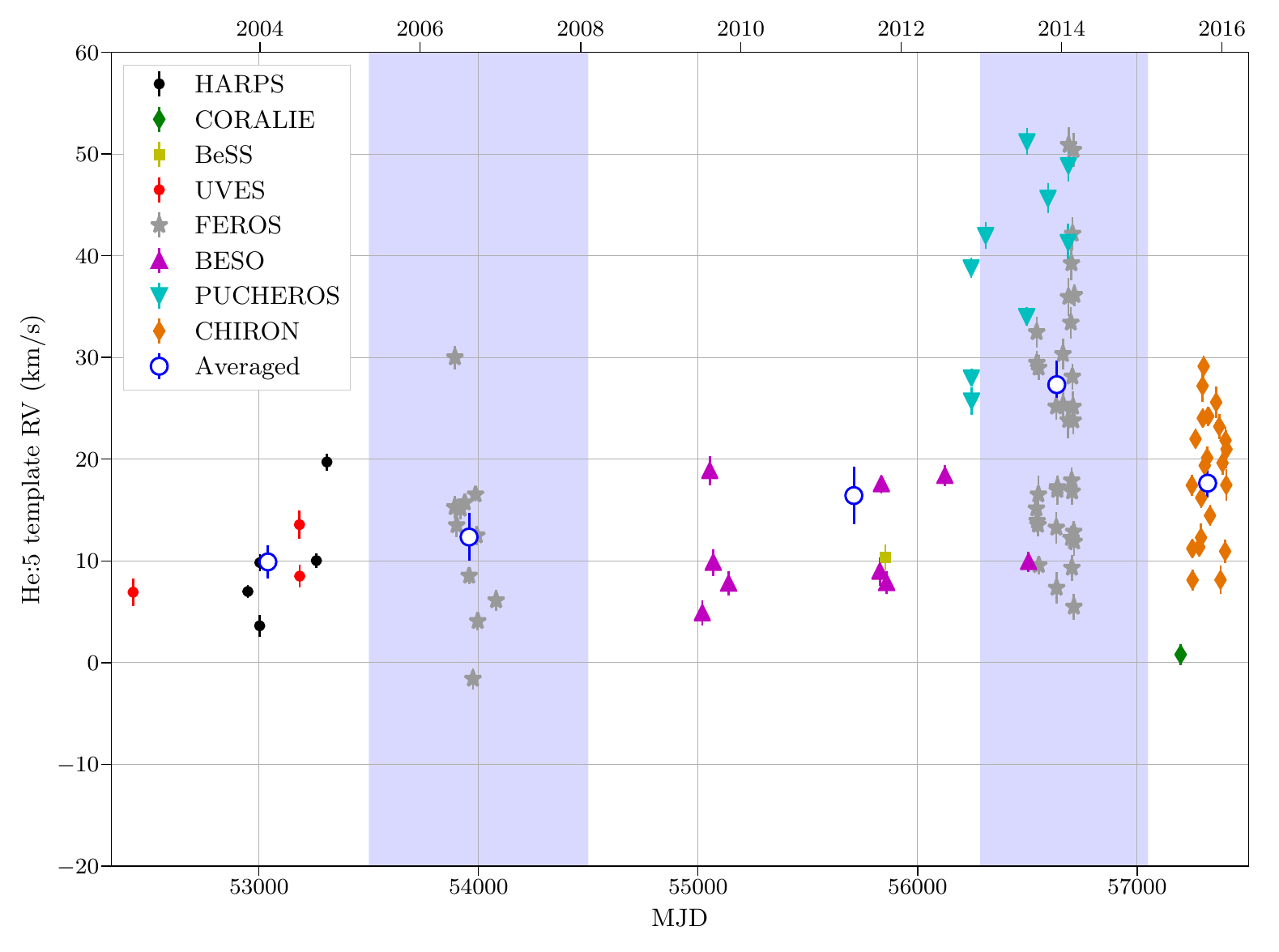} \ \ 
     \includegraphics[width=0.4\hsize]{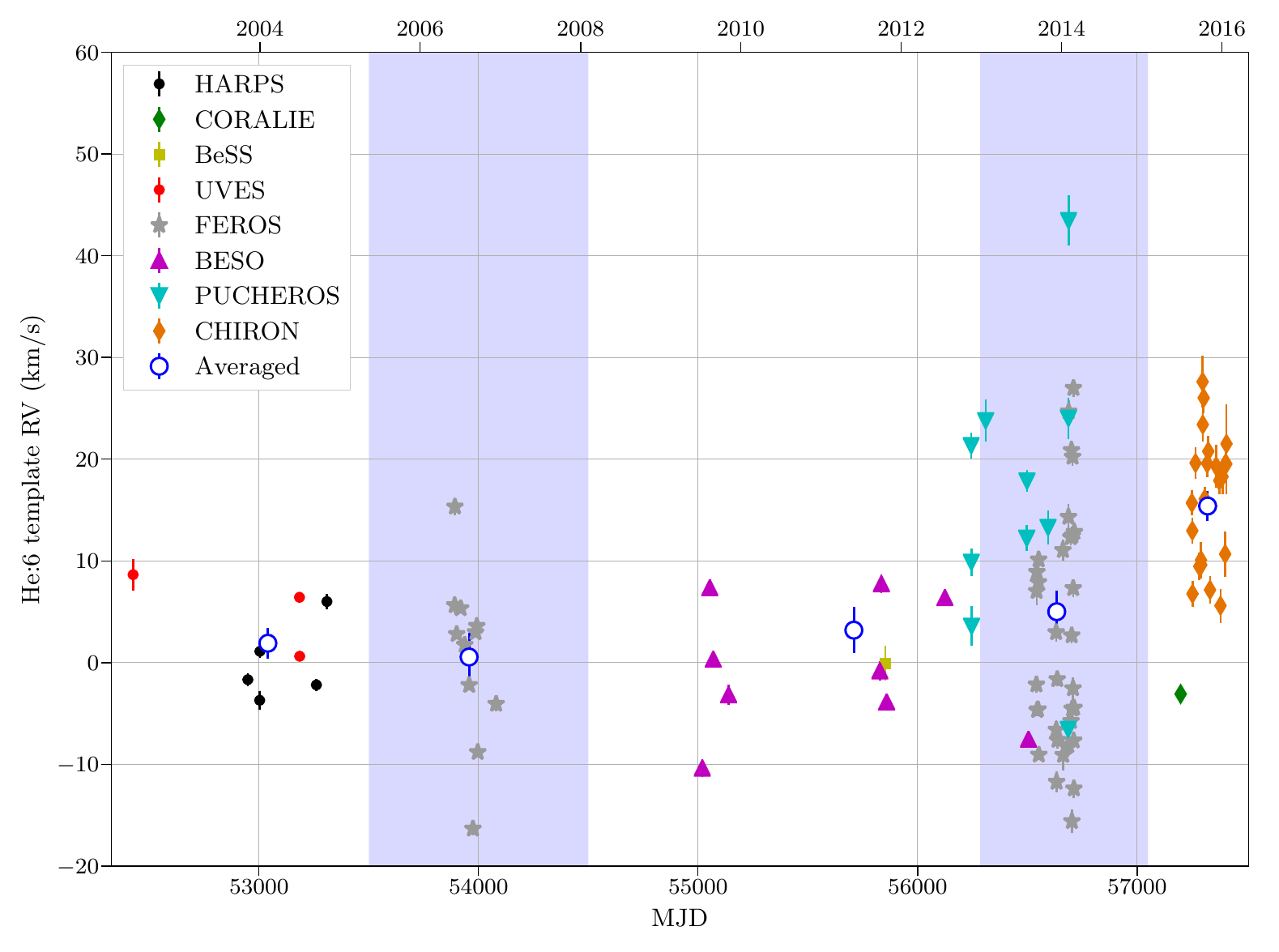} \ \ 
     \includegraphics[width=0.4\hsize]{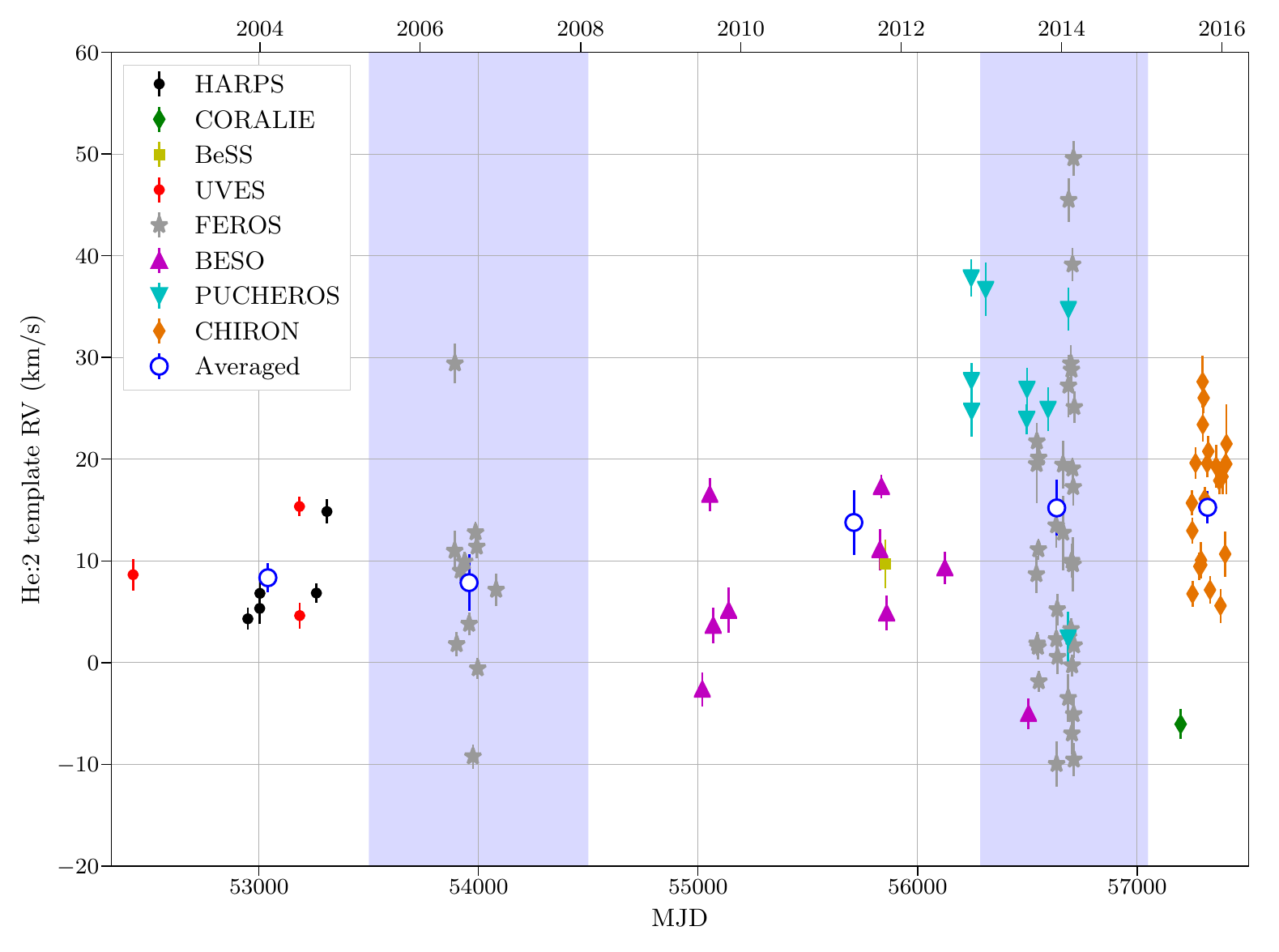}
     \caption{Radial velocity of Achernar A from the cross correlation technique. {\it Top left}: a CCF profile of Achernar A, built from a CHIRON stacked spectrum cross-correlated with a binary template including the five reddest He~I lines (line set 3), and fit with a generalized normal distribution (GND, see text). {\it Other panels}: resulting RV time series from the same five-line binary template ({\it top right}), from a binary template including the six less asymmetric He~I lines ({\it bottom left}), and from a binary template including only the two He~I lines common to both previous templates ({\it bottom right}), respectively.}\label{fig:ccf_rv}
\end{figure*}

\subsubsection{Cross-correlation RV data:}

We then combined the RV information from several of the photospheric lines at once by resorting to the commonly used cross-correlation function (CCF) framework (see, e.g., \citeads{borgniet2019}) to achieve a better RV accuracy. To that end, we built several binary templates including different combinations of some of the thirteen lines from Table~\ref{tab:spectro}.
Each binary template consisted in a set of boxcar functions with a value of one over the wavelength ranges corresponding to the selected absorption lines, and zero elsewhere. Instead of having a box width of one pixel, we adopted a larger box width around the line rest wavelength (with a total box width of 0.25 \AA), similarly to the approach detailed by \cite{borgniet2019}.
This allows to reduce both the noise in the wings of the resulting CCF profile, and the line-induced asymmetry of the CCF core. We cross-correlated each template with our Achernar stacked spectra over a wide RV range chosen to fully cover the CCF peak: from -600 to 600 \kms, with a 5~\kms~step. 
The resulting CCFs are typically very broad, and differ significantly from a Gaussian due to their ``flattened'' core (Fig.~\ref{fig:ccf_rv}, top left panel). For this reason, we decided to fit the CCFs with a Generalized Normal Function (GND) instead of a classical Gaussian model, as done and detailed by \cite{heitzmann2021} for young, active stars.

Our first cross-correlation template included the five reddest He~I lines (line set 3 in Table~\ref{tab:spectro}), that are the only lines to be covered by all spectrographs in our data set.
We display the resulting RV in Fig.~\ref{fig:ccf_rv} (upper right panel). Prior to 2013, the RV data exhibit a $\sim$6~\kms~dispersion with values mostly between 0 and 20~\kms. The late 2015 CHIRON RV data display the same characteristics, but are slightly redshifted compared to the RV prior 2013 (quite similarly to the \half~V/R ratio).
However, the intermediate RV data (mostly the PUCHEROS RV and the second FEROS data set) exhibit a much higher dispersion ($\sim$14~\kms) and are significantly redshifted.
We attribute the FEROS RV shift and increased dispersion to the strong asymmetry exhibited by some of the He~I lines included in the template during this period (mostly $\lambda$4921.93 and $\lambda$5015.68~\AA, see Fig.~\ref{fig:line_rv}, lower panel).
This also corresponds to the start of the second observed Be phase. 

Consequently, we selected for our second template the six less asymmetric He~I lines, \ie~that do not show a sudden increase in asymmetry at a given period. This includes the three bluest lines from line set 1, plus $\lambda$4471.48~\AA~from line set 2, and $\lambda$5875.66 and $\lambda$6678.15~\AA~only from line set 3. The resulting RV are not redshifted anymore over the 2013-2014 period, although the RV dispersion is still increased by a factor $\sim$2 over this period (Fig.~\ref{fig:ccf_rv}, lower left panel). However, the late CHIRON RV are now significantly redshifted (by $\sim$10~\kms) compared to the prior data. Although it could be tempting to deem this CHIRON data redshift the result of the companion passage at periastron, we attribute it to the fact that only the two reddest lines selected in the template contribute to the CHIRON RV data, while the prior RV result from the contribution of more or all of the six template lines. 

Finally, we tried to reconcile the advantages of the two previous templates by considering only the two reddest He~I lines ($\lambda$5875.66 and $\lambda$6678.15~\AA) for our third template. These two lines are the only ones that are both covered by all spectrographs and that do not exhibit too much asymmetry. The resulting RV are displayed in Fig.~\ref{fig:ccf_rv}, lower right panel. While the 2013-2014 RV data still exhibit a higher dispersion than the prior RV, they do not show a significant redshift in average anymore. The 2015-2016 CHIRON RV are now only very slightly redshifted (in average) compared to the prior RV data. This may or may not be the result of the companion-induced Doppler shift. We consider this final RV data set to be the more adequate for an orbital model fitting. For all three cross-correlation templates, individual GND RV uncertainties are on the order of 1~\kms.
The resulting spectroscopic radial velocity measurements are listed in Table~\ref{tab:allRV1}, and they correspond to the velocities used for the orbital fit (Sect.~\ref{sect:massA}; Fig.~\ref{fig:ABorbit}).

\begin{table}
\caption{\label{tab:allRV1} Radial velocity of Achernar A.}
\centering \small
  \begin{tabular}{lclrc}
    \hline
    \hline
    Date & MJD & Instrument & RV  & $\sigma$\\
         &     &            & (km/s) & (km/s)   \\
\hline
2003-11-06 & 52949.1351 & HARPS & 4.29 & 0.89 \\
2003-12-30 & 53003.0411 & HARPS & 5.26 & 1.31 \\
2003-12-31 & 53004.0536 & HARPS & 6.78 & 1.20 \\
2004-09-13 & 53261.3969 & HARPS & 6.81 & 0.87 \\
2004-10-31 & 53309.2992 & HARPS & 14.83 & 1.07 \\
2015-06-25 & 57198.3478 & CORALIE & -6.05 & 1.16 \\
2011-10-18 & 55852.4931 & BeSS & 9.75 & 1.86 \\
2002-06-01 & 52426.4496 & UVES & 8.57 & 1.50 \\
2004-06-28 & 53184.3915 & UVES & 15.35 & 0.87 \\
2004-06-29 & 53185.3605 & UVES & 4.59 & 1.19 \\
2006-06-05 & 53891.3289 & FEROS & 10.99 & 1.55 \\
2006-06-06 & 53892.3971 & FEROS & 29.41 & 1.58 \\
2006-06-14 & 53900.4326 & FEROS & 1.80 & 1.11 \\
2006-07-02 & 53918.4192 & FEROS & 8.99 & 0.90 \\
2006-07-21 & 53937.2450 & FEROS & 9.92 & 0.83 \\
2006-08-10 & 53957.2805 & FEROS & 3.79 & 0.93 \\
2006-08-27 & 53974.4215 & FEROS & -9.22 & 1.09 \\
2006-09-08 & 53986.3004 & FEROS & 12.81 & 0.78 \\
2006-09-14 & 53992.0732 & FEROS & 11.37 & 1.03 \\
2006-09-18 & 53996.3182 & FEROS & -0.59 & 0.90 \\
2006-12-11 & 54080.0443 & FEROS & 7.14 & 1.27 \\
2013-09-06 & 56541.2948 & FEROS & 8.68 & 1.40 \\
2013-09-07 & 56542.3936 & FEROS & 19.51 & 3.09 \\
2013-09-08 & 56543.3207 & FEROS & 21.73 & 1.45 \\
2013-09-09 & 56544.3869 & FEROS & 1.87 & 0.96 \\
2013-09-13 & 56548.2309 & FEROS & 1.52 & 1.07 \\
2013-09-15 & 56550.1783 & FEROS & 11.10 & 0.93 \\
2013-09-16 & 56551.3214 & FEROS & 20.13 & 1.04 \\
2013-09-17 & 56552.1839 & FEROS & -1.87 & 0.96 \\
2013-12-05 & 56631.1612 & FEROS & 13.48 & 1.66 \\
2013-12-06 & 56632.0942 & FEROS & 2.28 & 1.78 \\
2013-12-07 & 56633.1781 & FEROS & -9.98 & 1.72 \\
2013-12-10 & 56636.1613 & FEROS & 5.19 & 1.30 \\
2013-12-11 & 56637.2017 & FEROS & 0.51 & 1.41 \\
2014-01-05 & 56662.1193 & FEROS & 19.46 & 1.79 \\
2014-01-06 & 56663.0505 & FEROS & 12.78 & 2.98 \\
2014-01-29 & 56686.0648 & FEROS & -3.49 & 1.88 \\
2014-01-30 & 56687.1305 & FEROS & 27.18 & 2.46 \\
2014-01-31 & 56688.0347 & FEROS & 45.52 & 1.79 \\
2014-02-10 & 56698.1147 & FEROS & 29.40 & 1.43 \\
2014-02-11 & 56699.0005 & FEROS & 3.36 & 0.91 \\
2014-02-13 & 56701.0023 & FEROS & 28.81 & 1.19 \\
2014-02-13 & 56701.9942 & FEROS & 9.99 & 1.34 \\
2014-02-15 & 56703.0054 & FEROS & -7.10 & 1.93 \\
2014-02-16 & 56704.0040 & FEROS & -0.25 & 0.93 \\
2014-02-16 & 56704.9969 & FEROS & 19.05 & 0.90 \\
2014-02-18 & 56706.0051 & FEROS & 39.10 & 1.36 \\
2014-02-19 & 56707.9944 & FEROS & 9.65 & 2.12 \\
2014-02-20 & 56708.9962 & FEROS & 17.23 & 1.41 \\
2014-02-21 & 56709.9976 & FEROS & 49.64 & 1.58 \\
2014-02-22 & 56710.9969 & FEROS & -5.06 & 1.28 \\
2014-02-23 & 56711.9912 & FEROS & -9.56 & 1.33 \\
2014-02-24 & 56712.9851 & FEROS & 1.68 & 1.08 \\
2014-02-25 & 56713.9921 & FEROS & 25.07 & 1.44 \\
    \hline
\end{tabular}
\end{table}

\begin{table}
\renewcommand\thetable{C.2}
\caption{continued.}
\centering \small
  \begin{tabular}{lclrc}
    \hline
    \hline
    Date & MJD & Instrument & RV  & $\sigma$\\
         &     &            & (km/s) & (km/s)   \\
\hline
2009-07-07 & 55019.4037 & BESO & -2.65 & 1.45 \\
2009-08-10 & 55053.3692 & BESO & 16.55 & 1.51 \\
2009-08-26 & 55069.2857 & BESO & 3.60 & 1.48 \\
2009-11-04 & 55139.1535 & BESO & 5.08 & 1.82 \\
2011-09-24 & 55828.1306 & BESO & 10.86 & 1.87 \\
2011-10-01 & 55835.1813 & BESO & 17.28 & 1.04 \\
2011-10-25 & 55859.0970 & BESO & 4.52 & 1.59 \\
2012-07-16 & 56124.3491 & BESO & 9.32 & 1.43 \\
2013-08-01 & 56505.2947 & BESO & -5.05 & 1.24 \\
2012-11-13 & 56244.1555 & PUCHEROS & 37.79 & 1.61 \\
2012-11-14 & 56245.0632 & PUCHEROS & 27.74 & 1.44 \\
2012-11-15 & 56246.1428 & PUCHEROS & 24.42 & 2.29 \\
2013-01-18 & 56310.1090 & PUCHEROS & 36.66 & 2.34 \\
2013-07-24 & 56497.3180 & PUCHEROS & 23.86 & 1.48 \\
2013-07-25 & 56498.3135 & PUCHEROS & 26.98 & 2.04 \\
2013-10-29 & 56594.1605 & PUCHEROS & 24.9 & 2.12 \\
2014-01-29 & 56686.0584 & PUCHEROS & 2.41 & 2.40 \\
2014-01-30 & 56687.0713 & PUCHEROS & 34.69 & 1.89 \\
2014-01-31 & 56688.0451 & PUCHEROS & 72.00 & 2.89 \\
2015-08-15 & 57249.3968 & CHIRON & 15.68 & 0.93 \\
2015-08-17 & 57251.4446 & CHIRON & 12.97 & 0.95 \\
2015-08-18 & 57252.4188 & CHIRON & 6.72 & 1.13 \\
2015-09-01 & 57266.1840 & CHIRON & 19.63 & 1.26 \\
2015-09-18 & 57283.3084 & CHIRON & 9.43 & 1.14 \\
2015-09-26 & 57291.1585 & CHIRON & 10.08 & 1.52 \\
2015-09-27 & 57292.1931 & CHIRON & 9.59 & 1.01 \\
2015-10-03 & 57298.3221 & CHIRON & 27.60 & 2.01 \\
2015-10-04 & 57299.2363 & CHIRON & 23.42 & 1.32 \\
2015-10-08 & 57303.2824 & CHIRON & 26.01 & 1.29 \\
2015-10-13 & 57308.1616 & CHIRON & 16.08 & 0.85 \\
2015-10-24 & 57319.1371 & CHIRON & 19.57 & 1.08 \\
2015-10-29 & 57324.2545 & CHIRON & 20.77 & 1.24 \\
2015-11-06 & 57332.1485 & CHIRON & 7.11 & 1.14 \\
2015-12-04 & 57360.0806 & CHIRON & 19.31 & 1.78 \\
2015-12-18 & 57374.0530 & CHIRON & 17.87 & 1.17 \\
2015-12-24 & 57380.0573 & CHIRON & 5.57 & 1.37 \\
2016-01-02 & 57389.0557 & CHIRON & 18.24 & 1.26 \\
2016-01-14 & 57401.0938 & CHIRON & 10.69 & 1.67 \\
2016-01-16 & 57403.0599 & CHIRON & 19.52 & 1.41 \\
2016-01-19 & 57406.0564 & CHIRON & 19.49 & 2.35 \\
2016-01-20 & 57407.0401 & CHIRON & 21.50 & 2.88 \\
    \hline
\end{tabular}
\end{table}

%% file: Achernar-binary-v3r2.bbl
\begin{thebibliography}{131}
\expandafter\ifx\csname natexlab\endcsname\relax\def\natexlab#1{#1}\fi

\bibitem[{{Albrecht} {et~al.}(2007){Albrecht}, {Reffert}, {Snellen},
  {Quirrenbach}, \& {Mitchell}}]{2007A&A...474..565A}
{Albrecht}, S., {Reffert}, S., {Snellen}, I., {Quirrenbach}, A., \& {Mitchell},
  D.~S. 2007, \aap, 474, 565

\bibitem[{{Albrecht} {et~al.}(2013){Albrecht}, {Setiawan}, {Torres},
  {Fabrycky}, \& {Winn}}]{2013ApJ...767...32A}
{Albrecht}, S., {Setiawan}, J., {Torres}, G., {Fabrycky}, D.~C., \& {Winn},
  J.~N. 2013, \apj, 767, 32

\bibitem[{{Albrecht} {et~al.}(2011){Albrecht}, {Winn}, {Carter}, {Snellen}, \&
  {de Mooij}}]{2011ApJ...726...68A}
{Albrecht}, S., {Winn}, J.~N., {Carter}, J.~A., {Snellen}, I. A.~G., \& {de
  Mooij}, E. J.~W. 2011, \apj, 726, 68

\bibitem[{{Anderson} \& {Francis}(2012)}]{2012AstL...38..331A}
{Anderson}, E. \& {Francis}, C. 2012, Astronomy Letters, 38, 331

\bibitem[{{Anderson} {et~al.}(2016){Anderson}, {Saio}, {Ekstr{\"o}m}, {Georgy},
  \& {Meynet}}]{2016A&A...591A...8A}
{Anderson}, R.~I., {Saio}, H., {Ekstr{\"o}m}, S., {Georgy}, C., \& {Meynet}, G.
  2016, \aap, 591, A8

\bibitem[{{Andrews} \& {Breger}(1966)}]{1966Obs....86..108A}
{Andrews}, P.~J. \& {Breger}, M. 1966, The Observatory, 86, 108

\bibitem[{{Arcos} {et~al.}(2017){Arcos}, {Jones}, {Sigut}, {Kanaan}, \&
  {Cur{\'e}}}]{arcos2017}
{Arcos}, C., {Jones}, C.~E., {Sigut}, T.~A.~A., {Kanaan}, S., \& {Cur{\'e}}, M.
  2017, \apj, 842, 48

\bibitem[{{Astropy Collaboration} {et~al.}(2018){Astropy Collaboration},
  {Price-Whelan}, {Sip{\H o}cz}, {G{\"u}nther}, {Lim}, {Crawford}, {Conseil},
  {Shupe}, {Craig}, {Dencheva}, {Ginsburg}, {VanderPlas}, {Bradley},
  {P{\'e}rez-Su{\'a}rez}, {de Val-Borro}, {Aldcroft}, {Cruz}, {Robitaille},
  {Tollerud}, {Ardelean}, {Babej}, {Bach}, {Bachetti}, {Bakanov}, {Bamford},
  {Barentsen}, {Barmby}, {Baumbach}, {Berry}, {Biscani}, {Boquien}, {Bostroem},
  {Bouma}, {Brammer}, {Bray}, {Breytenbach}, {Buddelmeijer}, {Burke},
  {Calderone}, {Cano Rodr{\'{\i}}guez}, {Cara}, {Cardoso}, {Cheedella},
  {Copin}, {Corrales}, {Crichton}, {D'Avella}, {Deil}, {Depagne}, {Dietrich},
  {Donath}, {Droettboom}, {Earl}, {Erben}, {Fabbro}, {Ferreira}, {Finethy},
  {Fox}, {Garrison}, {Gibbons}, {Goldstein}, {Gommers}, {Greco}, {Greenfield},
  {Groener}, {Grollier}, {Hagen}, {Hirst}, {Homeier}, {Horton}, {Hosseinzadeh},
  {Hu}, {Hunkeler}, {Ivezi{\'c}}, {Jain}, {Jenness}, {Kanarek}, {Kendrew},
  {Kern}, {Kerzendorf}, {Khvalko}, {King}, {Kirkby}, {Kulkarni}, {Kumar},
  {Lee}, {Lenz}, {Littlefair}, {Ma}, {Macleod}, {Mastropietro}, {McCully},
  {Montagnac}, {Morris}, {Mueller}, {Mumford}, {Muna}, {Murphy}, {Nelson},
  {Nguyen}, {Ninan}, {N{\"o}the}, {Ogaz}, {Oh}, {Parejko}, {Parley}, {Pascual},
  {Patil}, {Patil}, {Plunkett}, {Prochaska}, {Rastogi}, {Reddy Janga},
  {Sabater}, {Sakurikar}, {Seifert}, {Sherbert}, {Sherwood-Taylor}, {Shih},
  {Sick}, {Silbiger}, {Singanamalla}, {Singer}, {Sladen}, {Sooley},
  {Sornarajah}, {Streicher}, {Teuben}, {Thomas}, {Tremblay}, {Turner},
  {Terr{\'o}n}, {van Kerkwijk}, {de la Vega}, {Watkins}, {Weaver}, {Whitmore},
  {Woillez}, {Zabalza}, \& {Astropy Contributors}}]{2018AJ....156..123A}
{Astropy Collaboration}, {Price-Whelan}, A.~M., {Sip{\H o}cz}, B.~M., {et~al.}
  2018, \aj, 156, 123

\bibitem[{{Astropy Collaboration} {et~al.}(2013){Astropy Collaboration},
  {Robitaille}, {Tollerud}, {Greenfield}, {Droettboom}, {Bray}, {Aldcroft},
  {Davis}, {Ginsburg}, {Price-Whelan}, {Kerzendorf}, {Conley}, {Crighton},
  {Barbary}, {Muna}, {Ferguson}, {Grollier}, {Parikh}, {Nair}, {Unther},
  {Deil}, {Woillez}, {Conseil}, {Kramer}, {Turner}, {Singer}, {Fox}, {Weaver},
  {Zabalza}, {Edwards}, {Azalee Bostroem}, {Burke}, {Casey}, {Crawford},
  {Dencheva}, {Ely}, {Jenness}, {Labrie}, {Lim}, {Pierfederici}, {Pontzen},
  {Ptak}, {Refsdal}, {Servillat}, \& {Streicher}}]{2013A&A...558A..33A}
{Astropy Collaboration}, {Robitaille}, T.~P., {Tollerud}, E.~J., {et~al.} 2013,
  \aap, 558, A33

\bibitem[{{Banyard} {et~al.}(2022){Banyard}, {Sana}, {Mahy}, {Bodensteiner},
  {Villase{\~n}or}, \& {Evans}}]{2022A&A...658A..69B}
{Banyard}, G., {Sana}, H., {Mahy}, L., {et~al.} 2022, \aap, 658, A69

\bibitem[{{Barraza} {et~al.}(2022){Barraza}, {Gomes}, {Messias}, {Le{\~a}o},
  {Almeida}, {Janot-Pacheco}, {Brito}, {Brito}, {Santana}, {Gon{\c{c}}alves},
  {das Chagas}, {Teixeira}, {De Medeiros}, \& {Canto
  Martins}}]{2022ApJ...924..117B}
{Barraza}, L.~F., {Gomes}, R.~L., {Messias}, Y.~S., {et~al.} 2022, \apj, 924,
  117

\bibitem[{{Beichman} {et~al.}(1988){Beichman}, {Neugebauer}, {Habing}, {Clegg},
  \& {Chester}}]{1988iras....1.....B}
{Beichman}, C.~A., {Neugebauer}, G., {Habing}, H.~J., {Clegg}, P.~E., \&
  {Chester}, T.~J., eds. 1988, {Infrared astronomical satellite (IRAS) catalogs
  and atlases. Volume 1: Explanatory supplement}, Vol.~1

\bibitem[{{Bell} {et~al.}(2015){Bell}, {Mamajek}, \&
  {Naylor}}]{2015MNRAS.454..593B}
{Bell}, C. P.~M., {Mamajek}, E.~E., \& {Naylor}, T. 2015, \mnras, 454, 593

\bibitem[{{Berger} {et~al.}(2010){Berger}, {Zins}, {Lazareff}, {Lebouquin},
  {Jocou}, {Kern}, {Millan-Gabet}, {Traub}, {Haguenauer}, {Absil}, {Augereau},
  {Benisty}, {Blind}, {Bonfils}, {Delboulbe}, {Feautrier}, {Germain},
  {Gillier}, {Gitton}, {Kiekebusch}, {Knudstrup}, {Lizon}, {Magnard}, {Malbet},
  {Maurel}, {Menard}, {Micallef}, {Michaud}, {Morel}, {Moulin}, {Popovic},
  {Perraut}, {Rabou}, {Rochat}, {Roussel}, {Roux}, {Stadler}, \&
  {Tatulli}}]{2010SPIE.7734E..99B}
{Berger}, J.-P., {Zins}, G., {Lazareff}, B., {et~al.} 2010, in Society of
  Photo-Optical Instrumentation Engineers (SPIE) Conference Series, Vol. 7734,
  Society of Photo-Optical Instrumentation Engineers (SPIE) Conference Series

\bibitem[{{Beuzit} {et~al.}(2019){Beuzit}, {Vigan}, {Mouillet}, {Dohlen},
  {Gratton}, {Boccaletti}, {Sauvage}, {Schmid}, {Langlois}, {Petit},
  {Baruffolo}, {Feldt}, {Milli}, {Wahhaj}, {Abe}, {Anselmi}, {Antichi},
  {Barette}, {Baudrand}, {Baudoz}, {Bazzon}, {Bernardi}, {Blanchard}, {Brast},
  {Bruno}, {Buey}, {Carbillet}, {Carle}, {Cascone}, {Chapron}, {Charton},
  {Chauvin}, {Claudi}, {Costille}, {De Caprio}, {de Boer}, {Delboulb{\'e}},
  {Desidera}, {Dominik}, {Downing}, {Dupuis}, {Fabron}, {Fantinel}, {Farisato},
  {Feautrier}, {Fedrigo}, {Fusco}, {Gigan}, {Ginski}, {Girard}, {Giro},
  {Gisler}, {Gluck}, {Gry}, {Henning}, {Hubin}, {Hugot}, {Incorvaia}, {Jaquet},
  {Kasper}, {Lagadec}, {Lagrange}, {Le Coroller}, {Le Mignant}, {Le Ruyet},
  {Lessio}, {Lizon}, {Llored}, {Lundin}, {Madec}, {Magnard}, {Marteaud},
  {Martinez}, {Maurel}, {M{\'e}nard}, {Mesa}, {M{\"o}ller-Nilsson}, {Moulin},
  {Moutou}, {Orign{\'e}}, {Parisot}, {Pavlov}, {Perret}, {Pragt}, {Puget},
  {Rabou}, {Ramos}, {Reess}, {Rigal}, {Rochat}, {Roelfsema}, {Rousset}, {Roux},
  {Saisse}, {Salasnich}, {Santambrogio}, {Scuderi}, {Segransan}, {Sevin},
  {Siebenmorgen}, {Soenke}, {Stadler}, {Suarez}, {Tiph{\`e}ne}, {Turatto},
  {Udry}, {Vakili}, {Waters}, {Weber}, {Wildi}, {Zins}, \&
  {Zurlo}}]{2019A&A...631A.155B}
{Beuzit}, J.~L., {Vigan}, A., {Mouillet}, D., {et~al.} 2019, \aap, 631, A155

\bibitem[{{Bodensteiner} {et~al.}(2020){Bodensteiner}, {Shenar}, \&
  {Sana}}]{2020A&A...641A..42B}
{Bodensteiner}, J., {Shenar}, T., \& {Sana}, H. 2020, \aap, 641, A42

\bibitem[{{Borgniet} {et~al.}(2019){Borgniet}, {Kervella}, {Nardetto},
  {Gallenne}, {M{\'e}rand}, {Anderson}, {Aufdenberg}, {Breuval}, {Gieren},
  {Hocd{\'e}}, {Javanmardi}, {Lagadec}, {Pietrzy{\'n}ski}, \&
  {Trahin}}]{borgniet2019}
{Borgniet}, S., {Kervella}, P., {Nardetto}, N., {et~al.} 2019, \aap, 631, A37

\bibitem[{{Carciofi} {et~al.}(2008){Carciofi}, {Domiciano de Souza},
  {Magalh{\~a}es}, {Bjorkman}, \& {Vakili}}]{2008ApJ...676L..41C}
{Carciofi}, A.~C., {Domiciano de Souza}, A., {Magalh{\~a}es}, A.~M.,
  {Bjorkman}, J.~E., \& {Vakili}, F. 2008, \apjl, 676, L41

\bibitem[{{Carciofi} {et~al.}(2007){Carciofi}, {Magalh{\~a}es}, {Leister},
  {Bjorkman}, \& {Levenhagen}}]{carciofi2007}
{Carciofi}, A.~C., {Magalh{\~a}es}, A.~M., {Leister}, N.~V., {Bjorkman}, J.~E.,
  \& {Levenhagen}, R.~S. 2007, \apjl, 671, L49

\bibitem[{{Castelli} \& {Kurucz}(2003)}]{2003IAUS..210P.A20C}
{Castelli}, F. \& {Kurucz}, R.~L. 2003, in Modelling of Stellar Atmospheres,
  ed. N.~{Piskunov}, W.~W. {Weiss}, \& D.~F. {Gray}, Vol. 210, A20

\bibitem[{{Chelli} {et~al.}(2009){Chelli}, {Utrera}, \&
  {Duvert}}]{Chelli2009_v502p705}
{Chelli}, A., {Utrera}, O.~H., \& {Duvert}, G. 2009, \aap, 502, 705

\bibitem[{{Cutri} {et~al.}(2012){Cutri}, {Wright}, {Conrow}, {Bauer},
  {Benford}, {Brandenburg}, {Dailey}, {Eisenhardt}, {Evans}, {Fajardo-Acosta},
  {Fowler}, {Gelino}, {Grillmair}, {Harbut}, {Hoffman}, {Jarrett},
  {Kirkpatrick}, {Leisawitz}, {Liu}, {Mainzer}, {Marsh}, {Masci}, {McCallon},
  {Padgett}, {Ressler}, {Royer}, {Skrutskie}, {Stanford}, {Wyatt}, {Tholen},
  {Tsai}, {Wachter}, {Wheelock}, {Yan}, {Alles}, {Beck}, {Grav}, {Masiero},
  {McCollum}, {McGehee}, {Papin}, \& {Wittman}}]{2012yCat.2311....0C}
{Cutri}, R.~M., {Wright}, E.~L., {Conrow}, T., {et~al.} 2012, VizieR Online
  Data Catalog, 2311

\bibitem[{{Dalla Vedova} {et~al.}(2017){Dalla Vedova}, {Millour}, {Domiciano de
  Souza}, {Petrov}, {Moser Faes}, {Carciofi}, {Kervella}, \&
  {Rivinius}}]{vedova2017}
{Dalla Vedova}, G., {Millour}, F., {Domiciano de Souza}, A., {et~al.} 2017,
  Astronomy and Astrophysics, 601, A118

\bibitem[{{Dekker} {et~al.}(2000){Dekker}, {D'Odorico}, {Kaufer}, {Delabre}, \&
  {Kotzlowski}}]{dekker2000}
{Dekker}, H., {D'Odorico}, S., {Kaufer}, A., {Delabre}, B., \& {Kotzlowski}, H.
  2000, in \procspie, Vol. 4008, Optical and IR Telescope Instrumentation and
  Detectors, ed. M.~{Iye} \& A.~F. {Moorwood}, 534--545

\bibitem[{{Domiciano de Souza} {et~al.}(2012){Domiciano de Souza}, {Hadjara},
  {Vakili}, {Bendjoya}, {Millour}, {Abe}, {Carciofi}, {Faes}, {Kervella},
  {Lagarde}, {Marconi}, {Monin}, {Niccolini}, {Petrov}, \&
  {Weigelt}}]{2012A&A...545A.130D}
{Domiciano de Souza}, A., {Hadjara}, M., {Vakili}, F., {et~al.} 2012, \aap,
  545, A130

\bibitem[{{Domiciano de Souza} {et~al.}(2003){Domiciano de Souza}, {Kervella},
  {Jankov}, {Abe}, {Vakili}, {di Folco}, \& {Paresce}}]{2003A&A...407L..47D}
{Domiciano de Souza}, A., {Kervella}, P., {Jankov}, S., {et~al.} 2003, \aap,
  407, L47

\bibitem[{{Domiciano de Souza} {et~al.}(2014){Domiciano de Souza}, {Kervella},
  {Moser Faes}, {Dalla Vedova}, {M{\'e}rand}, {Le Bouquin}, {Espinosa Lara},
  {Rieutord}, {Bendjoya}, {Carciofi}, {Hadjara}, {Millour}, \&
  {Vakili}}]{desouza2014}
{Domiciano de Souza}, A., {Kervella}, P., {Moser Faes}, D., {et~al.} 2014,
  Astronomy and Astrophysics, 569, A10

\bibitem[{{Eggleton}(1983)}]{1983ApJ...268..368E}
{Eggleton}, P.~P. 1983, \apj, 268, 368

\bibitem[{{Ekstr{\"o}m} {et~al.}(2012){Ekstr{\"o}m}, {Georgy}, {Eggenberger},
  {Meynet}, {Mowlavi}, {Wyttenbach}, {Granada}, {Decressin}, {Hirschi},
  {Frischknecht}, {Charbonnel}, \& {Maeder}}]{2012A&A...537A.146E}
{Ekstr{\"o}m}, S., {Georgy}, C., {Eggenberger}, P., {et~al.} 2012, \aap, 537,
  A146

\bibitem[{{El-Badry} {et~al.}(2022){El-Badry}, {Conroy}, {Quataert}, {Rix},
  {Labadie-Bartz}, {Jayasinghe}, {Thompson}, {Cargile}, {Stassun}, \&
  {Ilyin}}]{2022arXiv220105614E}
{El-Badry}, K., {Conroy}, C., {Quataert}, E., {et~al.} 2022, arXiv e-prints,
  arXiv:2201.05614

\bibitem[{{ESA}(1997)}]{1997ESASP1200.....E}
{ESA}, ed. 1997, ESA Special Publication, Vol. 1200, {The HIPPARCOS and TYCHO
  catalogues. Astrometric and photometric star catalogues derived from the ESA
  HIPPARCOS Space Astrometry Mission}

\bibitem[{{Foreman-Mackey} {et~al.}(2013){Foreman-Mackey}, {Hogg}, {Lang}, \&
  {Goodman}}]{Foreman-Mackey2013_v125p306}
{Foreman-Mackey}, D., {Hogg}, D.~W., {Lang}, D., \& {Goodman}, J. 2013, \pasp,
  125, 306

\bibitem[{{Gagn{\'e}} {et~al.}(2015){Gagn{\'e}}, {Lafreni{\`e}re}, {Doyon},
  {Malo}, \& {Artigau}}]{2015ApJ...798...73G}
{Gagn{\'e}}, J., {Lafreni{\`e}re}, D., {Doyon}, R., {Malo}, L., \& {Artigau},
  {\'E}. 2015, \apj, 798, 73

\bibitem[{{Gaia Collaboration} {et~al.}(2021){Gaia Collaboration}, {Brown, A.
  G. A.}, {Vallenari, A.}, {Prusti, T.}, {de Bruijne, J. H. J.}, {Babusiaux,
  C.}, {Biermann, M.}, {Creevey, O. L.}, {Evans, D. W.}, {Eyer, L.}, {Hutton,
  A.}, {Jansen, F.}, {Jordi, C.}, {Klioner, S. A.}, {Lammers, U.}, {Lindegren,
  L.}, {Luri, X.}, {Mignard, F.}, {Panem, C.}, {Pourbaix, D.}, {Randich, S.},
  {Sartoretti, P.}, {Soubiran, C.}, {Walton, N. A.}, {Arenou, F.},
  {Bailer-Jones, C. A. L.}, {Bastian, U.}, {Cropper, M.}, {Drimmel, R.}, {Katz,
  D.}, {Lattanzi, M. G.}, {van Leeuwen, F.}, {Bakker, J.}, {Cacciari, C.},
  {Casta\~neda, J.}, {De Angeli, F.}, {Ducourant, C.}, {Fabricius, C.},
  {Fouesneau, M.}, {Fr\'emat, Y.}, {Guerra, R.}, {Guerrier, A.}, {Guiraud, J.},
  {Jean-Antoine Piccolo, A.}, {Masana, E.}, {Messineo, R.}, {Mowlavi, N.},
  {Nicolas, C.}, {Nienartowicz, K.}, {Pailler, F.}, {Panuzzo, P.}, {Riclet,
  F.}, {Roux, W.}, {Seabroke, G. M.}, {Sordo, R.}, {Tanga, P.}, {Th\'evenin,
  F.}, {Gracia-Abril, G.}, {Portell, J.}, {Teyssier, D.}, {Altmann, M.},
  {Andrae, R.}, {Bellas-Velidis, I.}, {Benson, K.}, {Berthier, J.}, {Blomme,
  R.}, {Brugaletta, E.}, {Burgess, P. W.}, {Busso, G.}, {Carry, B.}, {Cellino,
  A.}, {Cheek, N.}, {Clementini, G.}, {Damerdji, Y.}, {Davidson, M.},
  {Delchambre, L.}, {Dell\'{}Oro, A.}, {Fern\'andez-Hern\'andez, J.},
  {Galluccio, L.}, {Garc\'{\i}a-Lario, P.}, {Garcia-Reinaldos, M.},
  {Gonz\'alez-N\'u\~nez, J.}, {Gosset, E.}, {Haigron, R.}, {Halbwachs, J.-L.},
  {Hambly, N. C.}, {Harrison, D. L.}, {Hatzidimitriou, D.}, {Heiter, U.},
  {Hern\'andez, J.}, {Hestroffer, D.}, {Hodgkin, S. T.}, {Holl, B.},
  {Jan\ss{}en, K.}, {Jevardat de Fombelle, G.}, {Jordan, S.}, {Krone-Martins,
  A.}, {Lanzafame, A. C.}, {L\"offler, W.}, {Lorca, A.}, {Manteiga, M.},
  {Marchal, O.}, {Marrese, P. M.}, {Moitinho, A.}, {Mora, A.}, {Muinonen, K.},
  {Osborne, P.}, {Pancino, E.}, {Pauwels, T.}, {Petit, J.-M.}, {Recio-Blanco,
  A.}, {Richards, P. J.}, {Riello, M.}, {Rimoldini, L.}, {Robin, A. C.},
  {Roegiers, T.}, {Rybizki, J.}, {Sarro, L. M.}, {Siopis, C.}, {Smith, M.},
  {Sozzetti, A.}, {Ulla, A.}, {Utrilla, E.}, {van Leeuwen, M.}, {van Reeven,
  W.}, {Abbas, U.}, {Abreu Aramburu, A.}, {Accart, S.}, {Aerts, C.}, {Aguado,
  J. J.}, {Ajaj, M.}, {Altavilla, G.}, {\'Alvarez, M. A.}, {\'Alvarez
  Cid-Fuentes, J.}, {Alves, J.}, {Anderson, R. I.}, {Anglada Varela, E.},
  {Antoja, T.}, {Audard, M.}, {Baines, D.}, {Baker, S. G.},
  {Balaguer-N\'u\~nez, L.}, {Balbinot, E.}, {Balog, Z.}, {Barache, C.},
  {Barbato, D.}, {Barros, M.}, {Barstow, M. A.}, {Bartolom\'e, S.}, {Bassilana,
  J.-L.}, {Bauchet, N.}, {Baudesson-Stella, A.}, {Becciani, U.}, {Bellazzini,
  M.}, {Bernet, M.}, {Bertone, S.}, {Bianchi, L.}, {Blanco-Cuaresma, S.},
  {Boch, T.}, {Bombrun, A.}, {Bossini, D.}, {Bouquillon, S.}, {Bragaglia, A.},
  {Bramante, L.}, {Breedt, E.}, {Bressan, A.}, {Brouillet, N.}, {Bucciarelli,
  B.}, {Burlacu, A.}, {Busonero, D.}, {Butkevich, A. G.}, {Buzzi, R.}, {Caffau,
  E.}, {Cancelliere, R.}, {C\'anovas, H.}, {Cantat-Gaudin, T.}, {Carballo, R.},
  {Carlucci, T.}, {Carnerero, M. I}, {Carrasco, J. M.}, {Casamiquela, L.},
  {Castellani, M.}, {Castro-Ginard, A.}, {Castro Sampol, P.}, {Chaoul, L.},
  {Charlot, P.}, {Chemin, L.}, {Chiavassa, A.}, {Cioni, M.-R. L.}, {Comoretto,
  G.}, {Cooper, W. J.}, {Cornez, T.}, {Cowell, S.}, {Crifo, F.}, {Crosta, M.},
  {Crowley, C.}, {Dafonte, C.}, {Dapergolas, A.}, {David, M.}, {David, P.}, {de
  Laverny, P.}, {De Luise, F.}, {De March, R.}, {De Ridder, J.}, {de Souza,
  R.}, {de Teodoro, P.}, {de Torres, A.}, {del Peloso, E. F.}, {del Pozo, E.},
  {Delbo, M.}, {Delgado, A.}, {Delgado, H. E.}, {Delisle, J.-B.}, {Di Matteo,
  P.}, {Diakite, S.}, {Diener, C.}, {Distefano, E.}, {Dolding, C.}, {Eappachen,
  D.}, {Edvardsson, B.}, {Enke, H.}, {Esquej, P.}, {Fabre, C.}, {Fabrizio, M.},
  {Faigler, S.}, {Fedorets, G.}, {Fernique, P.}, {Fienga, A.}, {Figueras, F.},
  {Fouron, C.}, {Fragkoudi, F.}, {Fraile, E.}, {Franke, F.}, {Gai, M.},
  {Garabato, D.}, {Garcia-Gutierrez, A.}, {Garc\'{\i}a-Torres, M.}, {Garofalo,
  A.}, {Gavras, P.}, {Gerlach, E.}, {Geyer, R.}, {Giacobbe, P.}, {Gilmore, G.},
  {Girona, S.}, {Giuffrida, G.}, {Gomel, R.}, {Gomez, A.},
  {Gonzalez-Santamaria, I.}, {Gonz\'alez-Vidal, J. J.}, {Granvik, M.},
  {Guti\'errez-S\'anchez, R.}, {Guy, L. P.}, {Hauser, M.}, {Haywood, M.},
  {Helmi, A.}, {Hidalgo, S. L.}, {Hilger, T.}, {Hladczuk, N.}, {Hobbs, D.},
  {Holland, G.}, {Huckle, H. E.}, {Jasniewicz, G.}, {Jonker, P. G.}, {Juaristi
  Campillo, J.}, {Julbe, F.}, {Karbevska, L.}, {Kervella, P.}, {Khanna, S.},
  {Kochoska, A.}, {Kontizas, M.}, {Kordopatis, G.}, {Korn, A. J.},
  {Kostrzewa-Rutkowska, Z.}, {Kruszy\'{}nska, K.}, {Lambert, S.}, {Lanza, A.
  F.}, {Lasne, Y.}, {Le Campion, J.-F.}, {Le Fustec, Y.}, {Lebreton, Y.},
  {Lebzelter, T.}, {Leccia, S.}, {Leclerc, N.}, {Lecoeur-Taibi, I.}, {Liao,
  S.}, {Licata, E.}, {Lindstr\o{}m, E. P.}, {Lister, T. A.}, {Livanou, E.},
  {Lobel, A.}, {Madrero Pardo, P.}, {Managau, S.}, {Mann, R. G.}, {Marchant, J.
  M.}, {Marconi, M.}, {Marcos Santos, M. M. S.}, {Marinoni, S.}, {Marocco, F.},
  {Marshall, D. J.}, {Martin Polo, L.}, {Mart\'{\i}n-Fleitas, J. M.}, {Masip,
  A.}, {Massari, D.}, {Mastrobuono-Battisti, A.}, {Mazeh, T.}, {McMillan, P.
  J.}, {Messina, S.}, {Michalik, D.}, {Millar, N. R.}, {Mints, A.}, {Molina,
  D.}, {Molinaro, R.}, {Moln\'ar, L.}, {Montegriffo, P.}, {Mor, R.},
  {Morbidelli, R.}, {Morel, T.}, {Morris, D.}, {Mulone, A. F.}, {Munoz, D.},
  {Muraveva, T.}, {Murphy, C. P.}, {Musella, I.}, {Noval, L.}, {Ord\'enovic,
  C.}, {Orr\`u, G.}, {Osinde, J.}, {Pagani, C.}, {Pagano, I.}, {Palaversa, L.},
  {Palicio, P. A.}, {Panahi, A.}, {Pawlak, M.}, {Pe\~nalosa Esteller, X.},
  {Penttil\"a, A.}, {Piersimoni, A. M.}, {Pineau, F.-X.}, {Plachy, E.}, {Plum,
  G.}, {Poggio, E.}, {Poretti, E.}, {Poujoulet, E.}, {Prsa, A.}, {Pulone, L.},
  {Racero, E.}, {Ragaini, S.}, {Rainer, M.}, {Raiteri, C. M.}, {Rambaux, N.},
  {Ramos, P.}, {Ramos-Lerate, M.}, {Re Fiorentin, P.}, {Regibo, S.}, {Reyl\'e,
  C.}, {Ripepi, V.}, {Riva, A.}, {Rixon, G.}, {Robichon, N.}, {Robin, C.},
  {Roelens, M.}, {Rohrbasser, L.}, {Romero-G\'omez, M.}, {Rowell, N.}, {Royer,
  F.}, {Rybicki, K. A.}, {Sadowski, G.}, {Sagrist\`a Sell\'es, A.}, {Sahlmann,
  J.}, {Salgado, J.}, {Salguero, E.}, {Samaras, N.}, {Sanchez Gimenez, V.},
  {Sanna, N.}, {Santove\~na, R.}, {Sarasso, M.}, {Schultheis, M.}, {Sciacca,
  E.}, {Segol, M.}, {Segovia, J. C.}, {S\'egransan, D.}, {Semeux, D.}, {Shahaf,
  S.}, {Siddiqui, H. I.}, {Siebert, A.}, {Siltala, L.}, {Slezak, E.}, {Smart,
  R. L.}, {Solano, E.}, {Solitro, F.}, {Souami, D.}, {Souchay, J.}, {Spagna,
  A.}, {Spoto, F.}, {Steele, I. A.}, {Steidelm\"uller, H.}, {Stephenson, C.
  A.}, {S\"uveges, M.}, {Szabados, L.}, {Szegedi-Elek, E.}, {Taris, F.},
  {Tauran, G.}, {Taylor, M. B.}, {Teixeira, R.}, {Thuillot, W.}, {Tonello, N.},
  {Torra, F.}, {Torra, J.}, {Turon, C.}, {Unger, N.}, {Vaillant, M.}, {van
  Dillen, E.}, {Vanel, O.}, {Vecchiato, A.}, {Viala, Y.}, {Vicente, D.},
  {Voutsinas, S.}, {Weiler, M.}, {Wevers, T.}, {Wyrzykowski, L.}, {Yoldas, A.},
  {Yvard, P.}, {Zhao, H.}, {Zorec, J.}, {Zucker, S.}, {Zurbach, C.}, \&
  {Zwitter, T.}}]{GaiaEDR3content}
{Gaia Collaboration}, {Brown, A. G. A.}, {Vallenari, A.}, {et~al.} 2021, A\&A,
  649, A1

\bibitem[{{Gaia Collaboration} {et~al.}(2022){Gaia Collaboration}, {Vallenari},
  {Brown}, {Prusti}, {de Bruijne}, {Arenou}, {Babusiaux}, {Biermann},
  {Creevey}, {Ducourant}, {Evans}, {Eyer}, {Guerra}, {Hutton}, {Jordi},
  {Klioner}, {Lammers}, {Lindegren}, {Luri}, {Mignard}, {Panem}, {Pourbaix},
  {Randich}, {Sartoretti}, {Soubiran}, {Tanga}, {Walton}, {Bailer-Jones},
  {Bastian}, {Drimmel}, {Jansen}, {Katz}, {Lattanzi}, {van Leeuwen}, {Bakker},
  {Cacciari}, {Casta{\~n}eda}, {De Angeli}, {Fabricius}, {Fouesneau},
  {Fr{\'e}mat}, {Galluccio}, {Guerrier}, {Heiter}, {Masana}, {Messineo},
  {Mowlavi}, {Nicolas}, {Nienartowicz}, {Pailler}, {Panuzzo}, {Riclet}, {Roux},
  {Seabroke}, {Sordo{\o}rcit}, {Th{\'e}venin}, {Gracia-Abril}, {Portell},
  {Teyssier}, {Altmann}, {Andrae}, {Audard}, {Bellas-Velidis}, {Benson},
  {Berthier}, {Blomme}, {Burgess}, {Busonero}, {Busso}, {C{\'a}novas}, {Carry},
  {Cellino}, {Cheek}, {Clementini}, {Damerdji}, {Davidson}, {de Teodoro},
  {Nu{\~n}ez Campos}, {Delchambre}, {Dell'Oro}, {Esquej},
  {Fern{\'a}ndez-Hern{\'a}ndez}, {Fraile}, {Garabato}, {Garc{\'\i}a-Lario},
  {Gosset}, {Haigron}, {Halbwachs}, {Hambly}, {Harrison}, {Hern{\'a}ndez},
  {Hestroffer}, {Hodgkin}, {Holl}, {Jan{\ss}en}, {Jevardat de Fombelle},
  {Jordan}, {Krone-Martins}, {Lanzafame}, {L{\"o}ffler}, {Marchal}, {Marrese},
  {Moitinho}, {Muinonen}, {Osborne}, {Pancino}, {Pauwels}, {Recio-Blanco},
  {Reyl{\'e}}, {Riello}, {Rimoldini}, {Roegiers}, {Rybizki}, {Sarro}, {Siopis},
  {Smith}, {Sozzetti}, {Utrilla}, {van Leeuwen}, {Abbas}, {{\'A}brah{\'a}m},
  {Abreu Aramburu}, {Aerts}, {Aguado}, {Ajaj}, {Aldea-Montero}, {Altavilla},
  {{\'A}lvarez}, {Alves}, {Anders}, {Anderson}, {Anglada Varela}, {Antoja},
  {Baines}, {Baker}, {Balaguer-N{\'u}{\~n}ez}, {Balbinot}, {Balog}, {Barache},
  {Barbato}, {Barros}, {Barstow}, {Bartolom{\'e}}, {Bassilana}, {Bauchet},
  {Becciani}, {Bellazzini}, {Berihuete}, {Bernet}, {Bertone}, {Bianchi},
  {Binnenfeld}, {Blanco-Cuaresma}, {Blazere}, {Boch}, {Bombrun}, {Bossini},
  {Bouquillon}, {Bragaglia}, {Bramante}, {Breedt}, {Bressan}, {Brouillet},
  {Brugaletta}, {Bucciarelli}, {Burlacu}, {Butkevich}, {Buzzi}, {Caffau},
  {Cancelliere}, {Cantat-Gaudin}, {Carballo}, {Carlucci}, {Carnerero},
  {Carrasco}, {Casamiquela}, {Castellani}, {Castro-Ginard}, {Chaoul},
  {Charlot}, {Chemin}, {Chiaramida}, {Chiavassa}, {Chornay}, {Comoretto},
  {Contursi}, {Cooper}, {Cornez}, {Cowell}, {Crifo}, {Cropper}, {Crosta},
  {Crowley}, {Dafonte}, {Dapergolas}, {David}, {David}, {de Laverny}, {De
  Luise}, {De March}, {De Ridder}, {de Souza}, {de Torres}, {del Peloso}, {del
  Pozo}, {Delbo}, {Delgado}, {Delisle}, {Demouchy}, {Dharmawardena}, {Di
  Matteo}, {Diakite}, {Diener}, {Distefano}, {Dolding}, {Edvardsson}, {Enke},
  {Fabre}, {Fabrizio}, {Faigler}, {Fedorets}, {Fernique}, {Fienga}, {Figueras},
  {Fournier}, {Fouron}, {Fragkoudi}, {Gai}, {Garcia-Gutierrez},
  {Garcia-Reinaldos}, {Garc{\'\i}a-Torres}, {Garofalo}, {Gavel}, {Gavras},
  {Gerlach}, {Geyer}, {Giacobbe}, {Gilmore}, {Girona}, {Giuffrida}, {Gomel},
  {Gomez}, {Gonz{\'a}lez-N{\'u}{\~n}ez}, {Gonz{\'a}lez-Santamar{\'\i}a},
  {Gonz{\'a}lez-Vidal}, {Granvik}, {Guillout}, {Guiraud},
  {Guti{\'e}rrez-S{\'a}nchez}, {Guy}, {Hatzidimitriou}, {Hauser}, {Haywood},
  {Helmer}, {Helmi}, {Sarmiento}, {Hidalgo}, {Hilger}, {H{\l}adczuk}, {Hobbs},
  {Holland}, {Huckle}, {Jardine}, {Jasniewicz}, {Jean-Antoine Piccolo},
  {Jim{\'e}nez-Arranz}, {Jorissen}, {Juaristi Campillo}, {Julbe}, {Karbevska},
  {Kervella}, {Khanna}, {Kontizas}, {Kordopatis}, {Korn}, {K{\'o}sp{\'a}l},
  {Kostrzewa-Rutkowska}, {Kruszy{\'n}ska}, {Kun}, {Laizeau}, {Lambert},
  {Lanza}, {Lasne}, {Le Campion}, {Lebreton}, {Lebzelter}, {Leccia}, {Leclerc},
  {Lecoeur-Taibi}, {Liao}, {Licata}, {Lindstr{\o}m}, {Lister}, {Livanou},
  {Lobel}, {Lorca}, {Loup}, {Madrero Pardo}, {Magdaleno Romeo}, {Managau},
  {Mann}, {Manteiga}, {Marchant}, {Marconi}, {Marcos}, {Marcos Santos},
  {Mar{\'\i}n Pina}, {Marinoni}, {Marocco}, {Marshall}, {Polo},
  {Mart{\'\i}n-Fleitas}, {Marton}, {Mary}, {Masip}, {Massari},
  {Mastrobuono-Battisti}, {Mazeh}, {McMillan}, {Messina}, {Michalik}, {Millar},
  {Mints}, {Molina}, {Molinaro}, {Moln{\'a}r}, {Monari}, {Mongui{\'o}},
  {Montegriffo}, {Montero}, {Mor}, {Mora}, {Morbidelli}, {Morel}, {Morris},
  {Muraveva}, {Murphy}, {Musella}, {Nagy}, {Noval}, {Oca{\~n}a}, {Ogden},
  {Ordenovic}, {Osinde}, {Pagani}, {Pagano}, {Palaversa}, {Palicio},
  {Pallas-Quintela}, {Panahi}, {Payne-Wardenaar}, {Pe{\~n}alosa Esteller},
  {Penttil{\"a}}, {Pichon}, {Piersimoni}, {Pineau}, {Plachy}, {Plum}, {Poggio},
  {Pr{\v{s}}a}, {Pulone}, {Racero}, {Ragaini}, {Rainer}, {Raiteri}, {Rambaux},
  {Ramos}, {Ramos-Lerate}, {Re Fiorentin}, {Regibo}, {Richards}, {Rios Diaz},
  {Ripepi}, {Riva}, {Rix}, {Rixon}, {Robichon}, {Robin}, {Robin}, {Roelens},
  {Rogues}, {Rohrbasser}, {Romero-G{\'o}mez}, {Rowell}, {Royer}, {Ruz Mieres},
  {Rybicki}, {Sadowski}, {S{\'a}ez N{\'u}{\~n}ez}, {Sagrist{\`a} Sell{\'e}s},
  {Sahlmann}, {Salguero}, {Samaras}, {Sanchez Gimenez}, {Sanna},
  {Santove{\~n}a}, {Sarasso}, {Schultheis}, {Sciacca}, {Segol}, {Segovia},
  {S{\'e}gransan}, {Semeux}, {Shahaf}, {Siddiqui}, {Siebert}, {Siltala},
  {Silvelo}, {Slezak}, {Slezak}, {Smart}, {Snaith}, {Solano}, {Solitro},
  {Souami}, {Souchay}, {Spagna}, {Spina}, {Spoto}, {Steele},
  {Steidelm{\"u}ller}, {Stephenson}, {S{\"u}veges}, {Surdej}, {Szabados},
  {Szegedi-Elek}, {Taris}, {Taylo}, {Teixeira}, {Tolomei}, {Tonello}, {Torra},
  {Torra}, {Torralba Elipe}, {Trabucchi}, {Tsounis}, {Turon}, {Ulla}, {Unger},
  {Vaillant}, {van Dillen}, {van Reeven}, {Vanel}, {Vecchiato}, {Viala},
  {Vicente}, {Voutsinas}, {Weiler}, {Wevers}, {Wyrzykowski}, {Yoldas}, {Yvard},
  {Zhao}, {Zorec}, {Zucker}, \& {Zwitter}}]{2022arXiv220800211G}
{Gaia Collaboration}, {Vallenari}, A., {Brown}, A.~G.~A., {et~al.} 2022, arXiv
  e-prints, arXiv:2208.00211

\bibitem[{{Gallenne} {et~al.}(2019){Gallenne}, {Kervella}, {Borgniet},
  {M{\'e}rand}, {Pietrzy{\'n}ski}, {Gieren}, {Monnier}, {Schaefer}, {Evans},
  {Anderson}, {Baron}, {Roettenbacher}, \& {Karczmarek}}]{2019A&A...622A.164G}
{Gallenne}, A., {Kervella}, P., {Borgniet}, S., {et~al.} 2019, \aap, 622, A164

\bibitem[{{Gallenne} {et~al.}(2014){Gallenne}, {M{\'e}rand}, {Kervella},
  {Breitfelder}, {Le Bouquin}, {Monnier}, {Gieren}, {Pilecki}, \&
  {Pietrzy{\'n}ski}}]{2014A&A...561L...3G}
{Gallenne}, A., {M{\'e}rand}, A., {Kervella}, P., {et~al.} 2014, \aap, 561, L3

\bibitem[{{Gallenne} {et~al.}(2015){Gallenne}, {M{\'e}rand}, {Kervella},
  {Monnier}, {Schaefer}, {Baron}, {Breitfelder}, {Le Bouquin}, {Roettenbacher},
  {Gieren}, {Pietrzy{\'n}ski}, {McAlister}, {ten Brummelaar}, {Sturmann},
  {Sturmann}, {Turner}, {Ridgway}, \& {Kraus}}]{gallenne2015}
{Gallenne}, A., {M{\'e}rand}, A., {Kervella}, P., {et~al.} 2015, \aap, 579, A68

\bibitem[{{Gallenne} {et~al.}(2013){Gallenne}, {Monnier}, {M{\'e}rand},
  {Kervella}, {Kraus}, {Schaefer}, {Gieren}, {Pietrzy{\'n}ski}, {Szabados},
  {Che}, {Baron}, {Pedretti}, {McAlister}, {ten Brummelaar}, {Sturmann},
  {Sturmann}, {Turner}, {Farrington}, \& {Vargas}}]{2013A&A...552A..21G}
{Gallenne}, A., {Monnier}, J.~D., {M{\'e}rand}, A., {et~al.} 2013, \aap, 552,
  A21

\bibitem[{{Gauchet} {et~al.}(2016){Gauchet}, {Lacour}, {Lagrange},
  {Ehrenreich}, {Bonnefoy}, {Girard}, \& {Boccaletti}}]{2016A&A...595A..31G}
{Gauchet}, L., {Lacour}, S., {Lagrange}, A.~M., {et~al.} 2016, \aap, 595, A31

\bibitem[{{Georgy} {et~al.}(2013){Georgy}, {Ekstr{\"o}m}, {Granada}, {Meynet},
  {Mowlavi}, {Eggenberger}, \& {Maeder}}]{2013A&A...553A..24G}
{Georgy}, C., {Ekstr{\"o}m}, S., {Granada}, A., {et~al.} 2013, \aap, 553, A24

\bibitem[{{Gieren} {et~al.}(2015){Gieren}, {Pilecki}, {Pietrzy{\'n}ski},
  {Graczyk}, {Udalski}, {Soszy{\'n}ski}, {Thompson}, {Prada Moroni}, {Smolec},
  {Konorski}, {G{\'o}rski}, {Karczmarek}, {Suchomska}, {Taormina}, {Gallenne},
  {Storm}, {Bono}, {Catelan}, {Szyma{\'n}ski}, {Koz{\l}owski}, {Pietrukowicz},
  {Wyrzykowski}, {Poleski}, {Skowron}, {Minniti}, {Ulaczyk}, {Mr{\'o}z},
  {Pawlak}, \& {Nardetto}}]{2015ApJ...815...28G}
{Gieren}, W., {Pilecki}, B., {Pietrzy{\'n}ski}, G., {et~al.} 2015, \apj, 815,
  28

\bibitem[{{Ginsburg} {et~al.}(2019){Ginsburg}, {Sip{\H{o}}cz}, {Brasseur},
  {Cowperthwaite}, {Craig}, {Deil}, {Guillochon}, {Guzman}, {Liedtke}, {Lian
  Lim}, {Lockhart}, {Mommert}, {Morris}, {Norman}, {Parikh}, {Persson},
  {Robitaille}, {Segovia}, {Singer}, {Tollerud}, {de Val-Borro}, {Valtchanov},
  {Woillez}, {Astroquery Collaboration}, \& {a subset of astropy
  Collaboration}}]{2019AJ....157...98G}
{Ginsburg}, A., {Sip{\H{o}}cz}, B.~M., {Brasseur}, C.~E., {et~al.} 2019, \aj,
  157, 98

\bibitem[{{Goss} {et~al.}(2011){Goss}, {Karoff}, {Chaplin}, {Elsworth}, \&
  {Stevens}}]{2011MNRAS.411..162G}
{Goss}, K.~J.~F., {Karoff}, C., {Chaplin}, W.~J., {Elsworth}, Y., \& {Stevens},
  I.~R. 2011, \mnras, 411, 162

\bibitem[{{Gravity Collaboration} {et~al.}(2017){Gravity Collaboration},
  {Abuter}, {Accardo}, {Amorim}, {Anugu}, {{\'A}vila}, {Azouaoui}, {Benisty},
  {Berger}, {Blind}, {Bonnet}, {Bourget}, {Brandner}, {Brast}, {Buron},
  {Burtscher}, {Cassaing}, {Chapron}, {Choquet}, {Cl{\'e}net}, {Collin},
  {Coud{\'e} Du Foresto}, {de Wit}, {de Zeeuw}, {Deen},
  {Delplancke-Str{\"o}bele}, {Dembet}, {Derie}, {Dexter}, {Duvert}, {Ebert},
  {Eckart}, {Eisenhauer}, {Esselborn}, {F{\'e}dou}, {Finger}, {Garcia}, {Garcia
  Dabo}, {Garcia Lopez}, {Gendron}, {Genzel}, {Gillessen}, {Gonte}, {Gordo},
  {Grould}, {Gr{\"o}zinger}, {Guieu}, {Haguenauer}, {Hans}, {Haubois}, {Haug},
  {Haussmann}, {Henning}, {Hippler}, {Horrobin}, {Huber}, {Hubert}, {Hubin},
  {Hummel}, {Jakob}, {Janssen}, {Jochum}, {Jocou}, {Kaufer}, {Kellner},
  {Kendrew}, {Kern}, {Kervella}, {Kiekebusch}, {Klein}, {Kok}, {Kolb}, {Kulas},
  {Lacour}, {Lapeyr{\`e}re}, {Lazareff}, {Le Bouquin}, {L{\`e}na}, {Lenzen},
  {L{\'e}v{\^e}que}, {Lippa}, {Magnard}, {Mehrgan}, {Mellein}, {M{\'e}rand},
  {Moreno-Ventas}, {Moulin}, {M{\"u}ller}, {M{\"u}ller}, {Neumann}, {Oberti},
  {Ott}, {Pallanca}, {Panduro}, {Pasquini}, {Paumard}, {Percheron}, {Perraut},
  {Perrin}, {Pfl{\"u}ger}, {Pfuhl}, {Phan Duc}, {Plewa}, {Popovic}, {Rabien},
  {Ram{\'{\i}}rez}, {Ramos}, {Rau}, {Riquelme}, {Rohloff}, {Rousset},
  {Sanchez-Bermudez}, {Scheithauer}, {Sch{\"o}ller}, {Schuhler}, {Spyromilio},
  {Straubmeier}, {Sturm}, {Suarez}, {Tristram}, {Ventura}, {Vincent},
  {Waisberg}, {Wank}, {Weber}, {Wieprecht}, {Wiest}, {Wiezorrek}, {Wittkowski},
  {Woillez}, {Wolff}, {Yazici}, {Ziegler}, \& {Zins}}]{2017A&A...602A..94G}
{Gravity Collaboration}, {Abuter}, R., {Accardo}, M., {et~al.} 2017, \aap, 602,
  A94

\bibitem[{{Haguenauer} {et~al.}(2010){Haguenauer}, {Alonso}, {Bourget},
  {Brillant}, {Gitton}, {Guisard}, {Poupar}, {Schuhler}, {Abuter}, {Andolfato},
  {Blanchard}, {Berger}, {Cortes}, {D{\'e}rie}, {Delplancke}, {di Lieto},
  {Dupuy}, {Gilli}, {Glindemann}, {Guniat}, {Huedepohl}, {Kaufer}, {Le
  Bouquin}, {L{\'e}v{\^e}que}, {M{\'e}nardi}, {M{\'e}rand}, {Morel},
  {Percheron}, {Phan Duc}, {Pino}, {Ramirez}, {Rengaswamy}, {Richichi},
  {Rivinius}, {Sahlmann}, {Schoeller}, {Schmid}, {Stefl}, {Valdes}, {van
  Belle}, {Wehner}, \& {Wittkowski}}]{Haguenauer2010_v7734p1}
{Haguenauer}, P., {Alonso}, J., {Bourget}, P., {et~al.} 2010, in Society of
  Photo-Optical Instrumentation Engineers (SPIE) Conference Series, Vol. 7734,
  Society of Photo-Optical Instrumentation Engineers (SPIE) Conference Series

\bibitem[{Harris {et~al.}(2020)Harris, Millman, van~der Walt, Gommers,
  Virtanen, Cournapeau, Wieser, Taylor, Berg, Smith, Kern, Picus, Hoyer, van
  Kerkwijk, Brett, Haldane, del R{\'\i}o, Wiebe, Peterson, G{\'e}rard-Marchant,
  Sheppard, Reddy, Weckesser, Abbasi, Gohlke, \& Oliphant}]{Harris20}
Harris, C.~R., Millman, K.~J., van~der Walt, S.~J., {et~al.} 2020, Nature, 585,
  357

\bibitem[{{Hastings} {et~al.}(2021){Hastings}, {Langer}, {Wang},
  {Schootemeijer}, \& {Milone}}]{2021A&A...653A.144H}
{Hastings}, B., {Langer}, N., {Wang}, C., {Schootemeijer}, A., \& {Milone},
  A.~P. 2021, \aap, 653, A144

\bibitem[{{Haubois} {et~al.}(2020){Haubois}, {Abuter}, {Aller-Carpentier},
  {Alonso}, {Beltran}, {Berger}, {Bourget}, {Bristow}, {Caniguante},
  {Chazelas}, {Cid}, {Conzelmann}, {Cortes}, {Darr{\'e}}, {Delboulb{\'e}},
  {Delplancke-Str{\"o}bele}, {Del Valle}, {Dembet}, {Donoso}, {Dupuy}, {Egner},
  {Eisenhauer}, {Faundez}, {Fuenteseca}, {Frahm}, {Gaytan}, {Gil},
  {Glindemann}, {Gont{\'e}}, {Gonzales}, {Guajardo}, {Guerlet}, {Guieu},
  {Gutierrez}, {Haguenauer}, {van der Heyden}, {Huber}, {Hubin}, {Hummel},
  {Jochum}, {Jocou}, {Kirchbauer}, {Kolb}, {Kosmalski}, {Krempl}, {Lacour}, {Le
  Bouquin}, {Leclercq}, {Lizon}, {Lopez}, {Magnard}, {Meilland}, {Meister},
  {M{\'e}rand}, {Mieske}, {Moulin}, {Osorio}, {Ott}, {Paladini}, {Pallanca},
  {Pavez}, {Pasquini}, {Pelluet}, {Percheron}, {Pettazzi}, {Pino}, {Poupar},
  {Ram{\'\i}rez}, {Reyes}, {Riquelme}, {Rivinius}, {Rochat}, {Salgado},
  {Sch{\"o}ller}, {Schuhler}, {Shchekaturov}, {Stephan}, {Suarez}, {Smette},
  {Tamblay}, {Tapia}, {Tristram}, {Valdes}, {Verinaud}, {Wittkowski},
  {Woillez}, \& {Zins}}]{2020SPIE11446E..06H}
{Haubois}, X., {Abuter}, R., {Aller-Carpentier}, E., {et~al.} 2020, in Society
  of Photo-Optical Instrumentation Engineers (SPIE) Conference Series, Vol.
  11446, Society of Photo-Optical Instrumentation Engineers (SPIE) Conference
  Series, 1144606

\bibitem[{{Heitzmann} {et~al.}(2021){Heitzmann}, {Marsden}, {Petit}, {Mengel},
  {Wright}, {Clerte}, {Millburn}, {Folsom}, {Addison}, {Wittenmyer}, \&
  {Waite}}]{heitzmann2021}
{Heitzmann}, A., {Marsden}, S.~C., {Petit}, P., {et~al.} 2021, \mnras, 505,
  4989

\bibitem[{Hunter(2007)}]{Hunter:2007}
Hunter, J.~D. 2007, Computing in Science \& Engineering, 9, 90

\bibitem[{{Hutter} {et~al.}(2021){Hutter}, {Tycner}, {Zavala}, {Benson},
  {Hummel}, \& {Zirm}}]{2021ApJS..257...69H}
{Hutter}, D.~J., {Tycner}, C., {Zavala}, R.~T., {et~al.} 2021, \apjs, 257, 69

\bibitem[{{Ishihara} {et~al.}(2010){Ishihara}, {Onaka}, {Kataza}, {Salama},
  {Alfageme}, {Cassatella}, {Cox}, {Garc{\'{\i}}a-Lario}, {Stephenson},
  {Cohen}, {Fujishiro}, {Fujiwara}, {Hasegawa}, {Ita}, {Kim}, {Matsuhara},
  {Murakami}, {M{\"u}ller}, {Nakagawa}, {Ohyama}, {Oyabu}, {Pyo}, {Sakon},
  {Shibai}, {Takita}, {Tanab{\'e}}, {Uemizu}, {Ueno}, {Usui}, {Wada},
  {Watarai}, {Yamamura}, \& {Yamauchi}}]{2010A&A...514A...1I}
{Ishihara}, D., {Onaka}, T., {Kataza}, H., {et~al.} 2010, \aap, 514, A1

\bibitem[{{Jackson} {et~al.}(2004){Jackson}, {MacGregor}, \&
  {Skumanich}}]{2004ApJ...606.1196J}
{Jackson}, S., {MacGregor}, K.~B., \& {Skumanich}, A. 2004, \apj, 606, 1196

\bibitem[{{Janson} {et~al.}(2017){Janson}, {Durkan}, {Hippler}, {Dai},
  {Brandner}, {Schlieder}, {Bonnefoy}, \& {Henning}}]{2017A&A...599A..70J}
{Janson}, M., {Durkan}, S., {Hippler}, S., {et~al.} 2017, \aap, 599, A70

\bibitem[{{Jones} {et~al.}(2017){Jones}, {Brahm}, {Wittenmyer}, {Drass},
  {Jenkins}, {Melo}, {Vos}, \& {Rojo}}]{jones2017}
{Jones}, M.~I., {Brahm}, R., {Wittenmyer}, R.~A., {et~al.} 2017, \aap, 602, A58

\bibitem[{{Kanaan} {et~al.}(2008){Kanaan}, {Meilland}, {Stee}, {Zorec},
  {Domiciano de Souza}, {Fr{\'e}mat}, \& {Briot}}]{2008A&A...486..785K}
{Kanaan}, S., {Meilland}, A., {Stee}, P., {et~al.} 2008, \aap, 486, 785

\bibitem[{{Kaufer} \& {Pasquini}(1998)}]{kaufer1998}
{Kaufer}, A. \& {Pasquini}, L. 1998, in \procspie, Vol. 3355, Optical
  Astronomical Instrumentation, ed. S.~{D'Odorico}, 844--854

\bibitem[{{Kervella} {et~al.}(2022){Kervella}, {Arenou}, \&
  {Th{\'e}venin}}]{2022A&A...657A...7K}
{Kervella}, P., {Arenou}, F., \& {Th{\'e}venin}, F. 2022, \aap, 657, A7

\bibitem[{{Kervella} \& {Domiciano de Souza}(2006)}]{2006A&A...453.1059K}
{Kervella}, P. \& {Domiciano de Souza}, A. 2006, \aap, 453, 1059

\bibitem[{{Kervella} \& {Domiciano de Souza}(2007)}]{2007A&A...474L..49K}
{Kervella}, P. \& {Domiciano de Souza}, A. 2007, \aap, 474, L49

\bibitem[{{Kervella} {et~al.}(2008){Kervella}, {Domiciano de Souza}, \&
  {Bendjoya}}]{kervella2008}
{Kervella}, P., {Domiciano de Souza}, A., \& {Bendjoya}, P. 2008, Astronomy and
  Astrophysics, 484, L13

\bibitem[{{Kervella} {et~al.}(2009){Kervella}, {Domiciano de Souza}, {Kanaan},
  {Meilland}, {Spang}, \& {Stee}}]{2009A&A...493L..53K}
{Kervella}, P., {Domiciano de Souza}, A., {Kanaan}, S., {et~al.} 2009, \aap,
  493, L53

\bibitem[{{Kervella} {et~al.}(2019{\natexlab{a}}){Kervella}, {Gallenne},
  {Evans}, {Szabados}, {Arenou}, {M{\'e}rand }, {Nardetto}, {Gieren}, \&
  {Pietrzynski}}]{2019A&A...623A.117K}
{Kervella}, P., {Gallenne}, A., {Evans}, N.~R., {et~al.} 2019{\natexlab{a}},
  \aap, 623, A117

\bibitem[{{Kervella} {et~al.}(2019{\natexlab{b}}){Kervella}, {Gallenne},
  {Evans}, {Szabados}, {Arenou}, {M{\'e}rand }, {Proto}, {Karczmarek},
  {Nardetto}, {Gieren}, \& {Pietrzynski}}]{2019A&A...623A.116K}
{Kervella}, P., {Gallenne}, A., {Evans}, N.~R., {et~al.} 2019{\natexlab{b}},
  \aap, 623, A116

\bibitem[{{Klement} {et~al.}(2019){Klement}, {Carciofi}, {Rivinius}, {Ignace},
  {Matthews}, {Torstensson}, {Gies}, {Vieira}, {Richardson}, {Domiciano de
  Souza}, {Bjorkman}, {Hallinan}, {Faes}, {Mota}, {Gullingsrud}, {de Breuck},
  {Kervella}, {Cur{\'e}}, \& {Gunawan}}]{2019ApJ...885..147K}
{Klement}, R., {Carciofi}, A.~C., {Rivinius}, T., {et~al.} 2019, \apj, 885, 147

\bibitem[{{Klement} {et~al.}(2022){Klement}, {Schaefer}, {Gies}, {Wang},
  {Baade}, {Rivinius}, {Gallenne}, {Carciofi}, {Monnier}, {M{\'e}rand},
  {Anugu}, {Kraus}, {Davies}, {Lanthermann}, {Gardner}, {Wysocki}, {Ennis},
  {Labdon}, {Setterholm}, \& {Le Bouquin}}]{2022ApJ...926..213K}
{Klement}, R., {Schaefer}, G.~H., {Gies}, D.~R., {et~al.} 2022, \apj, 926, 213

\bibitem[{{Kraus} {et~al.}(2014){Kraus}, {Shkolnik}, {Allers}, \&
  {Liu}}]{2014AJ....147..146K}
{Kraus}, A.~L., {Shkolnik}, E.~L., {Allers}, K.~N., \& {Liu}, M.~C. 2014, \aj,
  147, 146

\bibitem[{{Lacour} {et~al.}(2011){Lacour}, {Tuthill}, {Amico}, {Ireland},
  {Ehrenreich}, {Huelamo}, \& {Lagrange}}]{2011A&A...532A..72L}
{Lacour}, S., {Tuthill}, P., {Amico}, P., {et~al.} 2011, \aap, 532, A72

\bibitem[{{Lapeyrere} {et~al.}(2014){Lapeyrere}, {Kervella}, {Lacour},
  {Azouaoui}, {Garcia-Dabo}, {Perrin}, {Eisenhauer}, {Perraut}, {Straubmeier},
  {Amorim}, \& {Brandner}}]{2014SPIE.9146E..2DL}
{Lapeyrere}, V., {Kervella}, P., {Lacour}, S., {et~al.} 2014, in Society of
  Photo-Optical Instrumentation Engineers (SPIE) Conference Series, Vol. 9146,
  Optical and Infrared Interferometry IV, ed. J.~K. {Rajagopal}, M.~J.
  {Creech-Eakman}, \& F.~{Malbet}, 91462D

\bibitem[{{Le Bouquin} {et~al.}(2011){Le Bouquin}, {Berger}, {Lazareff},
  {Zins}, {Haguenauer}, {Jocou}, {Kern}, {Millan-Gabet}, {Traub}, {Absil},
  {Augereau}, {Benisty}, {Blind}, {Bonfils}, {Bourget}, {Delboulbe},
  {Feautrier}, {Germain}, {Gitton}, {Gillier}, {Kiekebusch}, {Kluska},
  {Knudstrup}, {Labeye}, {Lizon}, {Monin}, {Magnard}, {Malbet}, {Maurel},
  {M{\'e}nard}, {Micallef}, {Michaud}, {Montagnier}, {Morel}, {Moulin},
  {Perraut}, {Popovic}, {Rabou}, {Rochat}, {Rojas}, {Roussel}, {Roux},
  {Stadler}, {Stefl}, {Tatulli}, \& {Ventura}}]{lebouquin2011}
{Le Bouquin}, J.~B., {Berger}, J.~P., {Lazareff}, B., {et~al.} 2011, \aap, 535,
  A67

\bibitem[{{Leister} {et~al.}(2000){Leister}, {Janot-Pacheco}, {Leyton Z.~J.},
  {Hubert}, \& {Floquet}}]{2000ASPC..214..272L}
{Leister}, N.~V., {Janot-Pacheco}, E., {Leyton Z.~J.}, {Hubert}, A.~M., \&
  {Floquet}, M. 2000, in Astronomical Society of the Pacific Conference Series,
  Vol. 214, IAU Colloq. 175: The Be Phenomenon in Early-Type Stars, ed. M.~A.
  {Smith}, H.~F. {Henrichs}, \& J.~{Fabregat}, 272

\bibitem[{{Lenzen} {et~al.}(2003){Lenzen}, {Hartung}, {Brandner}, {Finger},
  {Hubin}, {Lacombe}, {Lagrange}, {Lehnert}, {Moorwood}, \&
  {Mouillet}}]{2003SPIE.4841..944L}
{Lenzen}, R., {Hartung}, M., {Brandner}, W., {et~al.} 2003, in Society of
  Photo-Optical Instrumentation Engineers (SPIE) Conference Series, Vol. 4841,
  Instrument Design and Performance for Optical/Infrared Ground-based
  Telescopes, ed. M.~{Iye} \& A.~F.~M. {Moorwood}, 944--952

\bibitem[{{Lenzen} {et~al.}(1998){Lenzen}, {Hofmann}, {Bizenberger}, \&
  {Tusche}}]{1998SPIE.3354..606L}
{Lenzen}, R., {Hofmann}, R., {Bizenberger}, P., \& {Tusche}, A. 1998, in
  Society of Photo-Optical Instrumentation Engineers (SPIE) Conference Series,
  ed. A.~M. {Fowler}, Vol. 3354, 606--614

\bibitem[{{Lin} \& {Ogilvie}(2017)}]{2017MNRAS.468.1387L}
{Lin}, Y. \& {Ogilvie}, G.~I. 2017, \mnras, 468, 1387

\bibitem[{{Lopez} {et~al.}(2022){Lopez}, {Lagarde}, {Petrov}, {Jaffe},
  {Antonelli, P.}, {Allouche, F.}, {Berio, P.}, {Matter, A.}, {Meilland, A.},
  {Millour, F.}, {Robbe-Dubois, S.}, {Henning, Th.}, {Weigelt, G.},
  {Glindemann, A.}, {Agocs, T.}, {Bailet, Ch.}, {Beckmann, U.}, {Bettonvil,
  F.}, {van Boekel, R.}, {Bourget, P.}, {Bresson, Y.}, {Bristow, P.},
  {Cruzal\`ebes, P.}, {Eldswijk, E.}, {Fante\"{\i} Caujolle, Y.}, {Gonz\'alez
  Herrera, J. C.}, {Graser, U.}, {Guajardo, P.}, {Heininger, M.}, {Hofmann,
  K.-H.}, {Kroes, G.}, {Laun, W.}, {Lehmitz, M.}, {Leinert, C.}, {Meisenheimer,
  K.}, {Morel, S.}, {Neumann, U.}, {Paladini, C.}, {Percheron, I.}, {Riquelme,
  M.}, {Schoeller, M.}, {Stee, Ph.}, {Venema, L.}, {Woillez, J.}, {Zins, G.},
  {\'Abrah\'am, P.}, {Abadie, S.}, {Abuter, R.}, {Accardo, M.}, {Adler, T.},
  {Alonso, J.}, {Augereau, J.-C.}, {B\"ohm, A.}, {Bazin, G.}, {Beltran, J.},
  {Bensberg, A.}, {Boland, W.}, {Brast, R.}, {Burtscher, L.}, {Castillo, R.},
  {Chelli, A.}, {Cid, C.}, {Clausse, J.-M.}, {Connot, C.}, {Conzelmann, R. D.},
  {Danchi, W.-C.}, {Delbo, M.}, {Drevon, J.}, {Dominik, C.}, {van Duin, A.},
  {Ebert, M.}, {Eisenhauer, F.}, {Flament, S.}, {Frahm, R.}, {G\'amez Rosas,
  V.}, {Gabasch, A.}, {Gallenne, A.}, {Garces, E.}, {Girard, P.}, {Glazenborg,
  A.}, {Gont\'e, F. Y. J.}, {Guitton, F.}, {de Haan, M.}, {Hanenburg, H.},
  {Haubois, X.}, {Hocd\'e, V.}, {Hogerheijde, M.}, {ter Horst, R.}, {Hron, J.},
  {Hummel, C. A.}, {Hubin, N.}, {Huerta, R.}, {Idserda, J.}, {Isbell, J. W.},
  {Ives, D.}, {Jakob, G.}, {Jask\'o, A.}, {Jochum, L.}, {Klarmann, L.}, {Klein,
  R.}, {Kragt, J.}, {Kuindersma, S.}, {Kokoulina, E.}, {Labadie, L.}, {Lacour,
  S.}, {Leftley, J.}, {Le Poole, R.}, {Lizon, J.-L.}, {Lopez, M.}, {Lykou, F.},
  {M\'erand, A.}, {Marcotto, A.}, {Mauclert, N.}, {Maurer, T.}, {Mehrgan, L.
  H.}, {Meisner, J.}, {Meixner, K.}, {Mellein, M.}, {Menut, J. L.}, {Mohr, L.},
  {Mosoni, L.}, {Navarro, R.}, {Nu\ss{}baum, E.}, {Pallanca, L.}, {Pantin, E.},
  {Pasquini, L.}, {Phan Duc, T.}, {Pott, J.-U.}, {Pozna, E.}, {Richichi, A.},
  {Ridinger, A.}, {Rigal, F.}, {Rivinius, Th.}, {Roelfsema, R.}, {Rohloff,
  R.-R.}, {Rousseau, S.}, {Salabert, D.}, {Schertl, D.}, {Schuhler, N.},
  {Schuil, M.}, {Shabun, K.}, {Soulain, A.}, {Stephan, C.}, {Toledo, P.},
  {Tristram, K.}, {Tromp, N.}, {Vakili, F.}, {Varga, J.}, {Vinther, J.},
  {Waters, L. B. F. M.}, {Wittkowski, M.}, {Wolf, S.}, {Wrhel, F.}, \& {Yoffe,
  G.}}]{matisse22}
{Lopez}, B., {Lagarde}, S., {Petrov}, R.~G., {et~al.} 2022, A\&A, 659, A192

\bibitem[{{Mann} {et~al.}(2015){Mann}, {Feiden}, {Gaidos}, {Boyajian}, \& {von
  Braun}}]{2015ApJ...804...64M}
{Mann}, A.~W., {Feiden}, G.~A., {Gaidos}, E., {Boyajian}, T., \& {von Braun},
  K. 2015, \apj, 804, 64

\bibitem[{{Meilland} {et~al.}(2012){Meilland}, {Millour}, {Kanaan}, {Stee},
  {Petrov}, {Hofmann}, {Natta}, \& {Perraut}}]{2012A&A...538A.110M}
{Meilland}, A., {Millour}, F., {Kanaan}, S., {et~al.} 2012, \aap, 538, A110

\bibitem[{{M{\'e}rand}(2022{\natexlab{a}})}]{2022arXiv220711047M}
{M{\'e}rand}, A. 2022{\natexlab{a}}, arXiv e-prints, arXiv:2207.11047

\bibitem[{{M{\'e}rand}(2022{\natexlab{b}})}]{2022ascl.soft05001M}
{M{\'e}rand}, A. 2022{\natexlab{b}}, {PMOIRED: Parametric Modeling of Optical
  Interferometric Data}, Astrophysics Source Code Library, record ascl:2205.001

\bibitem[{{M{\'e}rand} {et~al.}(2014){M{\'e}rand}, {Abuter},
  {Aller-Carpentier}, {Andolfato}, {Alonso}, {Berger}, {Blanchard}, {Boffin},
  {Bourget}, {Bristow}, {Cid}, {de Wit}, {del Valle},
  {Delplancke-Str{\"o}bele}, {Derie}, {Faundez}, {Ertel}, {Grellmann},
  {Gitton}, {Glindemann}, {Guajardo}, {Guieu}, {Guisard}, {Guniat},
  {Haguenauer}, {Herrera}, {Hummel}, {La Fuente}, {Lopez}, {Mardones}, {Morel},
  {M{\"u}ller}, {Percheron}, {Duc}, {Pino}, {Poupar}, {Pozna}, {Ramirez},
  {Rengaswamy}, {Rivas}, {Rivinius}, {Segovia}, {Schmid}, {Sch{\"o}ller},
  {Schuhler}, {Woillez}, \& {Wittkowski}}]{2014SPIE.9146E..0JM}
{M{\'e}rand}, A., {Abuter}, R., {Aller-Carpentier}, E., {et~al.} 2014, in
  \procspie, Vol. 9146, Optical and Infrared Interferometry IV, 91460J

\bibitem[{{Morel} \& {Magnenat}(1978)}]{1978A&AS...34..477M}
{Morel}, M. \& {Magnenat}, P. 1978, \aaps, 34, 477

\bibitem[{{Mourard} {et~al.}(2015){Mourard}, {Monnier}, {Meilland}, {Gies},
  {Millour}, {Benisty}, {Che}, {Grundstrom}, {Ligi}, {Schaefer}, {Baron},
  {Kraus}, {Zhao}, {Pedretti}, {Berio}, {Clausse}, {Nardetto}, {Perraut},
  {Spang}, {Stee}, {Tallon-Bosc}, {McAlister}, {ten Brummelaar}, {Ridgway},
  {Sturmann}, {Sturmann}, {Turner}, \& {Farrington}}]{2015A&A...577A..51M}
{Mourard}, D., {Monnier}, J.~D., {Meilland}, A., {et~al.} 2015, \aap, 577, A51

\bibitem[{{Nardetto} {et~al.}(2006){Nardetto}, {Mourard}, {Kervella},
  {Mathias}, {M{\'e}rand}, \& {Bersier}}]{nardetto2006}
{Nardetto}, N., {Mourard}, D., {Kervella}, P., {et~al.} 2006, \aap, 453, 309

\bibitem[{{Neiner} {et~al.}(2011){Neiner}, {de Batz}, {Cochard}, {Floquet},
  {Mekkas}, \& {Desnoux}}]{2011AJ....142..149N}
{Neiner}, C., {de Batz}, B., {Cochard}, F., {et~al.} 2011, \aj, 142, 149

\bibitem[{{Neuh{\"a}user} {et~al.}(2008){Neuh{\"a}user}, {Mugrauer},
  {Seifahrt}, {Schmidt}, \& {Vogt}}]{2008A&A...484..281N}
{Neuh{\"a}user}, R., {Mugrauer}, M., {Seifahrt}, A., {Schmidt}, T.~O.~B., \&
  {Vogt}, N. 2008, \aap, 484, 281

\bibitem[{{Ochsenbein} {et~al.}(2000){Ochsenbein}, {Bauer}, \&
  {Marcout}}]{2000A&AS..143...23O}
{Ochsenbein}, F., {Bauer}, P., \& {Marcout}, J. 2000, \aaps, 143, 23

\bibitem[{{Oudmaijer} \& {Parr}(2010)}]{2010MNRAS.405.2439O}
{Oudmaijer}, R.~D. \& {Parr}, A.~M. 2010, \mnras, 405, 2439

\bibitem[{{Pecaut} \& {Mamajek}(2013)}]{2013ApJS..208....9P}
{Pecaut}, M.~J. \& {Mamajek}, E.~E. 2013, \apjs, 208, 9

\bibitem[{{Pecaut} {et~al.}(2012){Pecaut}, {Mamajek}, \&
  {Bubar}}]{2012ApJ...746..154P}
{Pecaut}, M.~J., {Mamajek}, E.~E., \& {Bubar}, E.~J. 2012, \apj, 746, 154

\bibitem[{{Pepe} {et~al.}(2002){Pepe}, {Mayor}, {Rupprecht}, {Avila},
  {Ballester}, {Beckers}, {Benz}, {Bertaux}, {Bouchy}, {Buzzoni}, {Cavadore},
  {Deiries}, {Dekker}, {Delabre}, {D'Odorico}, {Eckert}, {Fischer}, {Fleury},
  {George}, {Gilliotte}, {Gojak}, {Guzman}, {Koch}, {Kohler}, {Kotzlowski},
  {Lacroix}, {Le Merrer}, {Lizon}, {Lo Curto}, {Longinotti}, {Megevand},
  {Pasquini}, {Petitpas}, {Pichard}, {Queloz}, {Reyes}, {Richaud}, {Sivan},
  {Sosnowska}, {Soto}, {Udry}, {Ureta}, {van Kesteren}, {Weber}, {Weilenmann},
  {Wicenec}, {Wieland}, {Christensen-Dalsgaard}, {Dravins}, {Hatzes},
  {K{\"u}rster}, {Paresce}, \& {Penny}}]{pepe2002}
{Pepe}, F., {Mayor}, M., {Rupprecht}, G., {et~al.} 2002, The Messenger, 110, 9

\bibitem[{{Perryman} {et~al.}(1997){Perryman}, {Lindegren}, {Kovalevsky},
  {Hoeg}, {Bastian}, {Bernacca}, {Cr{\'e}z{\'e}}, {Donati}, {Grenon},
  {Grewing}, {van Leeuwen}, {van der Marel}, {Mignard}, {Murray}, {Le Poole},
  {Schrijver}, {Turon}, {Arenou}, {Froeschl{\'e}}, \&
  {Petersen}}]{1997A&A...323L..49P}
{Perryman}, M.~A.~C., {Lindegren}, L., {Kovalevsky}, J., {et~al.} 1997, \aap,
  323, L49

\bibitem[{{Petrov} {et~al.}(2007){Petrov}, {Malbet}, {Weigelt}, {Antonelli},
  {Beckmann}, {Bresson}, {Chelli}, {Dugu{\'e}}, {Duvert}, {Gennari},
  {Gl{\"u}ck}, {Kern}, {Lagarde}, {Le Coarer}, {Lisi}, {Millour}, {Perraut},
  {Puget}, {Rantakyr{\"o}}, {Robbe-Dubois}, {Roussel}, {Salinari}, {Tatulli},
  {Zins}, {Accardo}, {Acke}, {Agabi}, {Altariba}, {Arezki}, {Aristidi},
  {Baffa}, {Behrend}, {Bl{\"o}cker}, {Bonhomme}, {Busoni}, {Cassaing},
  {Clausse}, {Colin}, {Connot}, {Delboulb{\'e}}, {Domiciano de Souza},
  {Driebe}, {Feautrier}, {Ferruzzi}, {Forveille}, {Fossat}, {Foy},
  {Fraix-Burnet}, {Gallardo}, {Giani}, {Gil}, {Glentzlin}, {Heiden},
  {Heininger}, {Hernandez Utrera}, {Hofmann}, {Kamm}, {Kiekebusch}, {Kraus},
  {Le Contel}, {Le Contel}, {Lesourd}, {Lopez}, {Lopez}, {Magnard}, {Marconi},
  {Mars}, {Martinot-Lagarde}, {Mathias}, {M{\`e}ge}, {Monin}, {Mouillet},
  {Mourard}, {Nussbaum}, {Ohnaka}, {Pacheco}, {Perrier}, {Rabbia}, {Rebattu},
  {Reynaud}, {Richichi}, {Robini}, {Sacchettini}, {Schertl}, {Sch{\"o}ller},
  {Solscheid}, {Spang}, {Stee}, {Stefanini}, {Tallon}, {Tallon-Bosc}, {Tasso},
  {Testi}, {Vakili}, {von der L{\"u}he}, {Valtier}, {Vannier}, \&
  {Ventura}}]{Petrov2007_v464p1}
{Petrov}, R.~G., {Malbet}, F., {Weigelt}, G., {et~al.} 2007, \aap, 464, 1

\bibitem[{{Pietrzy{\'n}ski} {et~al.}(2010){Pietrzy{\'n}ski}, {Thompson},
  {Gieren}, {Graczyk}, {Bono}, {Udalski}, {Soszy{\'n}ski}, {Minniti}, \&
  {Pilecki}}]{2010Natur.468..542P}
{Pietrzy{\'n}ski}, G., {Thompson}, I.~B., {Gieren}, W., {et~al.} 2010, \nat,
  468, 542

\bibitem[{{Pilecki} {et~al.}(2018){Pilecki}, {Gieren}, {Pietrzy{\'n}ski},
  {Thompson}, {Smolec}, {Graczyk}, {Taormina}, {Udalski}, {Storm}, {Nardetto},
  {Gallenne}, {Kervella}, {Soszy{\'n}ski}, {G{\'o}rski}, {Wielg{\'o}rski},
  {Suchomska}, {Karczmarek}, \& {Zgirski}}]{2018ApJ...862...43P}
{Pilecki}, B., {Gieren}, W., {Pietrzy{\'n}ski}, G., {et~al.} 2018, \apj, 862,
  43

\bibitem[{{Pilecki} {et~al.}(2017){Pilecki}, {Gieren}, {Smolec},
  {Pietrzy{\'n}ski}, {Thompson}, {Anderson}, {Bono}, {Soszy{\'n}ski},
  {Kervella}, {Nardetto}, {Taormina}, {St{\c{e}}pie{\'n}}, \&
  {Wielg{\'o}rski}}]{2017ApJ...842..110P}
{Pilecki}, B., {Gieren}, W., {Smolec}, R., {et~al.} 2017, \apj, 842, 110

\bibitem[{{Pilecki} {et~al.}(2015){Pilecki}, {Graczyk}, {Gieren},
  {Pietrzy{\'n}ski}, {Thompson}, {Smolec}, {Udalski}, {Soszy{\'n}ski},
  {Konorski}, {Taormina}, {Gallenne}, {Minniti}, \&
  {Catelan}}]{2015ApJ...806...29P}
{Pilecki}, B., {Graczyk}, D., {Gieren}, W., {et~al.} 2015, \apj, 806, 29

\bibitem[{{Pilecki} {et~al.}(2013){Pilecki}, {Graczyk}, {Pietrzy{\'n}ski},
  {Gieren}, {Thompson}, {Freedman}, {Scowcroft}, {Madore}, {Udalski},
  {Soszy{\'n}ski}, {Konorski}, {Smolec}, {Nardetto}, {Bono}, {Prada Moroni},
  {Storm}, \& {Gallenne}}]{2013MNRAS.436..953P}
{Pilecki}, B., {Graczyk}, D., {Pietrzy{\'n}ski}, G., {et~al.} 2013, \mnras,
  436, 953

\bibitem[{{Porter} \& {Rivinius}(2003)}]{2003PASP..115.1153P}
{Porter}, J.~M. \& {Rivinius}, T. 2003, \pasp, 115, 1153

\bibitem[{{Queloz} {et~al.}(2001){Queloz}, {Mayor}, {Udry}, {Burnet},
  {Carrier}, {Eggenberger}, {Naef}, {Santos}, {Pepe}, {Rupprecht}, {Avila},
  {Baeza}, {Benz}, {Bertaux}, {Bouchy}, {Cavadore}, {Delabre}, {Eckert},
  {Fischer}, {Fleury}, {Gilliotte}, {Goyak}, {Guzman}, {Kohler}, {Lacroix},
  {Lizon}, {Megevand}, {Sivan}, {Sosnowska}, \& {Weilenmann}}]{queloz2001}
{Queloz}, D., {Mayor}, M., {Udry}, S., {et~al.} 2001, The Messenger, 105, 1

\bibitem[{{Rivinius} {et~al.}(2013{\natexlab{a}}){Rivinius}, {Baade},
  {Townsend}, {Carciofi}, \& {{\v{S}}tefl}}]{2013A&A...559L...4R}
{Rivinius}, T., {Baade}, D., {Townsend}, R.~H.~D., {Carciofi}, A.~C., \&
  {{\v{S}}tefl}, S. 2013{\natexlab{a}}, \aap, 559, L4

\bibitem[{{Rivinius} {et~al.}(2003){Rivinius}, {Baade}, \&
  {{\v{S}}tefl}}]{2003A&A...411..229R}
{Rivinius}, T., {Baade}, D., \& {{\v{S}}tefl}, S. 2003, \aap, 411, 229

\bibitem[{{Rivinius} {et~al.}(2013{\natexlab{b}}){Rivinius}, {Carciofi}, \&
  {Martayan}}]{2013A&ARv..21...69R}
{Rivinius}, T., {Carciofi}, A.~C., \& {Martayan}, C. 2013{\natexlab{b}}, \aapr,
  21, 69

\bibitem[{{Rousset} {et~al.}(2003){Rousset}, {Lacombe}, {Puget}, {Hubin},
  {Gendron}, {Fusco}, {Arsenault}, {Charton}, {Feautrier}, {Gigan}, {Kern},
  {Lagrange}, {Madec}, {Mouillet}, {Rabaud}, {Rabou}, {Stadler}, \&
  {Zins}}]{2003SPIE.4839..140R}
{Rousset}, G., {Lacombe}, F., {Puget}, P., {et~al.} 2003, in Society of
  Photo-Optical Instrumentation Engineers (SPIE) Conference Series, ed. P.~L.
  {Wizinowich} \& D.~{Bonaccini}, Vol. 4839, 140--149

\bibitem[{{Royer} {et~al.}(2007){Royer}, {Zorec}, \&
  {G{\'o}mez}}]{2007A&A...463..671R}
{Royer}, F., {Zorec}, J., \& {G{\'o}mez}, A.~E. 2007, \aap, 463, 671

\bibitem[{{Saad} {et~al.}(2021){Saad}, {Nouh}, {Shokry}, \& {Zead}}]{saad2020}
{Saad}, S.~M., {Nouh}, M.~I., {Shokry}, A., \& {Zead}, I. 2021, \rmxaa, 57, 91

\bibitem[{{Sahlmann} {et~al.}(2018){Sahlmann}, {Mora}, {Mart{\'\i}n-Fleitas},
  {Abreu}, {Crowley}, \& {Fink}}]{2018IAUS..330..343S}
{Sahlmann}, J., {Mora}, A., {Mart{\'\i}n-Fleitas}, J.~M., {et~al.} 2018, in
  Astrometry and Astrophysics in the Gaia Sky, ed. A.~{Recio-Blanco}, P.~{de
  Laverny}, A.~G.~A. {Brown}, \& T.~{Prusti}, Vol. 330, 343--344

\bibitem[{Sansonetti \& Martin(2005)}]{sansonetti2005}
Sansonetti, J.~E. \& Martin, W.~C. 2005, Journal of Physical and Chemical
  Reference Data, 34, 1559

\bibitem[{{Shao} \& {Li}(2021)}]{2021ApJ...908...67S}
{Shao}, Y. \& {Li}, X.-D. 2021, \apj, 908, 67

\bibitem[{{Skrutskie} {et~al.}(2006){Skrutskie}, {Cutri}, {Stiening},
  {Weinberg}, {Schneider}, {Carpenter}, {Beichman}, {Capps}, {Chester},
  {Elias}, {Huchra}, {Liebert}, {Lonsdale}, {Monet}, {Price}, {Seitzer},
  {Jarrett}, {Kirkpatrick}, {Gizis}, {Howard}, {Evans}, {Fowler}, {Fullmer},
  {Hurt}, {Light}, {Kopan}, {Marsh}, {McCallon}, {Tam}, {Van Dyk}, \&
  {Wheelock}}]{2006AJ....131.1163S}
{Skrutskie}, M.~F., {Cutri}, R.~M., {Stiening}, R., {et~al.} 2006, \aj, 131,
  1163

\bibitem[{{Steiner} {et~al.}(2006){Steiner}, {Seifert}, {Stahl}, {Lemke},
  {Chini}, \& {Appenzeller}}]{steiner2006}
{Steiner}, I., {Seifert}, W., {Stahl}, O., {et~al.} 2006, in \procspie, Vol.
  6269, Society of Photo-Optical Instrumentation Engineers (SPIE) Conference
  Series, 62692W

\bibitem[{{Suffak} {et~al.}(2022){Suffak}, {Jones}, \&
  {Carciofi}}]{2022MNRAS.509..931S}
{Suffak}, M., {Jones}, C.~E., \& {Carciofi}, A.~C. 2022, \mnras, 509, 931

\bibitem[{{Sybilski} {et~al.}(2018){Sybilski}, {Paw{\l}aszek}, {Sybilska},
  {Konacki}, {He{\l}miniak}, {Koz{\l}owski}, \&
  {Ratajczak}}]{2018MNRAS.478.1942S}
{Sybilski}, P., {Paw{\l}aszek}, R.~K., {Sybilska}, A., {et~al.} 2018, \mnras,
  478, 1942

\bibitem[{{Tallon-Bosc} {et~al.}(2008){Tallon-Bosc}, {Tallon}, {Thi{\'e}baut},
  {B{\'e}chet}, {Mella}, {Lafrasse}, {Chesneau}, {Domiciano de Souza},
  {Duvert}, {Mourard}, {Petrov}, \& {Vannier}}]{2008SPIE.7013E..1JT}
{Tallon-Bosc}, I., {Tallon}, M., {Thi{\'e}baut}, E., {et~al.} 2008, in Society
  of Photo-Optical Instrumentation Engineers (SPIE) Conference Series, Vol.
  7013, Optical and Infrared Interferometry, ed. M.~{Sch{\"o}ller}, W.~C.
  {Danchi}, \& F.~{Delplancke}, 70131J

\bibitem[{{Tatulli} {et~al.}(2007){Tatulli}, {Millour}, {Chelli}, {Duvert},
  {Acke}, {Hernandez Utrera}, {Hofmann}, {Kraus}, {Malbet}, {M{\`e}ge},
  {Petrov}, {Vannier}, {Zins}, {Antonelli}, {Beckmann}, {Bresson}, {Dugu{\'e}},
  {Gennari}, {Gl{\"u}ck}, {Kern}, {Lagarde}, {Le Coarer}, {Lisi}, {Perraut},
  {Puget}, {Rantakyr{\"o}}, {Robbe-Dubois}, {Roussel}, {Weigelt}, {Accardo},
  {Agabi}, {Altariba}, {Arezki}, {Aristidi}, {Baffa}, {Behrend}, {Bl{\"o}cker},
  {Bonhomme}, {Busoni}, {Cassaing}, {Clausse}, {Colin}, {Connot},
  {Delboulb{\'e}}, {Domiciano de Souza}, {Driebe}, {Feautrier}, {Ferruzzi},
  {Forveille}, {Fossat}, {Foy}, {Fraix-Burnet}, {Gallardo}, {Giani}, {Gil},
  {Glentzlin}, {Heiden}, {Heininger}, {Kamm}, {Kiekebusch}, {Le Contel}, {Le
  Contel}, {Lesourd}, {Lopez}, {Lopez}, {Magnard}, {Marconi}, {Mars},
  {Martinot-Lagarde}, {Mathias}, {Monin}, {Mouillet}, {Mourard}, {Nussbaum},
  {Ohnaka}, {Pacheco}, {Perrier}, {Rabbia}, {Rebattu}, {Reynaud}, {Richichi},
  {Robini}, {Sacchettini}, {Schertl}, {Sch{\"o}ller}, {Solscheid}, {Spang},
  {Stee}, {Stefanini}, {Tallon}, {Tallon-Bosc}, {Tasso}, {Testi}, {Vakili},
  {von der L{\"u}he}, {Valtier}, \& {Ventura}}]{Tatulli2007_v464p29}
{Tatulli}, E., {Millour}, F., {Chelli}, A., {et~al.} 2007, \aap, 464, 29

\bibitem[{{Tokovinin} {et~al.}(2013){Tokovinin}, {Fischer}, {Bonati},
  {Giguere}, {Moore}, {Schwab}, {Spronck}, \& {Szymkowiak}}]{tokovinin2013}
{Tokovinin}, A., {Fischer}, D.~A., {Bonati}, M., {et~al.} 2013, \pasp, 125,
  1336

\bibitem[{{Trahin} {et~al.}(2021){Trahin}, {Breuval}, {Kervella}, {M{\'e}rand},
  {Nardetto}, {Gallenne}, {Hocd{\'e}}, \& {Gieren}}]{2021A&A...656A.102T}
{Trahin}, B., {Breuval}, L., {Kervella}, P., {et~al.} 2021, \aap, 656, A102

\bibitem[{{Tuthill} {et~al.}(2010){Tuthill}, {Lacour}, {Amico}, {Ireland},
  {Norris}, {Stewart}, {Evans}, {Kraus}, {Lidman}, {Pompei}, \&
  {Kornweibel}}]{2010SPIE.7735E..1OT}
{Tuthill}, P., {Lacour}, S., {Amico}, P., {et~al.} 2010, in Society of
  Photo-Optical Instrumentation Engineers (SPIE) Conference Series, Vol. 7735,
  Ground-based and Airborne Instrumentation for Astronomy III, ed. I.~S.
  {McLean}, S.~K. {Ramsay}, \& H.~{Takami}, 77351O

\bibitem[{{Tuthill} {et~al.}(2000){Tuthill}, {Monnier}, {Danchi}, {Wishnow}, \&
  {Haniff}}]{2000PASP..112..555T}
{Tuthill}, P.~G., {Monnier}, J.~D., {Danchi}, W.~C., {Wishnow}, E.~H., \&
  {Haniff}, C.~A. 2000, \pasp, 112, 555

\bibitem[{{van Belle}(2012)}]{2012A&ARv..20...51V}
{van Belle}, G.~T. 2012, \aapr, 20, 51

\bibitem[{{van Leeuwen}(2007{\natexlab{a}})}]{2007ASSL..350.....V}
{van Leeuwen}, F., ed. 2007{\natexlab{a}}, Astrophysics and Space Science
  Library, Vol. 350, {Hipparcos, the New Reduction of the Raw Data}

\bibitem[{{van Leeuwen}(2007{\natexlab{b}})}]{2007A&A...474..653V}
{van Leeuwen}, F. 2007{\natexlab{b}}, \aap, 474, 653

\bibitem[{{Vanzi} {et~al.}(2012){Vanzi}, {Chacon}, {Helminiak}, {Baffico},
  {Rivinius}, {{\v{S}}tefl}, {Baade}, {Avila}, \& {Guirao}}]{vanzi2012}
{Vanzi}, L., {Chacon}, J., {Helminiak}, K.~G., {et~al.} 2012, \mnras, 424, 2770

\bibitem[{{Vinicius} {et~al.}(2006){Vinicius}, {Zorec}, {Leister}, \&
  {Levenhagen}}]{2006A&A...446..643V}
{Vinicius}, M.~M.~F., {Zorec}, J., {Leister}, N.~V., \& {Levenhagen}, R.~S.
  2006, \aap, 446, 643

\bibitem[{Virtanen {et~al.}(2020)Virtanen, Gommers, Oliphant, Haberland, Reddy,
  Cournapeau, Burovski, Peterson, Weckesser, Bright, {van der Walt}, Brett,
  Wilson, Millman, Mayorov, Nelson, Jones, Kern, Larson, Carey, Polat, Feng,
  Moore, {VanderPlas}, Laxalde, Perktold, Cimrman, Henriksen, Quintero, Harris,
  Archibald, Ribeiro, Pedregosa, {van Mulbregt}, \& {SciPy 1.0
  Contributors}}]{scipy}
Virtanen, P., Gommers, R., Oliphant, T.~E., {et~al.} 2020, Nature Methods, 17,
  261

\bibitem[{{Wade} {et~al.}(2017){Wade}, {Shultz}, {Sikora}, {Bernier},
  {Rivinius}, {Alecian}, {Petit}, {Grunhut}, \& {BinaMIcS
  Collaboration}}]{wade2017}
{Wade}, G.~A., {Shultz}, M., {Sikora}, J., {et~al.} 2017, Monthly Notices of
  the Royal Astronomical Society, 465, 2517

\bibitem[{Wiese \& Fuhr(2009)}]{wiese2009}
Wiese, W.~L. \& Fuhr, J.~R. 2009, Journal of Physical and Chemical Reference
  Data, 38, 565

\bibitem[{{Woillez} {et~al.}(2018){Woillez}, {Darr{\'e}}, {Egner}, {Gont{\'e}},
  {Haubois}, {M{\'e}rand}, {Schuhler}, {Abad}, {Abuter}, {Aller-Carpentier},
  {Alonso}, {Andolfato}, {Barriga}, {Beltran}, {Berger}, {Bourget}, {Brast},
  {Bristow}, {Caniguante}, {Conzelmann}, {Cortes}, {Delboulb{\'e}},
  {Delplancke-Str{\"o}bele}, {Del Valle}, {Dembet}, {Derie}, {Donoso},
  {Duhoux}, {Dupuy}, {F{\"o}rster}, {Fuenteseca}, {Frahm}, {Gaytan},
  {Glindemann}, {Gonzales}, {Guerlet}, {Guieu}, {Guisard}, {Gutierrez},
  {Haguenauer}, {Haimerl}, {van der Heyden}, {Huber}, {Hubin}, {Jochum},
  {Jocou}, {Jolley}, {Kirchbauer}, {Kolb}, {Kosmalski}, {Krempl}, {Le Bouquin},
  {L{\'e}v{\^e}que}, {Lilley}, {Lizon}, {Magnard}, {Mardones}, {Meister},
  {Moulin}, {Osorio}, {Ott}, {Pallanca}, {Paufique}, {Pavez}, {Pasquini},
  {Percheron}, {Pettazzi}, {Phan Duc}, {Pirard}, {Pino}, {Poupar}, {Quentin},
  {Ram{\'\i}rez}, {Reyes}, {Ridings}, {Rochat}, {Rivinius}, {Salgado},
  {Sch{\"o}ller}, {Shchekaturov}, {Stephan}, {Suarez}, {Tamblay}, {Tapia},
  {Tristram}, {Valdes}, {Verinaud}, {de Wit}, \& {Zins}}]{2018SPIE10701E..03W}
{Woillez}, J., {Darr{\'e}}, P., {Egner}, S., {et~al.} 2018, in Society of
  Photo-Optical Instrumentation Engineers (SPIE) Conference Series, Vol. 10701,
  Optical and Infrared Interferometry and Imaging VI, ed. M.~J.
  {Creech-Eakman}, P.~G. {Tuthill}, \& A.~{M{\'e}rand}, 1070103

\bibitem[{{Zahn} {et~al.}(2010){Zahn}, {Ranc}, \&
  {Morel}}]{2010A&A...517A...7Z}
{Zahn}, J.~P., {Ranc}, C., \& {Morel}, P. 2010, \aap, 517, A7

\bibitem[{{Zharikov} {et~al.}(2013){Zharikov}, {Miroshnichenko}, {Pollmann},
  {Danford}, {Bjorkman}, {Morrison}, {Favaro}, {Guarro Fl{\'o}}, {Terry},
  {Desnoux}, {Garrel}, {Martineau}, {Buchet}, {Ubaud}, {Mauclaire},
  {Kalbermatten}, {Buil}, {Sawicki}, {Blank}, \& {Garde}}]{2013A&A...560A..30Z}
{Zharikov}, S.~V., {Miroshnichenko}, A.~S., {Pollmann}, E., {et~al.} 2013,
  \aap, 560, A30

\bibitem[{{Zorec} \& {Royer}(2012)}]{2012A&A...537A.120Z}
{Zorec}, J. \& {Royer}, F. 2012, \aap, 537, A120

\end{thebibliography}
